\documentclass[11pt,a4paper]{article}
\pdfoutput=1
\usepackage{jheppub}
\usepackage{amsmath,amssymb,bm}
\usepackage[x11names]{xcolor}
\usepackage{slashed}
\usepackage{booktabs}
\usepackage[normalem]{ulem}

\hypersetup{
  pdftitle={Landau-Ginzburg description of an exceptional N=1 minimal model},
  pdfsubject={Landau-Ginzburg description of the (E6,D8) super-W3 minimal model},
  pdfkeywords={Landau-Ginzburg, super-W3, minimal model, epsilon expansion, fusion ring, emergent supersymmetry}}
\newcommand{\eps}{\epsilon}
\newcommand{\ed}{(E_6,D_8)}
\newcommand{\ZR}{\mathbb Z_2^{R}}
\newcommand{\Zex}{\mathbb Z_2^{\rm ex}}

\begin{document}

\newcommand{\lphi}{\ensuremath{\phi}}
\newcommand{\lpsi}{\ensuremath{\psi}}
\newcommand{\lpsit}{\ensuremath{\tilde\psi}}
\newcommand{\lF}{\ensuremath{F}}
\title{
Landau--Ginzburg description of\\
an exceptional ${\mathcal N}=1$ minimal model
}

\author[\lphi]{Yu Nakayama,}
\author[\lpsi]{Andrei Katsevich,}
\author[\lpsi,\lpsit]{Igor R. Klebanov,}
\author[\lF]{Zimo Sun}

\affiliation[\lphi]{Yukawa Institute for Theoretical Physics,
Kyoto University, Kitashirakawa Oiwakecho, Sakyo-ku, Kyoto 606-8502, Japan}
\affiliation[\lpsi]{Joseph Henry Laboratories and Leinweber Forum for Theoretical Physics, Princeton University, Princeton, NJ 08544, USA}
\affiliation[\lpsit]{Princeton Center for Theoretical Science, Princeton University, Princeton, NJ 08544, USA}
\affiliation[\lF]{Institute for Advanced Study, Princeton, NJ 08540, USA}

\emailAdd{yu.nakayama@yukawa.kyoto-u.ac.jp, akatsevich@princeton.edu, klebanov@princeton.edu, zimo@ias.edu}

\abstract{
The $\mathcal N=1$ superconformal minimal model with $m=12$ and the exceptional modular invariant $\ed$ is the unitary minimal model of the super-$W_3$ algebra. We propose its Landau--Ginzburg description using two real scalar superfields with the cubic superpotential ${\cal W}=g_1 XY^2/2 + g_2X^3/6$. For $g_1=g_2$, this superpotential is known to describe a product of two $m=3$ $\mathcal N=1$ superconformal minimal models, which is the $m=10$ model with the $(D_6,E_6)$ modular invariant. The exceptional $m=12$ superconformal minimal model is realized at a different fixed point of the same theory. Testing this Landau--Ginzburg description requires the fusion ring of the minimal model, which we obtain from the modular data of the extended algebra. The fusion ring has a $\mathbb Z_2$ grading by chiral fermion parity that the ordinary fusion coefficients do not determine. This grading, composed with conjugation, gives the generator of the R-parity $\ZR$ of the Landau--Ginzburg theory. We then treat the theory with superpotential $\cal W$ as a Gross--Neveu--Yukawa model in $d=4-\eps$ and find a weakly coupled infrared fixed point with $g_1/g_2=3/2+\mathcal O(\eps)$, at which supersymmetry emerges. We also describe the renormalization group flow from this fixed point to the decoupled fixed point with $g_1=g_2$. The operator dimensions at the coupled fixed point, continued to $d=2$, agree approximately with their values in the $m=12$ superconformal minimal model. Finally, we estimate the scaling dimensions in the new interacting $d=3$ $\mathcal N=1$ superconformal field theory.
}

\maketitle

\flushbottom

\newpage
\section{Introduction and summary}
\label{sec:intro}

Landau and Ginzburg taught us to describe a critical point by a Lagrangian for a local order parameter. Such a Lagrangian can be defined in a range of dimensions $d$, and the renormalization group (RG) treats $d$ as a parameter. Two Euclidean dimensions are special, because there conformal field theory (CFT) solves exactly an infinite family of critical points, the minimal models 
$M(p,q)$ discovered by Belavin, Polyakov and Zamolodchikov \cite{Belavin:1984vu} and their supersymmetric counterparts \cite{Friedan:1984rv,Bershadsky:1985dq}. A Landau--Ginzburg (LG) description 
can be tested against the exactly known $d=2$ conformal data and then continued to higher dimensions, where no exact solution is available.

The 
LG Lagrangian is obtained by a standard procedure. One assigns an order parameter to each of the most relevant primary fields, uses the operator algebra to identify the remaining relevant operators as composites of these order parameters, and obtains the equations of motion from the same algebra. Those equations fix the Lagrangian. Zamolodchikov applied this procedure to the unitary diagonal Virasoro minimal models $M(m,m+1)$ \cite{Zamolodchikov:1986db}. Each of them is described by a single scalar field with the potential $V\sim\phi^{2m-2}$. The same procedure applies to the unitary $\mathcal N=1$ superconformal minimal models $SM(m)$, labeled by an integer $m\ge3$. For the $A$ series, which has diagonal modular invariants, the description is the LG theory of one $\mathcal N=1$ scalar superfield $X$ with superpotential ${\cal W}\sim X^m$ \cite{Zamolodchikov:1986db,Kastor:1988ef}. The non-diagonal modular invariants, by contrast, require more than one field, as is well known for the three-state Potts model \cite{Amit:1979ev,Cappelli:2003ct}. LG descriptions with two superfields are required for the $D$ series and $E$ series $\mathcal N=2$ superconformal minimal models \cite{Martinec:1988zu, Vafa:1988uu, Lerche:1989uy}.
Similarly, Keke Li proposed two-superfield LG descriptions of several $\mathcal N=1$ superconformal minimal models \cite{Li:1988pj,Li:1989tq}. The fusion ring is the part of the operator algebra that specifies which modules appear in each operator product. In the procedure above, it determines which operators are composites. For these non-diagonal models, however, the fusion ring is harder to obtain.

In the non-supersymmetric case, theories of a single scalar field describe minimal models with diagonal modular
invariants. The simplest non-unitary example is the Yang--Lee edge
singularity.
Its
$i\phi^3$ effective field theory is due to Fisher \cite{Fisher:1978pf}, and its $d=2$
identification with the minimal model $M(2,5)$ is due to Cardy
\cite{Cardy:1985yy}.\footnote{For recent progress in matching quasi-primary operators, see \cite{ArguelloCruz:2025zuq}. An attempt to generalize this construction to multicritical Yang--Lee models $M(2,2n+1)$ using $i\phi^{2n-1}$ field theory was made in \cite{Katsevich:2025ojk}.}
Theories of two fields
describe minimal models with $D$ series modular invariants. They cannot describe minimal models with diagonal modular invariants \cite{Katsevich:2024jgq}. The reason is that two
fields $\sigma$ and $\phi$ form the spin-$1$ current
$\sigma\,\partial\phi-\phi\,\partial\sigma$, whereas the spectrum of a diagonal modular invariant contains
only spinless primaries. The two-field theory with the cubic potential $\frac{g_1}{2}\sigma\phi^2+\frac{g_2}{6}\sigma^3$ is the cubic $O(N)$ model \cite{Fei:2014yja} continued to $N=1$.\footnote{The ${\cal N}=2$ supersymmetric version of the $O(N)$ model was studied in \cite{Chester:2015qca}.}
The $\mathcal{PT}$-symmetric fixed point of this model was proposed as the LG description of $M(3,8)$ and continued to two dimensions from $d=6-\eps$ \cite{Fei:2014xta,Klebanov:2022syt}.
Later, the modular invariant realized at this fixed point was identified as the non-diagonal $D_5$ \cite{Katsevich:2024jgq}.

The unitary $\mathcal N=1$ superconformal minimal models $SM(m)$ \cite{Friedan:1984rv,Bershadsky:1985dq} have central charge $c(m)=\frac32(1-\frac{8}{m(m+2)})$.
For each $m$, the consistent modular invariants are classified by a pair of ADE labels \cite{Cappelli:1986ed,Kastor:1986ig}. At $m=12$, the central charge is $\frac{10}7$, and $\ed$ is the exceptional modular invariant whose two labels are both non-diagonal. The full chiral algebra of this theory is not the $\mathcal N=1$ super-Virasoro algebra but the larger minimal super-$W_3$ algebra $SW(3/2,5/2)$, generated by the super stress tensor together with a superconformal primary of dimension $\frac52$, so that its component currents have spins $\frac32,2,\frac52,3$. Its bosonic counterpart is the $W_3$ algebra of Zamolodchikov \cite{Zamolodchikov:1985wn}, whose minimal models were constructed in \cite{Zamolodchikov:1985wn,Fateev:1987zh}.

The super-$W_3$ algebra is associative at only two values of the central charge, $c=\frac{10}7$ and $c=-\frac52$, of which only the first admits unitary representations \cite{Inami:1988xy}. A coset realization and an analysis of its highest-weight
representations were given in \cite{Hornfeck:1990zw}, while its full component operator algebra was worked out in \cite{Ahn:1990nr}. The modular invariant $\ed$ of the $\mathcal N=1$ series at $c=\frac{10}7$ was conjectured to be the diagonal modular invariant of the super-$W_3$ algebra \cite{Bilal:1990dw}. This identification was subsequently established in \cite{Schoutens:1990xg} (see \cite{Bouwknegt:1992wg} for a review). The super-$W_3$ algebra has thus been understood in detail for more than three decades. The LG description of its minimal model, however, has remained open. The reason, noted by Li \cite{Li:1988pj}, is that the operator algebra alone does not determine the equations of motion uniquely.

The ambiguity involves a pair of modules with identical characters. A similar pair appears in the $D$ series modular invariants of even rank, whose fusion rings have long been known \cite{Moore:1988ss}. There, each fusion ring has a $\mathbb Z_2$ automorphism that exchanges the two members of the pair, and Li used it to fix their identification with the fields of the LG theory. In the $\ed$ model, the pair consists of two modules of the super-$W_3$ algebra, and neither the fusion ring of its modules nor such an automorphism has been determined. The fusion ring follows from the modular $S$-matrix through the Verlinde formula \cite{Verlinde:1988sn}. Since the characters do not distinguish the two modules, however, they leave one number in the $S$-matrix undetermined. The fusion rings of the rational models of the algebras $SW(3/2,\delta)$ were studied in \cite{Eholzer:1993ek}, where a generalized Verlinde formula for fermionic rational theories was presented. For our algebra, with $\delta=\frac52$, that work showed that the degeneracy can be resolved by an extension of the fusion algebra, but the extension itself could not be constructed without an explicit $S$-matrix. We determine the missing number here and find that conjugation exchanges the two members of the pair, exactly as the $\mathbb Z_2$ automorphism does in the $D$ series.

We propose an LG description that reflects the properties of the $\ed$ superconformal minimal model.
It involves two real scalar superfields with the cubic superpotential
\begin{equation}\label{eq:W}
    {\cal W}=\frac{g_1}{2} XY^2 +\frac{g_2}{6} X^3\,.
\end{equation}
This superpotential may be viewed as the $\mathcal N=1$ counterpart of the cubic potential of the bosonic two-field theory mentioned above, with $X$ and $Y$ in place of $\sigma$ and $\phi$. Our LG theory is invariant under three $\mathbb Z_2$ symmetries: the R-parity $\ZR$, the exchange parity $\Zex$ that reverses the sign of $Y$, and the fermion parity $\mathbb Z_2^F$. We support our proposal with two results, one in $d=2$ and the other in the $\eps$-expansion. The first is the fusion ring of the $\ed$ superconformal minimal model, which we construct from the modular data of the extended algebra. The fusion ring has a grading by chiral fermion parity that the ordinary fusion coefficients do not determine. Composed with conjugation, this grading becomes the chiral $\ZR$ symmetry of the LG theory, and the two together fix the $\ZR$ and $\Zex$ charges of all the fields that we compare with the LG theory. They also imply that seven three-point functions of the lowest states of the extended modules vanish. These are exactly the three-point functions that the ordinary fusion rules allow but the $\ZR$ and $\Zex$ charges of
the LG fields forbid.

The second result concerns the theory with superpotential \eqref{eq:W} regarded as a Gross--Neveu--Yukawa (GNY) model \cite{Gross:1974jv, Zinn-Justin:1991ksq, Hasenfratz:1991it}. To carry out the $\eps$-expansion, we follow the method of
\cite{Fei:2016sgs}. We replace each two-component Majorana spinor by $n$
four-component ones, and in the resulting $O(n)$-symmetric GNY model
formally continue $n$ to $\frac12$ (this is equivalent to evaluating the fermion loops using $\operatorname{tr}\mathbf1=2$). We then locate the infrared fixed
points of the beta functions in $d=4-\eps$. One of them, the coupled fixed point, lies at $g_1/g_2=3/2+\mathcal O(\eps)$, and we identify it, upon continuation to $d=2$, with the $\ed$ superconformal minimal model. We do not impose supersymmetry, but it emerges at this fixed point. At two loops, the quartic couplings contribute to the scalar anomalous dimensions but not to the fermion ones. The scalar and the fermion of each superfield nevertheless have equal anomalous dimensions. When we continue the operator dimensions to $d=2$, they agree approximately with the exact $d=2$ values of the $\ed$ superconformal minimal model. See Table~\ref{tab:web} for a summary of the operator spectrum.

In the $\eps$-expansion, exactly one combination of the two cubic superpotential couplings is relevant at the fixed point. The stability analysis identifies the corresponding composite operator. For one sign of its coupling, this deformation drives the RG flow to the neighboring exceptional model $(D_6,E_6)$ at $m=10$, which the superpotential \eqref{eq:W} realizes at the decoupled fixed point $g_1=g_2$ as a product of two $\mathcal N=1$ superconformal minimal models $SM(3)$ \cite{Li:1988pj,Kastor:1988ef,Gang:2008sz,Makabe:2017ygy}. For the other sign, the RG flow ends at the fixed point $g_1=-g_2$, where supersymmetry is enhanced to $\mathcal N=2$ and which we identify with the $m=4$ model $(A_3,D_4)$. We reproduce these RG flows in perturbation theory around four dimensions, where the fixed points are weakly coupled.

Although the spin-3 current $W$ of the extended algebra is conserved in the $\ed$ superconformal minimal model, the fixed point
near four dimensions has no conserved spin-3 current. At the fixed point, the two lightest
spin-3 operators acquire anomalous dimensions of order $\eps$, and their discrete
charges show that neither of them becomes $W$ upon continuation to $d=2$. The conserved spin-3 charge
must therefore emerge at $d=2$. There the lowest states of the pair of modules with identical characters
are degenerate and have opposite spin-3 charges. In the Lagrangian description, which has no such
charge, their two Hermitian combinations appear as the bottom component $\phi_Y$ of $Y$ and the quadratic composite $P=3\phi_X^2-2\phi_Y^2+\mathcal O(\eps)$, where $\phi_X$ is the bottom component of $X$.
Since the anomalous dimensions of $\phi_Y$ and $P$ agree at one loop but differ at two loops, the degeneracy at $d=2$ is a
constraint on the sum of the continued series rather than a term-by-term
identity.

Finally, continuation to $\eps=1$ gives a candidate interacting
three-dimensional $\mathcal N=1$ superconformal field theory of two
superfields. Resumming the two-loop dimensions with a two-sided $[2,1]$
Pad\'e that imposes the exact $d=2$ values, we predict
$\Delta_X\approx0.567$ and $\Delta_Y\approx0.612$ for the bottom
components of its two superfields $X$ and $Y$. Table~\ref{tab:d3} also gives the predictions for $P$ and for the product $XP$. These predictions can be tested by
the conformal bootstrap and by fuzzy sphere methods. Both have
already been applied successfully to the three-dimensional super-Ising
model, the theory of one $\mathcal N=1$ superfield with a cubic superpotential
\cite{Bashkirov:2013vya,Atanasov:2018kqw,Rong:2018okz,Atanasov:2022bpi,Tang:2025wtj}.

The paper is organized as follows. In Section~\ref{sec:model} we define the LG theory and its
three $\mathbb Z_2$ symmetries. In Section~\ref{sec:fusion} we present the $d=2$
construction, building the fusion ring of the $\ed$ superconformal minimal model from
the modular data of the extended algebra and extracting the chiral $\ZR$ symmetry.
In Section~\ref{sec:fixedpoint} we present the $d=4-\eps$ construction. In Section~\ref{sub:fp} we locate
the coupled fixed point, compute the two-loop anomalous dimensions, and propose the
dictionary. In Section~\ref{sub:deg} we show that the anomalous dimensions of $\phi_Y$ and $P$, equal at one loop, differ at two loops, and in Section~\ref{sub:flow} we reproduce the RG flows from $\ed$ to $(D_6,E_6)$ and to $(A_3,D_4)$. In Section~\ref{sub:spin3} we compute the anomalous dimensions of the twist-two operators, and in Section~\ref{sub:web} we match the low-lying operators of Table~\ref{tab:web} to the $\ed$ superconformal minimal model. In Section~\ref{sub:threed} we continue the $\eps$-expansion to $\eps=1$ and discuss the resulting $d=3$ theory. In Section~\ref{sec:discussion} we return to the origin of the spin-3 current and list some further directions.
In Appendix~\ref{app:currents} we construct the free twist-two currents,
obtain their mixing matrices by recombination, and verify the construction in three ways. In Appendix~\ref{app:twoloop} we record the two-loop RG functions that we use.

\section{The model and its symmetries}
\label{sec:model}

We propose that the $\ed$ superconformal minimal model is the infrared limit of
the two-dimensional $\mathcal N=1$ LG theory of two real scalar superfields, $X$ and $Y$, with the cubic superpotential
\eqref{eq:W}.\footnote{
We use the notation of \cite{Polchinski:1998rr} for the Euclidean $d=2$ $\mathcal N=1$ superfield,
$
\Phi(\theta, \tilde{\theta})=\phi_\Phi+i\theta\psi_\Phi+i\tilde{\theta}\tilde{\psi}_\Phi+\theta\tilde{\theta}F_\Phi$.
}
Throughout the paper, $\mathcal N=1$ supersymmetry means two supercharges in both two and three dimensions.\footnote{In two dimensions, the two supercharges have opposite chirality, and $\mathcal N=1$ is shorthand for the standard $\mathcal N=(1,1)$.} Each superfield contains a real scalar, a two-component Majorana fermion, and an auxiliary field. In components, the theory is a GNY model of two real scalars, $\phi_X$ and $\phi_Y$, and two Majorana spinors,
\begin{equation}\label{eq:spinors}
    \Psi_X=\begin{pmatrix}\psi_X\\\tilde\psi_X\end{pmatrix}\,,\qquad\Psi_Y=\begin{pmatrix}\psi_Y\\\tilde\psi_Y\end{pmatrix}\,.
\end{equation}
The Yukawa interaction is $\frac12\,h_{ijk}\,\phi_i\,\bar\Psi_j\Psi_k$ with the symmetric tensor $h_{ijk}=\partial_i\partial_j\partial_k {\cal W}$, where $i,j,k$ run over the two fields.\footnote{Here $\bar\Psi_k\Psi_l=i(\psi_k\tilde\psi_l+\psi_l\tilde\psi_k)$.} The scalar potential is $\frac12\sum_i(\partial_i {\cal W})^2$, whose nonzero quartic couplings in the normalization $V=g_{ijkl}\,\phi_i\phi_j\phi_k\phi_l/4!$ are $g_{XXXX}=3g_2^2$, $g_{XXYY}=2g_1^2+g_1g_2$ and $g_{YYYY}=3g_1^2$.

This theory has three commuting discrete symmetries. The first is the exchange parity $\Zex$,
\begin{equation}\label{eq:Zex}
    \Zex:\quad (X,Y)\ \longmapsto\ (X,-Y)\,,
\end{equation}
which reverses the sign of the odd field $Y$ and leaves the even field $X$ fixed.\footnote{In the non-unitary CFT case \cite{Katsevich:2024jgq}, the corresponding $\mathbb{Z}_2$ symmetry acts as $\phi\rightarrow -\phi$, while the $\ZR$ symmetry is replaced by the $\mathcal{PT}$ symmetry acting as $\sigma\rightarrow-\sigma$, $i\rightarrow-i$.}
The name of the symmetry comes from its action in another basis. We introduce two superfields $\Phi_1$ and $\Phi_2$, whose even and odd combinations are $X$ and $Y$,
\begin{equation}\label{eq:flavor}
    X=\frac{1}{\sqrt2}(\Phi_1+\Phi_2)\,,\qquad Y=\frac{1}{\sqrt2}(\Phi_1-\Phi_2)\,,
\end{equation}
and $\Zex$ interchanges $\Phi_1$ and $\Phi_2$. At a generic ratio $g_1/g_2$, the superpotential couples $\Phi_1$ and $\Phi_2$. At $g_1=g_2$, however, it becomes the sum ${\cal W}=\frac{g_1}{3\sqrt2}(\Phi_1^3+\Phi_2^3)$ of two copies of the single-superfield cubic superpotential. The theory at $g_1=g_2$ is therefore the LG description of the product $SM(3)^{\otimes2}$. This product is the superconformal minimal model $SM(10)_{(D_6,E_6)}$ \cite{Li:1988pj,Kastor:1988ef,Gang:2008sz,Makabe:2017ygy}.\footnote{A lattice realization of this \(SM(10)_{(D_6,E_6)}\) theory was constructed as a local spin-1 chain obtained by coupling two tricritical Blume--Capel models \cite{Lahtinen_2014}.} The superpotential also degenerates at $g_1=0$, where $Y$ decouples and is
free. The fixed point that we study lies at neither of the two values.

The second symmetry is the R-parity $\ZR$,
\begin{equation}\label{eq:Rparity}
    \ZR:\quad \Phi(x,\theta,\tilde\theta)\ \longmapsto\ -\Phi(x,-\theta,\tilde\theta),\quad\Phi=(X,Y)\,,
\end{equation}
which reverses the sign of both superfields while reflecting a single Grassmann coordinate. In components, $\ZR$ acts as $\phi_{X,Y}\rightarrow-\phi_{X,Y}$, $\psi_{X,Y}\rightarrow\psi_{X,Y}$, $\tilde{\psi}_{X,Y}\rightarrow-\tilde{\psi}_{X,Y}$, $F_{X,Y}\rightarrow F_{X,Y}$. It thus acts on the two-component fermions as $\Psi_i\rightarrow\gamma_5\Psi_i$, with the chirality matrix $\gamma_5=\mathrm{diag}(1,-1)$ in the basis of \eqref{eq:spinors}. Since the measure $d^2\theta$ and a cubic superpotential both change sign under it, $\int d^2\theta\, {\cal W}$ is invariant. The $\Zex$-even quadratics $X^2$ and $Y^2$ are $\ZR$-even, so they cannot compensate for the sign change of the measure. The mass terms of the superpotential are therefore forbidden by symmetry. Like the cubic superpotential, the linear superpotential $tX$ is $\ZR$-odd and $\Zex$-even, so $\int d^2\theta\,tX$ is invariant under both symmetries. It is a relevant operator, so we tune its coefficient $t$ to zero, as at any multicritical point.\footnote{Classically, when $g_1g_2>0$ and $t$ has the same sign as $g_2$, supersymmetry is spontaneously broken. This is consistent with the Witten index discussed in Section~\ref{sub:flow}.}

The third symmetry is the fermion parity $(-1)^{F+\tilde F}$,
\begin{equation}\label{eq:fparity}
    \mathbb{Z}_2^F:\quad\Phi(x,\theta,\tilde{\theta})\ \longmapsto\ \Phi(x,-\theta,-\tilde{\theta})\,,\quad\Phi=(X,Y)\,,
\end{equation}
which reflects both Grassmann coordinates. In components, it acts as $\phi_{X,Y}\rightarrow\phi_{X,Y}$, $\psi_{X,Y}\rightarrow-\psi_{X,Y}$, $\tilde{\psi}_{X,Y}\rightarrow-\tilde{\psi}_{X,Y}$, $F_{X,Y}\rightarrow F_{X,Y}$, so it counts the fermion number modulo two.

The superpotential \eqref{eq:W} is of $D_4$ type and not of $E_6$ type. The two uses of the ADE names must be distinguished, because the pair $\ed$ labels a single modular invariant rather than a catastrophe-theory singularity.\footnote{The two classifications can nevertheless coincide. In Section~\ref{sub:flow} we find a fixed point of \eqref{eq:W} at $g_1=-g_2$ and identify it as $SM(4)$ with the $(A_3,D_4)$ invariant. The $D_4$ of that label is the catastrophe type of \eqref{eq:W} as well.} A cubic superpotential of this type was considered by Li \cite{Li:1988pj}.
For the $(D_6,E_6)$ model at $m=10$, he obtained the superpotential $X^3+Y^3$,
up to coupling constants, from the operator algebra. He also found that an
alternative identification of the fields yields $X^3+XY^2$ instead. The two superpotentials are
related by a linear field redefinition, but the redefinition is not orthogonal
and therefore changes the kinetic term. For the $\ed$ model at $m=12$, he
obtained the same form $X^3+Y^3$. These are the two models whose equations of
motion he found not to be unique. He further noted that a single superpotential may
have several nontrivial fixed points, and that each of them is a different CFT.

The RG analysis confirms this last observation. In the theory with the superpotential \eqref{eq:W} and a canonical kinetic term,
the ratio $g_1/g_2$ cannot be redefined away. In $d=4-\eps$, the zeros of the beta functions fix the ratio,
and we list the values it takes at the fixed points in
Table~\ref{tab:fps}. Two of these values are $g_1=g_2$, at the decoupled fixed point $SM(3)^{\otimes2}=SM(10)_{(D_6,E_6)}$, and $g_1/g_2=3/2+\mathcal O(\eps)$, at the coupled fixed point that we identify with the $\ed$ superconformal minimal model. We stress that the two are different CFTs, since conformal data such as the anomalous dimensions differ. By \eqref{eq:flavor}, the first of the two has the superpotential $\Phi_1^3+\Phi_2^3$, Li's $X^3+Y^3$, with a canonical kinetic
term. In this form, Li's superpotential describes the $(D_6,E_6)$ superconformal
minimal model rather than $\ed$.

\section{The fusion ring and the chiral $\ZR$ symmetry}
\label{sec:fusion}

To constrain the LG description of the $\ed$ superconformal minimal model, we need its fusion ring. While $\ed$ is non-diagonal as a super-Virasoro modular invariant, it is the diagonal invariant of the extended super-$W_3$ algebra \cite{Schoutens:1990xg,Bouwknegt:1992wg}. The fusion ring is therefore naturally defined for the modules of the extended algebra rather than for the modules of the super-Virasoro algebra. Because two modules of the extended algebra have identical characters, no computation with characters alone can distinguish them. However, the Verlinde formula requires a modular matrix that distinguishes them. We construct this modular matrix in Section~\ref{sub:modular}.

\subsection{Warm-up: the super-Ising model $SM(3)$}
\label{sub:warmup}

The derivation of the graded fusion ring of the $\ed$ superconformal minimal model
in the rest of this section is quite technical, so we preview it here
with a discussion of the $\mathcal N=1$ superconformal minimal model $SM(3)$, often called the super-Ising model. It is the fermionic version of the tricritical Ising minimal model $M(4,5)$.\footnote{Fermionic versions of the Virasoro minimal
models are constructed in \cite{Runkel:2020zgg,Hsieh:2020uwb}.} 
Its LG description
uses one superfield with the superpotential ${\cal W}=X^3$
\cite{Zamolodchikov:1986db,Kastor:1988ef}. This theory has a $\ZR\times\mathbb{Z}_2^F$
symmetry.
The action of $\mathbb{Z}_2^F$ is specified in \eqref{eq:fparity}, while
the chiral $\ZR$ symmetry acts as
\begin{equation}\label{eq:wR}
    X(x,\theta,\tilde\theta)\ \to\ -X(x,-\theta,\tilde\theta)\,.
\end{equation}
In components, with $X=\phi+i\theta\psi+i\tilde\theta\tilde\psi+\theta\tilde\theta F$, this reads
\begin{equation}\label{eq:wRcomp}
    \phi\to-\phi\,,\qquad \psi\to\psi\,,\qquad\tilde\psi\to-\tilde\psi\,,\qquad F\to F\,.
\end{equation}
Since the scalar is odd and the auxiliary field is even, the $\ZR$ symmetry forbids a nonzero correlator $\langle\phi\phi\phi\rangle$ but allows a nonzero $\langle\phi\phi F\rangle$. From the graded fusion ring, we now show that $\langle\phi\phi\phi\rangle$ vanishes, while $\langle\phi\phi F\rangle$ does not have to vanish.

The super-Ising model $SM(3)$ is built on the $\mathcal N=1$ super-Virasoro algebra at $c=\frac7{10}$, generated by the modes of the stress tensor $T$ and of the supercurrent $G$. Since the theory is diagonal, the right module of every field equals its left module. The model has four super-Virasoro modules. Two of them are Neveu--Schwarz (NS), $[\mathbf1]$ and $[X]$, with primaries at $h=0$ and $h=\frac1{10}$, and the other two are Ramond, with primaries at $h=\frac7{16}$ and $h=\frac3{80}$.

Each NS module decomposes into two Virasoro modules,
\begin{equation}
[\mathbf1]=[0]\oplus[\tfrac32]\,,\qquad
[X]=[\tfrac1{10}]\oplus[\tfrac35]\,,
\label{eq:wvirdec}
\end{equation}
where $[h]$ is the Virasoro module of conformal weight $h$. We call the
first summand of each the bottom and the second the top, and their
primaries the bottom and top components. On $[X]$, the two are related by
$X_{\rm t}=G_{-1/2}X_{\rm b}$. The vacuum module is an exception, since
$G_{-1/2}|0\rangle=0$, and its top is the supercurrent $G=G_{-3/2}|0\rangle$.

The superfield $X$ is NS on both sides and is not the
identity, so its left module and its right module are both $[X]$. Its
four components are labeled by the choice of a bottom or a top on the left
and on the right,
\begin{equation}
\begin{aligned}
\phi&=(\text{bottom},\text{bottom})\,, & (h,\tilde h)&=(\tfrac1{10},\tfrac1{10})\,, & \Delta&=\tfrac15\,, & s&=0\,,\\
\psi&=(\text{top},\text{bottom})\,, & (h,\tilde h)&=(\tfrac35,\tfrac1{10})\,, & \Delta&=\tfrac7{10}\,, & s&=\tfrac12\,,\\
\tilde\psi&=(\text{bottom},\text{top})\,, & (h,\tilde h)&=(\tfrac1{10},\tfrac35)\,, & \Delta&=\tfrac7{10}\,, & s&=-\tfrac12\,,\\
F&=(\text{top},\text{top})\,, & (h,\tilde h)&=(\tfrac35,\tfrac35)\,, & \Delta&=\tfrac65\,, & s&=0\,.
\end{aligned}
\label{eq:wcomponents}
\end{equation}
In terms of modes, the last three are $\psi=G_{-1/2}\phi$, $\tilde\psi=\tilde G_{-1/2}\phi$
and $F=G_{-1/2}\tilde G_{-1/2}\phi$, where $\tilde G$ is the right supercurrent.
From here on, we work with one chirality at a time, since a correlator of
the two-dimensional theory is nonzero only if it is allowed on both
sides.

The product $a\times b$ contains the module $c$ with a multiplicity $N_{abc}$, which is also the dimension of the space of three-point functions of $a$, $b$ and the conjugate of $c$. The Verlinde formula expresses $N_{abc}$ in terms of a modular matrix. In the character $\operatorname{Tr}_a q^{\,L_0-c/24}$ of a module $a$, with $q=e^{2\pi i\tau}$, the fermions are antiperiodic around the time cycle of the torus, and in the NS sector they are antiperiodic around the space cycle as well. Since the transformation $\tau\to-1/\tau$ exchanges the two cycles and therefore preserves the boundary conditions, the NS characters $\chi^{\rm NS}_a$ transform among themselves,
\begin{equation}
    \chi^{\rm NS}_a(-1/\tau)=\sum_b \widetilde S_{ab}\,\chi^{\rm NS}_b(\tau)\,.
\label{eq:wSdef}
\end{equation}
The sum in the Verlinde formula \cite{Verlinde:1988sn}
\begin{equation}\label{eq:wverlinde}
N_{abc}=\sum_{\rho}\frac{\widetilde S_{a\rho}\widetilde S_{b\rho}\overline{\widetilde S_{c\rho}}}{\widetilde S_{\mathbf1\rho}}
\end{equation}
is then over the NS modules alone.

An NS character is the sum of the Virasoro characters of the
bottom and the top,
\begin{equation}\label{eq:wchitilde}
\chi^{\rm NS}_{\mathbf1}=\chi^{V}_{0}+\chi^{V}_{3/2}\,,\qquad
\chi^{\rm NS}_{X}=\chi^{V}_{1/10}+\chi^{V}_{3/5}\,,
\end{equation}
where $\chi^{V}_h$ is the trace over the Virasoro module $[h]$ of \eqref{eq:wvirdec}
in the Virasoro minimal model $M(4,5)$ at $c=\frac7{10}$. The modular matrix of $M(4,5)$
takes the pair \eqref{eq:wchitilde} into itself, and in the order
$(\mathbf1,X)$ the modular matrix of \eqref{eq:wSdef} is
\begin{equation}
\widetilde S=\frac{2}{\sqrt5}
\begin{pmatrix}
\sin\frac{4\pi}5 & \sin\frac{2\pi}5\\[2pt]
\sin\frac{2\pi}5 & -\sin\frac{4\pi}5
\end{pmatrix}\,,
\label{eq:wS}
\end{equation}
so that the Verlinde formula \eqref{eq:wverlinde} gives $N_{XX\mathbf1}=N_{XXX}=1$. The product
of $[X]$ with itself is therefore
\begin{equation}
[X]\times[X]=[\mathbf1]+[X]\,.
\label{eq:wfus}
\end{equation}
The terms of the product are called channels.

The components $\phi$ and $F$ are built from the same module, so by
\eqref{eq:wcomponents} the correlators $\langle\phi\phi\phi\rangle$ and
$\langle\phi\phi F\rangle$ belong to the same channel $[X]$. They differ
only in the components they involve. On one chirality, $\langle\phi\phi\phi\rangle$
involves three bottom components, while $\langle\phi\phi F\rangle$
involves two bottom components and one top component. The fusion coefficient $N_{XXX}=1$, however, counts three-point functions of
whole modules and not of single components, so the fusion ring does
not resolve the two correlators.

Each independent three-point function of three super-Virasoro modules is called a
three-point structure. In components, an even structure gives correlators
with an even number of top components and an odd structure gives
correlators with an odd number. More explicitly, the even correlators are
$(\mathrm{bottom},\mathrm{bottom},\mathrm{bottom})$ and the three
of the type $(\mathrm{top},\mathrm{top},\mathrm{bottom})$,
and the odd ones are $(\mathrm{top},\mathrm{top},\mathrm{top})$
and the three of the type $(\mathrm{top},\mathrm{bottom},\mathrm{bottom})$.

Let $N^+$ denote the number of even structures of a channel and $N^-$
the number of odd ones. The graded fusion ring is then the fusion ring
with these two numbers in place of $N_{abc}$. Since both fusion coefficients of the ordinary fusion ring
\eqref{eq:wfus} are one, each channel has a single structure. A nonzero
$\langle\phi\phi\phi\rangle$ therefore requires the structure of $[X]$ to
be even and a nonzero $\langle\phi\phi F\rangle$ requires it to be odd.

The characters \eqref{eq:wchitilde} count the bottom and the top with the
same sign, so the ordinary Verlinde formula \eqref{eq:wverlinde} gives
only $N^++N^-$. To resolve $N^+$
and $N^-$, we use the graded Verlinde formula
\cite{Aasen:2017ubm,Lou:2020gfq}, which requires one further modular matrix.
To define it, we introduce the graded character, marked by the superscript $\widetilde{\rm NS}$. It is the trace with an
insertion of $(-1)^F$, which gives $+1$ on the bottom of a super-Virasoro module and $-1$ on
its top,
\begin{equation}
\chi^{\widetilde{\rm NS}}_{\mathbf1}=\chi^{V}_{0}-\chi^{V}_{3/2}\,,\qquad
\chi^{\widetilde{\rm NS}}_{X}=\chi^{V}_{1/10}-\chi^{V}_{3/5}\,.
\label{eq:wchicheck}
\end{equation}
Since $G$ is odd and the top is one mode of $G$ above the bottom, the
insertion is chiral fermion parity up to an overall sign on each module.

Inserting fermion parity makes the fermions periodic around the time
cycle of the torus. The transformation $\tau\to-1/\tau$ maps the time
cycle to the space cycle, so it takes \eqref{eq:wchicheck} into the
sector with fermions periodic around the space cycle. An explicit computation gives the transformation of the graded
characters,
\begin{equation}
    \chi^{\widetilde{\rm NS}}_a(-1/\tau)=\sqrt2\,\sum_{\rho}U_{a\rho}\chi^{\rm R}_\rho(\tau)\,,\qquad U=\frac{2}{\sqrt5}\begin{pmatrix}
    \sin\frac{4\pi}5 & \sin\frac{2\pi}5\\[2pt]
    -\sin\frac{2\pi}5 & \sin\frac{4\pi}5
    \end{pmatrix}\,,
\label{eq:wnstor}
\end{equation}
where $\chi^{\rm R}_\rho$ are the two Ramond characters, which coincide with the Virasoro characters $\chi^{V}_{7/16}$ and $\chi^{V}_{3/80}$. The rows of $U$ are labeled $(\mathbf1,X)$ and its columns
$(\frac7{16},\frac3{80})$. Since both
sides of \eqref{eq:wnstor} are completely determined $q$-series, the factor $\sqrt2$ in
front is fixed by either row.

The graded Verlinde formula uses both modular matrices. The sum built from
$\widetilde S$ counts all the structures of a channel, and the sum built
from $U$ counts the even ones with $+1$ and the odd ones with $-1$,
\begin{equation}\label{eq:wgv}
N^{+}_{abc}+N^{-}_{abc}=\sum_\rho\frac{\widetilde S_{a\rho}\widetilde S_{b\rho}\overline{\widetilde S_{c\rho}}}{\widetilde S_{\mathbf1\rho}}\,,\quad
N^{+}_{abc}-N^{-}_{abc}=\sum_\rho\frac{U_{a\rho}U_{b\rho}\overline{U_{c\rho}}}{U_{\mathbf1\rho}}\,.
\end{equation}
The sum for $N^++N^-$ is over the two NS modules and the sum
for $N^+-N^-$ over the two Ramond ones.

The graded Verlinde formula \eqref{eq:wgv} is general, but the crucial observation
here is that in $SM(3)$ the two modular matrices differ only in the sign
of the second row,\footnote{The index $\rho$ labels a Ramond module on the left of \eqref{eq:wU} and an NS module on the right. The identity therefore depends on the labeling of \eqref{eq:wnstor}, which assigns $[\frac7{16}]$ to the column of $\mathbf1$ and $[\frac3{80}]$ to the column of $X$.}
\begin{equation}
U_{a\rho}=\omega_a\,\widetilde S_{a\rho}\,,\qquad\omega_{\mathbf1}=+1\,,\qquad \omega_X=-1\,.
\label{eq:wU}
\end{equation}
The sum for $N^+-N^-$ is then $\omega_a\omega_b\omega_c$ times the sum for
$N^++N^-$, so
\begin{equation}
N^{\pm}_{abc}=\frac12\,N_{abc}\bigl(1\pm\omega_a\omega_b\omega_c\bigr)\,,
\label{eq:wnpm}
\end{equation}
and in every channel $N^+$ or $N^-$ vanishes. This signals a
symmetry.

The symmetry is indeed defined by $\omega_a$. Let us consider the operation that multiplies a field of the module
$a$ by $+\omega_a$ when it has a bottom on the left in
\eqref{eq:wcomponents} and by $-\omega_a$ when it has a top. Since each
correlator of a structure involves one component from each of the three modules $a$, $b$ and $c$, the
operation multiplies an even structure by $\omega_a\omega_b\omega_c$ and
an odd one by $-\omega_a\omega_b\omega_c$. By \eqref{eq:wnpm}, a channel
has only even structures when $\omega_a\omega_b\omega_c=+1$ and only odd
ones when it is $-1$, so every surviving structure is multiplied by $+1$.
Since every nonvanishing correlator is uncharged, the operation defines a symmetry
of the theory. This operation is the origin of the $\ZR$ symmetry
\eqref{eq:wR} in $SM(3)$. We call $\omega_a$ the chiral fermion parity of the bottom
component of the module $a$.

Evaluating \eqref{eq:wgv} gives $(N^+,N^-)=(1,0)$ for the identity channel
and $(N^+,N^-)=(0,1)$ for the channel $[X]$, so the graded fusion ring now reads
\begin{equation}
[X]\times[X]=[\mathbf1]_{+}+[X]_{-}\,.
\label{eq:wring}
\end{equation}
By \eqref{eq:wring} and \eqref{eq:wcomponents}, the correlator
$\langle\phi\phi\phi\rangle$ requires an even structure on the left and on
the right, and therefore vanishes, while $\langle\phi\phi F\rangle$ requires the odd
structure and is allowed to be nonzero. The symmetry behind the two statements is
\eqref{eq:wRcomp} itself, since $\phi$ and $\tilde\psi$ have a bottom
on the left and are multiplied by $\omega_X=-1$, while $\psi$ and $F$
have a top on the left and are multiplied by $-\omega_X=+1$.

For the $\ed$ superconformal minimal model, the graded Verlinde formula \eqref{eq:wgv} still applies, but its two
modular matrices are harder to obtain. The NS and the Ramond sectors each have five extended modules, two of which have identical characters. Accordingly, the characters determine neither
one number in $\widetilde S$ nor the difference between the two
columns of $U$ labeled by the Ramond pair. Integrality of the ordinary fusion coefficients
fixes the square of the undetermined number, and its sign follows once the Ramond sector is included. Furthermore, requiring $N^+$ and $N^-$ to be
non-negative integers fixes this difference up to an overall sign, which does not affect $N^\pm$. The graded fusion ring then follows uniquely. As in $SM(3)$, the chiral fermion parities come from matching
$U$ against $\widetilde S$.

\subsection{The coset and the extended modules}
\label{sub:coset}

The $\mathcal N=1$ superconformal minimal models have the coset realization
\cite{Goddard:1984vk,Goddard:1986ee}
\begin{equation}
    SM(m)=\frac{\widehat{su}(2)_{m-2}\times\widehat{su}(2)_2}{\widehat{su}(2)_m}\,,
\label{eq:cosetgen}
\end{equation}
whose central charge is
$c(m)=3\frac{m-2}{m}+\frac32-3\frac{m}{m+2}
=\frac32\bigl(1-\frac{8}{m(m+2)}\bigr)$.
The
$\widehat{su}(2)_2$ factor is the theory of three free Majorana
fermions, and it provides the supercurrent. At $m=12$, the coset is
\begin{equation}
\frac{\widehat{su}(2)_{10}\times\widehat{su}(2)_2}{\widehat{su}(2)_{12}}\,.
\label{eq:coset}
\end{equation}
The torus partition
functions of $\widehat{su}(2)_k$ admit an ADE classification \cite{Cappelli:1987xt,Kato:1987td}. At level two, only the diagonal invariant exists, so a modular invariant of $SM(m)$ is fixed by one ADE label for each of the two $m$-dependent algebras, $\widehat{su}(2)_{m-2}$ and $\widehat{su}(2)_m$ \cite{Cappelli:1986ed}.

In the $\ed$ superconformal minimal model, the labels are $E_6$ for $\widehat{su}(2)_{10}$ and $D_8$ for $\widehat{su}(2)_{12}$. With the diagonal invariant for both algebras, the chiral algebra would be the super-Virasoro algebra itself. The choices in $\ed$, in contrast, are both non-diagonal, and each enlarges the chiral algebra, $E_6$ through a conformal embedding and $D_8$ through a simple-current extension. The enlarged algebra is the super-$W_3$ algebra $SW(3/2,5/2)$ mentioned in the Introduction.\footnote{The $W_3$ minimal model with the same central charge $c=\frac{10}{7}$ is $W_3(6,7)$ \cite{Zamolodchikov:1985wn,Fateev:1987zh,Inami:1988xy,Ahn:1990nr,Schoutens:1990xg,Bouwknegt:1992wg}.} We identify it from its vacuum module at the end of this subsection.

The label $E_6$ stands for the conformal embedding
$\widehat{su}(2)_{10}\subset\widehat{so}(5)_1$. The algebra $\widehat{so}(5)_1$ is the theory
of five free Majorana fermions, so its central charge is $\frac52$. The
embedding is conformal because $\widehat{su}(2)_{10}$ has the same central charge. A conformal embedding decomposes every module of the larger algebra into finitely many modules of the smaller one \cite{Kac:1988tf,Altschuler:1989nm}. Here $\widehat{so}(5)_1$ has three modules, the vacuum, the vector
with $h=\frac12$, and the spinor with $h=\frac5{16}$. Their
$\widehat{su}(2)_{10}$ content follows from matching conformal weights modulo
integers. With $r=2j+1$ and $h_j=j(j+1)/12$, the three decompose as
\begin{equation}
\begin{aligned}
\text{vacuum}&:\quad r\in\{1,7\}\,, &\qquad h&=0,\ 1\,,\\
\text{vector}&:\quad r\in\{5,11\}\,, &\qquad h&=\tfrac12,\ \tfrac52\,,\\
\text{spinor}&:\quad r\in\{4,8\}\,, &\qquad h&=\tfrac5{16},\ \tfrac{21}{16}\,.
\end{aligned}
\label{eq:so5content}
\end{equation}

The modular invariant $E_6$ of
$\widehat{su}(2)_{10}$ is the diagonal modular invariant of
$\widehat{so}(5)_1$ written in terms of the $\widehat{su}(2)_{10}$
characters,\footnote{Throughout, $\chi^{(k)}_a(\tau)$ denotes the
character of the integrable representation $a$ of $\widehat{su}(2)_k$.}
\begin{equation}
    Z_{E_6}=\bigl|\chi_1^{(10)}+\chi_7^{(10)}\bigr|^2+\bigl|\chi_4^{(10)}+\chi_8^{(10)}\bigr|^2+\bigl|\chi_5^{(10)}+\chi_{11}^{(10)}\bigr|^2\,.
\label{eq:ZE6}
\end{equation}
Choosing $E_6$ for $\widehat{su}(2)_{10}$ therefore replaces it by $\widehat{so}(5)_1$ throughout. As a result, the first Kac index $r$ appears only through the $\widehat{so}(5)_1$ class to which it belongs. The NS sector of the five fermions consists of the vacuum and the vector, and their Ramond sector is the spinor. Accordingly, in the NS sector of the coset the first Kac index takes the values $r\in\{1,5,7,11\}$, and in the Ramond sector $r\in\{4,8\}$. Moreover, the numerator of the coset is now $\widehat{so}(5)_1\times\widehat{su}(2)_2$, the theory of eight free Majorana fermions. From these free fermions we build the modular data in Section~\ref{sub:modular}, derive the chiral fermion parities in Section~\ref{sub:parity}, and construct the operator that realizes conjugation in Section~\ref{sub:rparity} and Section~\ref{sub:spinning}.

The label $D_8$ stands for a simple-current extension of
$\widehat{su}(2)_{12}$. A simple current is a primary whose fusion with
any primary yields a single primary \cite{Schellekens:1989am,Intriligator:1989zw}. In
$\widehat{su}(2)_{k}$, with $s=1,\dots,k+1$ labeling the integrable
representations, the nontrivial simple current $J$ is the primary with the
largest isospin, $s=k+1$. Its conformal weight is $h_J=k/4$, and its fusion acts on
the labels as $J:s\mapsto k+2-s$. Because the conformal weight $h_J=3$ is an integer at $k=12$, the simple current can be added to the
chiral algebra, and the
$D_8$ modular invariant is the diagonal invariant of the extended algebra.

The simple-current extension has three consequences, and all three follow from locality
with respect to $J$. First, only those representations survive whose monodromy charge
$Q(s)=h_s+h_J-h_{Js}$ vanishes modulo one. With $h^{(12)}_s=(s^2-1)/56$,
the monodromy charge is $(s-1)/2$ modulo one, so $s$ must be odd. Second, the surviving representations are identified within the orbits $\{s,14-s\}$, namely $\{1,13\}$,
$\{3,11\}$ and $\{5,9\}$, and each orbit becomes a single module of the
extended algebra with character $\chi^{(12)}_s+\chi^{(12)}_{14-s}$. Third, the value
$s=7$ is fixed by $J$, and a fixed point of a simple current yields two
modules rather than one \cite{Schellekens:1989am,Intriligator:1989zw,Schellekens:1990xy,Fuchs:1995tq,Fuchs:1996dd}. The
extension current maps the representation at $s=7$ to itself, and its action there squares to the identity because $J\times J=\mathbf1$. The two signs of the
action define two inequivalent modules with equal characters. The $D_8$
partition function of $\widehat{su}(2)_{12}$ shows all three
consequences \cite{Intriligator:1989zw},
\begin{equation}\label{eq:ZD8}
    Z_{D_8}=|\chi_1^{(12)}+\chi_{13}^{(12)}|^2+|\chi_3^{(12)}+\chi_{11}^{(12)}|^2+|\chi_5^{(12)}+\chi_9^{(12)}|^2+2|\chi_7^{(12)}|^2\,.
\end{equation}
The partition function \eqref{eq:ZD8} does not distinguish the two fixed-point modules.

Let us now construct the extended modules and their super-Virasoro content. A coset primary is labeled by a pair of Kac indices $(r,s)$, subject to the reflection $(r,s)\simeq(12-r,14-s)$. Its NS conformal weight is the Kac weight of $SM(m)$ at $m=12$,
\begin{equation}\label{eq:kac}
    h_{r,s}=\frac{((m+2)r-ms)^2-4}{8m(m+2)}=\frac{(7r-6s)^2-1}{336}\,,
\end{equation}
with $1\le r\le11$ and $1\le s\le13$. A Ramond conformal weight includes the additional offset $\frac1{16}$. We list the weights in Table~\ref{tab:kac}.

\begin{table}[!htbp]
\caption{\label{tab:kac}Conformal weights \eqref{eq:kac} of the $\ed$ invariant, which keeps only the odd $s$, with $r\in\{1,5,7,11\}$ in the NS sector and $r\in\{4,8\}$ in the Ramond sector. A Ramond weight includes the offset $\frac1{16}$, and the reflection $(r,s)\simeq(12-r,14-s)$ gives the omitted rows $r=7$, $8$ and $11$.}
\centering
\begin{tabular}{cc|*{7}{c}}
\toprule
$r$ & sector & $s=1$ & $3$ & $5$ & $7$ & $9$ & $11$ & $13$ \\
\midrule
$1$ & NS & $0$ & $\frac{5}{14}$ & $\frac{11}{7}$ & $\frac{51}{14}$ & $\frac{46}{7}$ & $\frac{145}{14}$ & $15$ \\
$5$ & NS & $\frac{5}{2}$ & $\frac{6}{7}$ & $\frac{1}{14}$ & $\frac{1}{7}$ & $\frac{15}{14}$ & $\frac{20}{7}$ & $\frac{11}{2}$ \\
$4$ & R & $\frac{3}{2}$ & $\frac{5}{14}$ & $\frac{1}{14}$ & $\frac{9}{14}$ & $\frac{29}{14}$ & $\frac{61}{14}$ & $\frac{15}{2}$ \\
\bottomrule
\end{tabular}
\end{table}

The extended module on the orbit $\{s,14-s\}$ contains one
super-Virasoro primary $h_{r,s}$ for each of the four labels
$r\in\{1,5,7,11\}$ of the vacuum and vector classes. Its character is
the sum of the four super-Virasoro characters, which is the branching
rule of \cite{Schoutens:1990xg}. The branching rule follows from combining the $\widehat{so}(5)_1$ decomposition \eqref{eq:so5content} with the character decomposition of the coset \eqref{eq:cosetgen}. In both the NS and the Ramond sectors, the product of an $\widehat{su}(2)_{10}$ character and the character of the three fermions is a sum of $\widehat{su}(2)_{12}$ characters whose coefficients are irreducible super-Virasoro characters \cite{Goddard:1986ee,Iohara:2003a,Iohara:2003b}.

The lowest state of an extended module is its super-Virasoro primary of
smallest conformal weight. We list the modules with their orbits and conformal weights in
Table~\ref{tab:modules}. The orbit $\{1,13\}$ gives
the vacuum module, the orbit $\{5,9\}$ a module with lowest state at
$h_{5,5}=\frac1{14}$, called $\widehat X$ because in Section~\ref{sec:fixedpoint} we
match its lowest state to $\phi_X$, and the orbit $\{3,11\}$
a module with lowest state at $h_{1,3}=\frac5{14}$, called $\widehat\Phi_{1,3}$. At the fixed point $s=7$, the reflection
acts within the orbit $\{7\}$, identifying $(1,7)\simeq(11,7)$ and $(5,7)\simeq(7,7)$. The four labels give the conformal weight $h_{5,7}=\frac17$ twice and $h_{1,7}=\frac{51}{14}$ twice. The fixed point therefore resolves into two modules, called $\Phi_+$ and $\Phi_-$, with equal characters, each containing $h_{5,7}=\frac17$ once and $h_{1,7}=\frac{51}{14}$ once. The
$\ed$ superconformal minimal model thus has five
NS modules.

The Ramond modules come from the same orbits
with the spinor class $r\in\{4,8\}$. The three length-two orbits give one
module each, with the two conformal weights listed in Table~\ref{tab:modules}. At
the fixed point, the reflection identifies $(4,7)\simeq(8,7)$, leaving the
single conformal weight $\frac9{14}$, and the fixed point resolves, exactly as in the NS sector, into a pair $R_{9/14,\pm}$ \cite{Schoutens:1990xg}. The assignment of $r\in\{1,5,7,11\}$ to the NS sector and $r\in\{4,8\}$ to the Ramond sector is consistent with the
Kac table, in which NS labels have $r-s$ even and Ramond
labels $r-s$ odd, since the surviving second index is always odd.

We identify the chiral algebra from the vacuum module. Its four
super-Virasoro primaries lie at $h_{r,1}=0$, $\frac52$, $\frac{11}2$
and $15$, and each generates a tower, the primary together with its
super-Virasoro descendants. The tower of the identity contains the
stress tensor and the supercurrent, of spins $2$ and $\frac32$. The primary at $h=\frac52$ is
a new fermionic current $U$, and its superpartner $W=G_{-1/2}U$ is a
bosonic current of spin $3$. The generating spins are therefore
$\frac32,2,\frac52,3$, which identifies the chiral algebra as
$SW(3/2,5/2)$. The
primaries at $h=\frac{11}2$ and $15$ are composites of these generators.
The zero mode $W_0$ of the spin-3 current takes opposite eigenvalues $\pm w$ on the two members of a resolved pair, $\Phi_\pm$ or $R_{9/14,\pm}$ \cite{Schoutens:1990xg}.

\begin{table}[!htbp]
\caption{\label{tab:modules}Modules of the extended algebra, labeled by the $D_8$ orbits of the second Kac index, with the conformal weights of their super-Virasoro primaries from Table~\ref{tab:kac}. The chiral fermion parity $\omega_a$ of an NS module follows from its lowest state by \eqref{eq:omegarule} of Section~\ref{sub:parity}. The orbit $\{7\}$ is the fixed point of the simple-current extension, and each of its two rows is a resolved pair.}
\centering
\begin{tabular}{llll}
\toprule
module & orbit & conformal weights & $\omega_a$\\
\midrule
$\mathbf1$ & $\{1,13\}$ & $0,\ \frac52,\ \frac{11}2,\ 15$ & $+$\\
$\widehat\Phi_{1,3}$ & $\{3,11\}$ & $\frac5{14},\ \frac67,\ \frac{20}7,\ \frac{145}{14}$ & $-$\\
$\widehat X$ & $\{5,9\}$ & $\frac1{14},\ \frac{15}{14},\ \frac{11}7,\ \frac{46}7$ & $-$\\
$\Phi_\pm$ & $\{7\}$ & $\frac17,\ \frac{51}{14}$ & $+$\\
\midrule
$R_{3/2}$ & $\{1,13\}$ & $\frac32,\ \frac{15}2$ & \\
$R_{5/14}$ & $\{3,11\}$ & $\frac5{14},\ \frac{61}{14}$ & \\
$R_{1/14}$ & $\{5,9\}$ & $\frac1{14},\ \frac{29}{14}$ & \\
$R_{9/14,\pm}$ & $\{7\}$ & $\frac9{14}$ & \\
\bottomrule
\end{tabular}
\end{table}

\subsection{Modular data and the fixed-point resolution}
\label{sub:modular}

The graded fusion ring is obtained from the graded
Verlinde formula \eqref{eq:wgv},
which requires the modular matrix of the five extended modules. Neither numerator factor distinguishes the extended modules, because every NS module of Section~\ref{sub:coset} contains both NS classes of $\widehat{so}(5)_1$ and both NS representations of the fermionic factor $\widehat{su}(2)_2$, isospin zero and isospin one. Consequently, the denominator $\widehat{su}(2)_{12}$ alone labels the extended
modules, and its labels determine their fusion. The modular matrix of the five extended modules is therefore obtained from that of the $D_8$ extension of $\widehat{su}(2)_{12}$.

Almost every component of the extended $S$-matrix follows from the characters.
The modular matrix of $\widehat{su}(2)_{12}$ is
$S^{(12)}_{ab}=\sqrt{1/7}\,\sin(\pi ab/14)$, which is real and symmetric and satisfies
$\chi^{(12)}_a(-1/\tau)=\sum_bS^{(12)}_{ab}\chi^{(12)}_b(\tau)$. The character
of an extended module on a length-two orbit is $\chi^{(12)}_a+\chi^{(12)}_{14-a}$, and under
$\tau\to-1/\tau$ it transforms with the sum of two rows of $S^{(12)}$.
Since the identity $S^{(12)}_{14-a,s}=(-1)^{s+1}S^{(12)}_{as}$ doubles the
terms with odd $s$ and cancels the terms with even $s$, only the representations that survive the simple-current extension appear. The same identity gives
$S^{(12)}_{a,14-b}=S^{(12)}_{ab}$ for the odd representatives $a$, which
groups the surviving terms into orbits,
\begin{equation}
\bigl(\chi^{(12)}_a+\chi^{(12)}_{14-a}\bigr)(-1/\tau)
=\sum_{b\in\{1,3,5\}}2S^{(12)}_{ab}\bigl(\chi^{(12)}_b+\chi^{(12)}_{14-b}\bigr)
+2S^{(12)}_{a7}\,\chi^{(12)}_7\,.
\label{eq:orbittransform}
\end{equation}

The two resolved modules share a single character, the full
$\chi^{(12)}_7$. The modules $\Phi_+$ and $\Phi_-$ are distinguished not by the character but by the resolved modular data, and physically by the sign of the spin-3 charge $w$. The factor $2$ in the term
$2|\chi^{(12)}_7|^2$ of \eqref{eq:ZD8} counts the two modules, each contributing $|\chi^{(12)}_7|^2$, with $\chi^{(12)}_{\Phi_+}=\chi^{(12)}_{\Phi_-}=\chi^{(12)}_7$. The last term of \eqref{eq:orbittransform} is then
$2S^{(12)}_{a7}\chi^{(12)}_7=S^{(12)}_{a7}(\chi^{(12)}_{\Phi_+}+\chi^{(12)}_{\Phi_-})$, and
the orbit rows of the extended $S$-matrix, those of the three length-two orbits, read
$\widetilde S^{(12)}_{ab}=2S^{(12)}_{ab}$ and
$\widetilde S^{(12)}_{a\Phi_\pm}=S^{(12)}_{a7}$.

The same computation for $\chi^{(12)}_7$ gives
\begin{equation}\label{eq:fixedtransform}
    \chi^{(12)}_7(-1/\tau)=\sum_{b\in\{1,3,5\}} S^{(12)}_{7b}(\chi^{(12)}_b+\chi^{(12)}_{14-b})+\frac12S^{(12)}_{77}(\chi^{(12)}_{\Phi_+}+\chi^{(12)}_{\Phi_-})\,,
\end{equation}
so the fixed-point rows have components $S^{(12)}_{7b}$ in the orbit columns, while inside the resolved pair $\Phi_\pm$ the characters determine only the sums, $\widetilde S^{(12)}_{\Phi_\pm\Phi_+}+\widetilde S^{(12)}_{\Phi_\pm\Phi_-}=S^{(12)}_{77}$. Since the extended $S$-matrix is symmetric and the two row sums are equal, the two diagonal components are equal as well. Only one number is then undetermined, the difference $\check S^{(12)}\equiv\widetilde S^{(12)}_{\Phi_+\Phi_+}-\widetilde S^{(12)}_{\Phi_+\Phi_-}$. In the basis $(\mathbf1,\widehat\Phi_{1,3},\widehat X,\Phi_+,\Phi_-)$, with the first three labeled by $a,b\in\{1,3,5\}$, the extended $S$-matrix is therefore \cite{Fuchs:1996dd}
\begin{equation}
\widetilde S^{(12)}=
\begin{pmatrix}
2S^{(12)}_{ab} & S^{(12)}_{a7} & S^{(12)}_{a7}\\[2pt]
S^{(12)}_{7b} & \frac12(S^{(12)}_{77}+\check S^{(12)})
              & \frac12(S^{(12)}_{77}-\check S^{(12)})\\[2pt]
S^{(12)}_{7b} & \frac12(S^{(12)}_{77}-\check S^{(12)})
              & \frac12(S^{(12)}_{77}+\check S^{(12)})
\end{pmatrix}\,.
\label{eq:Sext}
\end{equation}

Two relations of the modular group constrain $\check S^{(12)}$. The first is $(\widetilde S^{(12)})^2=\mathcal C$, where $\mathcal C$ is the
conjugation matrix, an order-two permutation of the extended modules. On the orbit rows and columns, the relation descends from the corresponding relation in $\widehat{su}(2)_{12}$ and holds automatically. To evaluate the fixed-point block, we use
$S^{(12)}_{7b}=\pm1/\sqrt7$ for the three orbit representatives and
$S^{(12)}_{77}=-1/\sqrt7$, and find
\begin{equation}
\bigl((\widetilde S^{(12)})^2\bigr)_{\Phi_\pm\Phi_\pm}=\frac12\bigl(1+(\check S^{(12)})^2\bigr)\,,
\qquad
\bigl((\widetilde S^{(12)})^2\bigr)_{\Phi_+\Phi_-}=\frac12\bigl(1-(\check S^{(12)})^2\bigr)\,.
\label{eq:S2block}
\end{equation}
A permutation matrix has components $0$ and $1$, so either the diagonal component is $1$ and the off-diagonal component $0$, which requires $(\check S^{(12)})^2=1$, or the diagonal component is $0$ and the off-diagonal component $1$, which requires $(\check S^{(12)})^2=-1$. The four values $\pm1$ and $\pm i$ survive, and the relation determines the action of conjugation in each case. For $\check S^{(12)}=\pm1$, the conjugation is the identity on
$\Phi_\pm$. For $\check S^{(12)}=\pm i$, it exchanges $\Phi_+$ with $\Phi_-$.

The second relation is $(\widetilde S^{(12)}T)^3=\mathcal C$. It depends on the conformal weights through the matrix $T$, which we must first identify. The coset theory itself has no diagonal $T$ on its NS sector, because it is fermionic, and $\tau\to\tau+1$ maps its NS
characters to the sector with the chiral fermion parity $(-1)^F$ inserted. However, $\widetilde S^{(12)}$
is the modular matrix of the $D_8$ extension of $\widehat{su}(2)_{12}$, which
is bosonic, because the extension current has integer spin $h_J=3$. This bosonic theory has central charge $c=\frac{18}7$ and its five modules are the orbits of Section~\ref{sub:coset} with the conformal weights $h^{(12)}_s$. Its matrix $T=\mathrm{diag}\,e^{2\pi i(h-c/24)}$ is well defined, because the conformal weights of the two members of each orbit differ by an integer. In this bosonic theory, the relation $(\widetilde S^{(12)}T)^3=\mathcal C$ fixes $\check S^{(12)}$ explicitly.

The antisymmetric combination $v=e_{\Phi_+}-e_{\Phi_-}$ of the two
fixed-point basis vectors of \eqref{eq:Sext} is a common eigenvector of
$\widetilde S^{(12)}$, of $T$ and of $\mathcal C$. Since in every row of \eqref{eq:Sext} the two fixed-point columns differ only through $\check S^{(12)}$, we have $\widetilde S^{(12)}v=\check S^{(12)}v$. The two resolved modules have the same conformal weight, which gives $Tv=t_7v$ with $t_7=e^{2\pi i(h^{(12)}_7-c/24)}$. The conjugation $\mathcal C$ fixes the three orbit modules, whose conformal weights differ, and maps the pair $\Phi_\pm$ to itself, either member by member, giving $\mathcal Cv=+v$, or by exchange, giving $\mathcal Cv=-v$. We denote the eigenvalue by $\sigma$, $\mathcal Cv=\sigma v$. The two modular relations then become
$(\check S^{(12)})^2=\sigma$ and $(\check S^{(12)}\,t_7)^3=\sigma$, and eliminating
$\sigma$ gives
\begin{equation}
\check S^{(12)}=t_7^{-3}\,,\qquad
t_7=e^{2\pi i\left(\frac67-\frac3{28}\right)}=-i\,,
\label{eq:checkS}
\end{equation}
so that $\check S^{(12)}=(-i)^{-3}=-i$. The same pair of equations determines
$\sigma=(\check S^{(12)})^2=-1$, so at level twelve the conjugation
exchanges $\Phi_+$ with $\Phi_-$, and of the four candidates of \eqref{eq:S2block}
only $\check S^{(12)}=-i$ survives.\footnote{For general level $k$, the fixed point of the simple current lies at $s=k/2+1$. The exponent is then $h^{(k)}_{k/2+1}-c/24=\frac{k(k+4)}{16(k+2)}-\frac{k}{8(k+2)}=\frac{k}{16}$, so that $\check S^{(k)}=e^{-3\pi ik/8}$. A $D$-type extension requires $h_J=k/4$ to be an integer, so $k$ is a multiple of four and $\check S^{(k)}$ reduces to $e^{i\pi k/8}$, which is $+i$ at $k=4$ and $-1$ at $k=8$.}

Although the construction so far resolves the fixed point of the $\widehat{su}(2)_{12}$ extension, the coset realization requires its own modular matrix. The coset characters are the
branching functions $b_{\Lambda s}$, defined by
$\chi^{\rm num}_\Lambda=\sum_s b_{\Lambda s}\,\chi^{(12)}_s$ with numerator
$\widehat{so}(5)_1\times\widehat{su}(2)_2$. Under $\tau\to-1/\tau$, the left side of the definition transforms with the modular matrix $S^{\rm num}$ of the numerator and the $\chi^{(12)}_s$ on the right with $S^{(12)}$. The transformation of the $b_{\Lambda s}$ therefore involves the inverse of $S^{(12)}$. Since $S^{(12)}$ is symmetric and unitary, its inverse is its complex conjugate. We obtain
\begin{equation}
b_{\Lambda s}(-1/\tau)
=\sum_{\Lambda',s'}S^{\rm num}_{\Lambda\Lambda'}\,
\overline{S^{(12)}_{ss'}}\;b_{\Lambda's'}(\tau)\,.
\label{eq:branchS}
\end{equation}

The five NS modules all descend from a single numerator
sector, the NS sector of the eight free fermions, in which
the fermions are antiperiodic around both cycles of the torus. The
transformation $\tau\to-1/\tau$ exchanges the two cycles and preserves
this spin structure, so the character of this sector is invariant and
$S^{\rm num}$ in \eqref{eq:branchS} reduces to the identity.
The complex conjugation in \eqref{eq:branchS} has no effect, because
the unextended $S^{(12)}$
is real. The matrix $S^{(12)}$ squares to one, since every $\widehat{su}(2)$ representation is self-conjugate. The branching functions therefore
transform with the same matrix as the parent characters, and in the modular matrix of the coset
every component that the characters determine is real and is inherited from \eqref{eq:Sext} unchanged.

The one number in $\widetilde S$ that no character identity determines is the difference $\check S$, because the two resolved modules have
identical characters. The coset realization therefore does not inherit the sign found for the
$\widehat{su}(2)_{12}$ extension. The ordinary
Verlinde formula \eqref{eq:wverlinde}, evaluated on
$\widetilde S$, gives half-integer fusion coefficients for
$\check S=\pm1$, so both real values are excluded and $\check S=\pm i$
with $\check S^{\,2}=-1$. By \eqref{eq:S2block}, conjugation exchanges $\Phi_+$ with $\Phi_-$, in agreement with $(\check S^{(12)})^2=-1$ found for the $\widehat{su}(2)_{12}$ extension.
Both signs obey $\widetilde S^{\,2}=\mathcal C$, because the block \eqref{eq:S2block}
involves $\check S$ only through $\check S^{\,2}$. Since the NS sector has no diagonal $T$, the relation $(ST)^3=\mathcal C$ that fixed $\check S^{(12)}$ can be imposed only together with the other spin structures. We find $\check S=+i$ at the end of this subsection and use this value in the modular matrix
$\widetilde S$ of the coset from here on.

These data fix the NS partition function. Since we treat the
super-$W_3$ algebra as the chiral algebra of the theory, its
representations are the five modules of Section~\ref{sub:coset}. The
exceptional modular invariant is the diagonal combination of their characters,
\begin{equation}\label{eq:ZNS}
\begin{aligned}
Z_{\rm NS}&=\sum_{a}\bigl|\hat\chi^{\rm NS}_a\bigr|^2
=\bigl|\hat\chi^{\rm NS}_{\mathbf1}\bigr|^2
+\bigl|\hat\chi^{\rm NS}_{\widehat\Phi_{1,3}}\bigr|^2
+\bigl|\hat\chi^{\rm NS}_{\widehat X}\bigr|^2
+\bigl|\hat\chi^{\rm NS}_{\Phi_+}\bigr|^2
+\bigl|\hat\chi^{\rm NS}_{\Phi_-}\bigr|^2\\
&=\bigl|\chi^{\rm NS}_{0}+\chi^{\rm NS}_{5/2}+\chi^{\rm NS}_{11/2}+\chi^{\rm NS}_{15}\bigr|^2
+\bigl|\chi^{\rm NS}_{5/14}+\chi^{\rm NS}_{6/7}+\chi^{\rm NS}_{20/7}+\chi^{\rm NS}_{145/14}\bigr|^2\\
&+\bigl|\chi^{\rm NS}_{1/14}+\chi^{\rm NS}_{15/14}+\chi^{\rm NS}_{11/7}+\chi^{\rm NS}_{46/7}\bigr|^2
+2\,\bigl|\chi^{\rm NS}_{1/7}+\chi^{\rm NS}_{51/14}\bigr|^2\,,
\end{aligned}
\end{equation}
where $\hat\chi^{\rm NS}_a(\tau)$ is the character of the extended module $a$ and $\chi^{\rm NS}_h(\tau)$ is the NS character of the super-Virasoro tower with lowest conformal weight $h$, so that each extended module is the sum of the towers listed in Table~\ref{tab:modules}.
This partition function is invariant under the modular transformations
that preserve the NS sector. Under
$\tau\to-1/\tau$, the extended characters transform among themselves through the unitary matrix $\widetilde S$, which leaves the diagonal sum unchanged. Under $\tau\to\tau+2$, each character acquires a
pure phase, and the phase cancels in $|\hat\chi^{\rm NS}_a|^2$.

The remaining modular images of $Z_{\rm NS}$ are the partition functions of the other two spin structures. One of them, $Z_{\widetilde{\rm NS}}$, has $(-1)^{F+\tilde F}$ inserted in the NS sector and comes from the same characters through the phases of $\tau\to\tau+1$. The other, $Z_{\rm R}$, has Ramond boundary conditions and is generated by the Ramond data of Section~\ref{sub:parity}, see \eqref{eq:ZR}. The fourth combination, the Ramond supertrace,
vanishes here, because only states at $h=c/24=5/84$ contribute
and the lowest Ramond conformal weight of Table~\ref{tab:modules} is $\frac1{14}$. The Witten index therefore vanishes, in agreement with the count of critical points of the superpotential in Section~\ref{sub:flow}.

We now determine the sign of $\check S$. The transformation $\tau\to\tau+1$ exchanges the NS characters with the $\widetilde{\rm NS}$ ones and multiplies each Ramond character by a phase, while $\tau\to-1/\tau$ exchanges the $\widetilde{\rm NS}$ characters with the Ramond ones, as we show in Section~\ref{sub:parity}. The resulting matrices $\bm S$ and $\bm T$ on the three sectors follow from the modular data of the bosonic theory with the partition function \eqref{eq:Ztot}, so they obey $\bm S^2=(\bm S\bm T)^3=\mathcal C$. As in the determination of $\check S^{(12)}$, we restrict them to the eigenspace of $\mathcal C$ with the eigenvalue $\sigma=-1$, which in the NS and $\widetilde{\rm NS}$ sectors is spanned by the antisymmetric combinations $v_{\rm NS}$ and $v_{\widetilde{\rm NS}}$ of $\Phi_+$ and $\Phi_-$. Since $\bm S$ maps $v_{\widetilde{\rm NS}}$ into the Ramond sector, $\mathcal C$ also exchanges two Ramond modules, necessarily the pair $R_{9/14,\pm}$ of equal conformal weights, and we denote their antisymmetric combination by $v_{\rm R}$. On these vectors, $\bm Sv_{\rm NS}=\check S\,v_{\rm NS}$, $\bm Sv_{\widetilde{\rm NS}}=\alpha_1v_{\rm R}$ and $\bm Sv_{\rm R}=\alpha_2v_{\widetilde{\rm NS}}$, with $\alpha_1\alpha_2=\sigma$ by $\bm S^2=\mathcal C$. The matrix $\bm T$ exchanges $v_{\rm NS}$ with $v_{\widetilde{\rm NS}}$ up to the phase $t_{\rm NS}=e^{2\pi i(1/7-5/84)}$ and multiplies $v_{\rm R}$ by $t_{\rm R}=e^{2\pi i(9/14-5/84)}$. The product $\bm S\bm T$ therefore permutes $v_{\rm NS}$, $v_{\rm R}$ and $v_{\widetilde{\rm NS}}$ cyclically up to factors, and $(\bm S\bm T)^3=\mathcal C$ requires $t_{\rm NS}^2\,t_{\rm R}\,\check S\,\alpha_1\alpha_2=\sigma$. We thus find
\begin{equation}
\check S=\bigl(t_{\rm NS}^2\,t_{\rm R}\bigr)^{-1}=+i\,,
\label{eq:checkScoset}
\end{equation}
the complex conjugate of $\check S^{(12)}=-i$.

\subsection{Chiral fermion parity and the Ramond sector}
\label{sub:parity}

A fusion product $a\times b$ is a sum of channels, and supersymmetry gives
each channel two kinds of three-point structure, an even one and an odd
one. In Section~\ref{sub:gradedring} we determine which kind occurs in each
channel, using the chiral fermion parity $\omega_a$ of the lowest state of
each extended module. In this subsection we determine the five parities $\omega_a$.

Let us first count the independent three-point structures in a single channel,
since supersymmetry doubles the usual number. On one chiral half, the superconformal
transformations form the supergroup $OSp(1|2)$. Its even part takes any three points to $0$, $1$ and $\infty$ with no transformation remaining, so there is
no even invariant, while its two odd parameters remove two of the three odd
coordinates and leave one. Since a function of one Grassmann variable has two
components, a channel admits two structures per chirality, one even and
one odd. A three-point function of the theory combines a left structure, in the channel $[a_3]$ of $a_1\times a_2$, with a right one, in the channel $[\tilde a_3]$ of $\tilde a_1\times\tilde a_2$. The two channels coincide because the
right module of every field equals its left module in \eqref{eq:ZNS}. All counts below are for a single chirality.

Here we consider the correlators in which each module contributes either its lowest state, which is a bottom component, or the top component that the supercurrent produces from the bottom. On one chirality, a correlator of
bottom and top components belongs to the even structure when the number
of top components is even and to the odd structure when it is odd.

As in Section~\ref{sub:warmup}, $N^+$ counts the even structures of a channel and $N^-$ the odd ones. Because the correlator of three bottom components lies in the even structure, a channel with $N^+=0$ has no three-point function of lowest states even when $N^-$ is not zero. The selection rules of Section~\ref{sub:selection} follow from
this vanishing.

Fermion parity distinguishes the even structure from the odd one. Let us define the parity product of a three-point structure
as the product of the fermion parities of its three states. Since the
supercurrent is parity-odd, a top component has the opposite fermion parity from
its bottom component. The even and the odd structure of a channel therefore
have opposite parity products, and exactly one of the two structures has
parity product $+1$. These signs alone do not decide whether the other structure occurs. In Section~\ref{sub:gradedring} we show from the modular data that no
channel admits both, which signals a symmetry.

We find that the five chiral fermion parities are
\begin{equation}
\bigl(\omega_{\mathbf1},\omega_{\widehat\Phi_{1,3}},
\omega_{\widehat X},\omega_{\Phi_+},\omega_{\Phi_-}\bigr)=(+,-,-,+,+)\,.
\label{eq:omega}
\end{equation}
In the rest of this subsection, we obtain them in two independent ways.

First, we derive the chiral fermion parities from the free numerator of the coset construction. We recall
from Section~\ref{sub:coset} that with the $E_6$ invariant the coset is $[\widehat{so}(5)_1\times\widehat{su}(2)_2]/\widehat{su}(2)_{12}$, and that of its eight free Majorana fermions five come from $\widehat{so}(5)_1$ and three from $\widehat{su}(2)_2$. In the NS
sector of a free-fermion theory, every mode is half-integer, so a state built from $n$ fermion modes has conformal weight $n/2$ modulo integers and fermion parity $(-1)^n=(-1)^{2h}$, which depends on the conformal weight only modulo one.

Consider a state of the coset module $a$ with conformal weight
$h$. Inside the numerator, this state appears in a tensor product with a state of a denominator
representation. We call the label of this representation $s_a$. The denominator state has conformal weight $h^{(12)}_{s_a}$ modulo
integers, so the coset state descends from numerator states of conformal weight $h+h^{(12)}_{s_a}$ modulo
integers. The denominator, in contrast, is bosonic and has no fermion parity. Consequently, the coset state inherits the fermion parity $(-1)^{2(h+h^{(12)}_{s_a})}$ of the numerator state from which it descends. Both members of the orbit of $a$ give the same $h^{(12)}_{s_a}$ modulo one, because $h^{(12)}_{14-s}-h^{(12)}_{s}=\frac{7-s}2$ is an integer at odd $s$. For the lowest state, whose conformal weight we denote $h_a$, the fermion parity is
\begin{equation}\label{eq:omegarule}
    \omega_a=(-1)^{2(h_a+h^{(12)}_{s_a})}\,.
\end{equation}
The five sums $h_a+h^{(12)}_{s_a}$ are $0$, $\frac5{14}+\frac17=\frac12$,
$\frac1{14}+\frac37=\frac12$, and $\frac17+\frac67=1$ twice, which
reproduces \eqref{eq:omega}. The same rule at general $h$ also fixes the relative fermion parities
within each extended module, since a shift of $h$ by one half reverses the
sign, in agreement with the fact that the supercurrent is parity-odd.

Second, we determine the chiral fermion parities from the characters of the theory itself, using the Ramond sector. The same computation also gives the modular matrix weighted by fermion parity, which we need in Section~\ref{sub:gradedring}. On the torus, a fermionic theory has four sectors, three of which enter here.
Besides the NS characters, there are the NS
characters with $(-1)^F$ inserted and the Ramond characters
$\hat\chi^{\rm R}_\rho$. In Section~\ref{sub:modular} we used the transformation $\tau\to\tau+1$, which exchanges the first two. Because the transformation $\tau\to-1/\tau$
exchanges the roles of space and time on the torus, it replaces the
insertion of $(-1)^F$ by a change of boundary condition and maps
the NS characters with $(-1)^F$ inserted to the Ramond characters. The modular matrix that implements this map is new modular data. Because
the $(-1)^F$ insertion contributes the overall sign $\omega_a$ to the
character of the module $a$, matching the two sides of the map fixes the
$\omega_a$.

To write the map explicitly, we fix conventions on both sides. The extended module $a$ has ordinary character $\hat\chi^{\rm NS}_a$, the sum of its super-Virasoro characters. The map is written in terms of the $(-1)^F$-graded version of $\hat\chi^{\rm NS}_a$. A state
at level $\ell$ above the lowest state has fermion parity $\omega_a(-1)^{2\ell}$, because
the odd generators $G$ and $U$ raise the level by half-integers while the
even generators $L$ and $W$ raise it by integers. The bottom and its
superpartner therefore have opposite fermion parity. Let us collect the
level factor $(-1)^{2\ell}$ into a graded character $\hat\chi^{\widetilde{\rm NS}}_a$, the
trace over the extended module in which a state at level $\ell$ above the
lowest state is counted with $(-1)^{2\ell}$, so that the lowest state contributes
$+1$. The towers listed in Table~\ref{tab:modules} determine $\hat\chi^{\widetilde{\rm NS}}_a$, since the level factor $(-1)^{2\ell}$ follows from the conformal weights. The $(-1)^F$-graded character is then $\omega_a\hat\chi^{\widetilde{\rm NS}}_a$, the unknown sign times a known $q$-series.

On the Ramond side,
the five modules were constructed in Section~\ref{sub:coset} and listed in
Table~\ref{tab:modules}. Since the two members of the resolved pair
$R_{9/14,\pm}$ have identical characters, the five modules give
four linearly independent $q$-series.

The Ramond sector has the two zero modes $G_0$ and $U_0$
\cite{Schoutens:1990xg}. The first satisfies $G_0^2=h-c/24$, so it
pairs the lowest states two by two unless the conformal weight is exactly
$c/24$ \cite{Cohn:1987hw}. The Ramond module $(r,s)=(\frac m2,\frac m2+1)$ of
$SM(m)$ has conformal weight $c/24$ for every even $m$ \cite{Chen:2019sif},
but it is absent from the $\ed$ superconformal minimal model, whose Ramond sector of
Section~\ref{sub:coset} admits only $r=4$ and $r=8$ while $\frac m2=6$ at
$m=12$. Since no Ramond conformal weight of Table~\ref{tab:modules} equals
$c/24=\frac5{84}$, the lowest level of a Ramond module is degenerate. It contains two states in
$SM(3)$, which has only $G_0$, and four states in the $\ed$ superconformal minimal model,
where $U_0$ acts as well \cite{Schoutens:1990xg}. With $\hat\chi^{\rm R}_\rho$ normalized to leading
coefficient one, the prefactor of the modular transformation is the $\sqrt2$ of
\eqref{eq:wnstor} in $SM(3)$ and $2$ in \eqref{eq:nstor} below.

We denote by $U$ the matrix that takes the graded characters to the Ramond
ones,
\begin{equation}\label{eq:nstor}
    \hat\chi^{\widetilde{\rm NS}}_a(-1/\tau)=2\sum_{\rho}U_{a\rho}\,\hat\chi^{\rm R}_\rho(\tau)\,,
\end{equation}
where $\rho$ runs over the five Ramond modules. Note that the rows of $U$ are labeled by NS modules and its columns by Ramond ones. We recall that each of the three orbits gives one module of each sector, and that the fixed point resolves into $\Phi_\pm$ in the NS sector and $R_{9/14,\pm}$ in the Ramond one. We assign the Ramond module of each orbit to the column of the NS module of the same orbit, and $R_{9/14,+}$ and $R_{9/14,-}$ to the columns of $\Phi_+$ and $\Phi_-$. Once the characters are normalized as above, both sides of \eqref{eq:nstor} are fixed $q$-series, and matching them row by row determines $U$. Because $\hat\chi^{\rm R}_{9/14,+}=\hat\chi^{\rm R}_{9/14,-}$, the identity involves
the two fixed-point columns only through their sum. The differences
\begin{equation}
d_a=U_{a,R_{9/14,+}}-U_{a,R_{9/14,-}}\,,
\label{eq:dadef}
\end{equation}
one for each of the five rows, are therefore left undetermined. Every component that the
identity does determine agrees with the parity-weighted $\widetilde S$,
\begin{equation}
U_{a\rho}=\omega_a\,\widetilde S_{a\rho}\,,
\label{eq:U}
\end{equation}
with the signs \eqref{eq:omega}.\footnote{The index $\rho$ labels a Ramond module on the left of \eqref{eq:U} and an NS module on the right. The relation \eqref{eq:U} therefore depends on the columns chosen below \eqref{eq:nstor}.} Since $\omega_{\mathbf1}=+1$, the vacuum row confirms the prefactor $2$ of \eqref{eq:nstor}, sign included.
Reversing any of the other four signs would multiply the whole right-hand
side of its row by $-1$, so the two sides would differ already in the
leading term. The characters therefore determine those four signs, and the values
agree with \eqref{eq:omega}.

The same map gives the Ramond partition function. Substituting \eqref{eq:nstor} into the NS partition function with $(-1)^{F+\tilde F}$ inserted, $Z_{\widetilde{\rm NS}}=\sum_a|\hat\chi^{\widetilde{\rm NS}}_a|^2$, and using the unitarity of $\widetilde S$, we find that $\tau\to-1/\tau$ takes it to
\begin{equation}\label{eq:ZR}
\begin{aligned}
    Z_{\rm R}&=4\sum_{\rho}\bigl|\hat\chi^{\rm R}_\rho\bigr|^2=4\bigl|\chi^{\rm R}_{3/2}+\chi^{\rm R}_{15/2}\bigr|^2+4\bigl|\chi^{\rm R}_{5/14}+\chi^{\rm R}_{61/14}\bigr|^2\\
    &+4\bigl|\chi^{\rm R}_{1/14}+\chi^{\rm R}_{29/14}\bigr|^2+8\bigl|\chi^{\rm R}_{9/14}\bigr|^2\,,
\end{aligned}
\end{equation}
where $\chi^{\rm R}_h(\tau)$ is the Ramond character of the super-Virasoro tower with lowest conformal weight $h$, normalized to leading coefficient one, like $\hat\chi^{\rm R}_\rho$. The factor $4$ is the square of the prefactor of \eqref{eq:nstor}, and the last term collects the two members of the resolved pair $R_{9/14,\pm}$, whose characters coincide. This partition function is invariant under $\tau\to\tau+1$ and is mapped back to $Z_{\widetilde{\rm NS}}$
under $\tau\to-1/\tau$. The complete modular invariant partition function of the $\ed$ model is
\begin{equation}\label{eq:Ztot}
Z=\frac12\bigl(Z_{\rm NS}+Z_{\widetilde{\rm NS}}+Z_{\rm R}\bigr)\,.
\end{equation}
The fourth spin structure, the Ramond sector with $(-1)^{F+\tilde F}$ inserted, would add the Witten index, which vanishes here.

The matrix $U$ can also be checked against the unextended theory $SM(12)$. Its modular matrix on the super-Virasoro characters is known explicitly \cite{Minces:1998vc}, and one block of it takes the NS characters with $(-1)^F$ inserted to the Ramond characters. Each extended module is a sum of super-Virasoro modules, listed in Table~\ref{tab:modules}. Summing that block over the super-Virasoro modules of each extended module, with the NS towers weighted by their relative fermion parities $(-1)^{2(h-h_a)}$, where $h$ is the lowest conformal weight of the tower, gives a matrix on the extended modules. The result agrees with $U$ of \eqref{eq:U}, signs $\omega_a$ included. The exception is $d_a$ of \eqref{eq:dadef}, which the sums cannot determine, because they are built from characters and the two resolved modules have the same character.

To determine $d_a$, we can use the graded Verlinde formula \eqref{eq:gv} of Section~\ref{sub:gradedring},
in the same way as the ordinary one fixed $\check S^{\,2}$ in Section~\ref{sub:modular}. Its multiplicities $N^\pm_{abc}$ count three-point structures, so they are non-negative integers. Imposing this condition on all $5^3$ components of $N^\pm_{abc}$ gives two solutions. Both have $d_a=0$ on the three orbit rows and
$d_{\Phi_+}=-d_{\Phi_-}$ on the other two, with $d_{\Phi_+}=i$ in one
solution and $d_{\Phi_+}=-i$ in the other. The two solutions are opposite in sign but give the same $N^\pm_{abc}$ on all components. Which of the two solutions holds is not determined here, but for definiteness we use the one that extends \eqref{eq:U} to the
fixed-point columns, so that $U_{a\rho}=\omega_a\widetilde S_{a\rho}$ holds on all five columns. The
results below do not depend on the choice.

\subsection{The graded fusion ring}
\label{sub:gradedring}

We can now compute the graded fusion ring from the modular matrix
$\widetilde S$ of Section~\ref{sub:modular} and the parity-weighted matrix $U$
of Section~\ref{sub:parity}. For a fermionic theory, the graded Verlinde formula uses
the modular matrix of each spin structure of the torus
\cite{Aasen:2017ubm,Lou:2020gfq}. Its two sums give
$N^++N^-$ from $\widetilde S$ and $N^+-N^-$ from $U$, as in \eqref{eq:wgv}, so that
\begin{equation}\label{eq:gv}
    N^{\pm}_{abc}=\frac12\left(\sum_\rho\frac{\widetilde S_{a\rho}\widetilde S_{b\rho}\overline{\widetilde S_{c\rho}}}{\widetilde S_{\mathbf1\rho}}\ \pm\ \sum_\rho\frac{U_{a\rho}U_{b\rho}\overline{U_{c\rho}}}{U_{\mathbf1\rho}}\right)\,.
\end{equation}
The two sums run over the five NS and the five Ramond modules, respectively, which we label by the same orbits in Table~\ref{tab:modules}, so one index $\rho$ serves both.

We list the graded fusion ring in
Table~\ref{tab:fusion}. For example, the product
$\Phi_+\times\Phi_-$ contains the channel $[\widehat X]_-$, so the three-point
function $\langle\Phi_+\Phi_-\widehat X\rangle$ of the three lowest
states vanishes, while the correlators of the odd structure, which
contain top components, can be nonzero.

Let us check the construction by applying it to the exceptional modular invariant $(D_6,E_6)$ of $SM(10)$, whose graded fusion ring is known because the model coincides with the tensor square
$SM(3)^{\otimes2}$ \cite{Li:1988pj,Kastor:1988ef,Gang:2008sz,Makabe:2017ygy}. At $m=10$, the two labels exchange roles, because $D_6$ extends the numerator factor $\widehat{su}(2)_8$ by a simple current while the conformal embedding $E_6$ replaces the denominator $\widehat{su}(2)_{10}$ by $\widehat{so}(5)_1$. There the lowest states of the extended NS modules have conformal weights $0$, $\frac1{10}$, $\frac1{10}$ and $\frac15$ at central charge $\frac75=2\times\frac7{10}$. These are the lowest weights and the central charge of two copies of $SM(3)$. A module of the tensor square is a pair, one module from each $SM(3)$ factor, and its chiral fermion parity is the product of those of the two modules. The graded fusion ring of the tensor square then follows from $[X]\times[X]=[\mathbf1]_++[X]_-$ of Section~\ref{sub:warmup}. Our construction at $m=10$ reproduces $N^+$ and $N^-$ in every channel. The identification is confirmed on the Lagrangian side in Section~\ref{sub:flow}.

\begin{table}[!htbp]
\caption{\label{tab:fusion}Graded fusion ring of the $\ed$ superconformal minimal model. A subscript $+$ or
$-$ shows whether the channel admits an even or an odd structure
in the sense of Section~\ref{sub:parity},
and square brackets denote extended modules. Products with the identity are omitted.}
\centering
\begin{tabular}{ll}
\toprule
$\widehat\Phi_{1,3}^{\,2}$ & $[\mathbf1]_++[\widehat\Phi_{1,3}]_-+[\widehat X]_-$\\
$\widehat\Phi_{1,3}\widehat X$ & $[\widehat\Phi_{1,3}]_-+[\widehat X]_-+[\Phi_+]_++[\Phi_-]_+$\\
$\widehat\Phi_{1,3}\Phi_\pm$ & $[\widehat X]_++[\Phi_\mp]_-$\\
$\widehat X^2$ & $[\mathbf1]_++[\widehat\Phi_{1,3}]_-+2[\widehat X]_-+[\Phi_+]_++[\Phi_-]_+$\\
$\widehat X\Phi_\pm$ & $[\widehat\Phi_{1,3}]_++[\widehat X]_++[\Phi_\pm]_-$\\
$\Phi_\pm^2$ & $[\widehat\Phi_{1,3}]_-+[\Phi_\mp]_+$\\
$\Phi_+\Phi_-$ & $[\mathbf1]_++[\widehat X]_-$\\
\bottomrule
\end{tabular}
\end{table}

\begin{sloppypar}
The rest of the paper uses two features of the graded fusion ring. The first feature is that the chiral fermion parities grade the fusion ring. With $U_{a\rho}=\omega_a\widetilde S_{a\rho}$, the second sum of \eqref{eq:gv} is
$\omega_a\omega_b\omega_c$ times the first, which is the ordinary fusion coefficient
$N_{abc}$, so
\begin{equation}
N^{\pm}_{abc}=\frac12N_{abc}\bigl(1\pm\omega_a\omega_b\omega_c\bigr)\,.
\label{eq:npm}
\end{equation}
As in \eqref{eq:wnpm}, each channel therefore admits structures of only one parity, the even
ones when $\omega_a\omega_b\omega_c=+1$ and the odd ones when it is $-1$.
\end{sloppypar}

The second feature concerns a single channel, the one that identifies $\widehat X$ with the $\ZR$-odd order parameter $\phi_X$ in Section~\ref{sub:selection}:
\begin{equation}
\bigl(N^+,N^-\bigr)_{\widehat X\widehat X\widehat X}=(0,2)\,,
\qquad\text{so}\qquad
\bigl\langle\widehat X_{\rm low}\widehat X_{\rm low}\widehat X_{\rm low}\bigr\rangle=0\,,
\label{eq:XXX}
\end{equation}
where the subscript marks the lowest states. The super-Virasoro fusion rules would allow a nonzero
three-point function of three $\widehat X$ lowest states. The
vanishing in \eqref{eq:XXX} is therefore a property of the extended algebra,
not of the super-Virasoro algebra. The channel itself, however, is not empty, since $N_{\widehat X\widehat X\widehat X}=2$. The correlator of the three lowest states vanishes because $\omega_{\widehat X}=-1$ gives
$\omega_{\widehat X}^3=-1$ and hence $N^+=0$ by \eqref{eq:npm}.

The multiplicity $N_{\widehat X\widehat X\widehat X}=2$ comes from the orbit structure. In
$\widehat{su}(2)_{12}$, the fusion of $s=5$ with itself is
$5\times5=1+3+5+7+9$. The extended module $\widehat X$ is the orbit $\{5,9\}$,
and the product contains each member of the orbit once, so the channel
$[\widehat X]$ appears with multiplicity two. The term $7$ is the fixed
point and produces the channels $[\Phi_+]$ and $[\Phi_-]$ of
Table~\ref{tab:fusion}.

The graded fusion coefficients \eqref{eq:npm} also give a conservation law. In every nonvanishing
three-point structure, the fermion parities of the three states multiply to $+1$. A state at level $\ell$ above the lowest state of module $a$ has
fermion parity $\omega_a(-1)^{2\ell}$ by Section~\ref{sub:parity}, so the three
states of a structure have fermion parities whose product is
$\omega_a\omega_b\omega_c\,(-1)^{n}$, where $n$ counts the top components. The graded fusion coefficients \eqref{eq:npm} keep the even structures, which have an even $n$, exactly when
$\omega_a\omega_b\omega_c=+1$, and the odd ones exactly when it is $-1$, so
the product is $+1$ in both cases. The conservation law extends from the bottom and
top components to all states, because every mode of the chiral algebra
has a definite fermion parity. We emphasize that the conserved charge is the fermion parity of the individual state, not the sign $\omega_a$ of its extended module.

\subsection{The $\ZR$ symmetry and conjugation}
\label{sub:rparity}

The exact symmetries of the $\ed$ superconformal minimal model can
now be derived. They act on the state space of the diagonal invariant
\eqref{eq:ZNS}. Let $B_a$ denote the extended module with character
$\hat\chi^{\rm NS}_a$. The state space is a sum of five sectors
$B_a\otimes\tilde B_a$. An
internal $\mathbb Z_2$ can act on the state space in three ways. It can
multiply each sector by a constant sign, it can act
on the individual states of each sector, or it can permute the sectors.
The first way is excluded below, and the $\ZR$ symmetry of the LG
description is built from the other two. We take the
three ways in turn.

Constant signs are excluded by the fusion ring. A constant assignment $q_a$ is conserved only if $q_aq_bq_c=1$ on every
channel with $N_{abc}\neq0$. Such an assignment is a grading of the ordinary
fusion ring, whose multiplicities are the sums $N_{abc}=N^+_{abc}+N^-_{abc}$.
Note, however, that the ordinary fusion ring admits no nontrivial grading, because the channel
$\widehat X\in\widehat X\times\widehat X$ of Table~\ref{tab:fusion}
gives $q_{\widehat X}=1$, the channel
$\widehat\Phi_{1,3}\in\widehat\Phi_{1,3}\times\widehat\Phi_{1,3}$ gives
$q_{\widehat\Phi_{1,3}}=1$, and the channels
$\Phi_\mp\in\Phi_\pm\times\Phi_\pm$ give $q_{\Phi_+}=q_{\Phi_-}=1$.

The second way, action on the individual states, is realized by the chiral fermion parity, which is conserved state by state rather than as
one sign per sector. As an operator, it acts on each chirality separately. On a state
whose level above the lowest state is $\ell$ in its left module and $\tilde\ell$
in its right module, the two chiral fermion parities are
\begin{equation}
\omega=\omega_a(-1)^{2\ell}\,,\qquad
\tilde\omega=\omega_a(-1)^{2\tilde\ell}\,,
\label{eq:grading}
\end{equation}
with the $\omega_a$ of \eqref{eq:omega}. A three-point
function
of the theory combines a left structure with a right structure in one
common channel, as we explained in Section~\ref{sub:parity}, and the conservation
law at the end of Section~\ref{sub:gradedring} applies to the left
and the right structure separately. The theory therefore has two chiral fermion
parity symmetries, $\omega$ and $\tilde\omega$, whose product is the fermion parity
$\mathbb{Z}_2^F$ of \eqref{eq:fparity},
since $\omega_a^2=1$ and the lowest conformal weight cancels in
$\omega\tilde\omega=(-1)^{2(\ell-\tilde\ell)}=(-1)^{2(h-\tilde h)}$.

Each of $\omega$ and $\tilde\omega$ is the chiral fermion parity that grades the fusion ring. On the chiral algebra itself, $\omega$ acts by $(G,U)\to(-G,-U)$, the sign change of the odd currents. Every superconformal algebra admits this automorphism at every central charge, but this does not mean that it is a symmetry of every theory.
Here, its eigenvalues on the lowest states are the chiral fermion parities $\omega_a$
of \eqref{eq:omega}, and it is conserved in every three-point structure
precisely because those five signs obey the conservation law of
Section~\ref{sub:gradedring}. An automorphism of the chiral algebra therefore becomes a symmetry of the theory only when combined with a consistent action on the extended modules.

The third way, permutation of the sectors, is realized by conjugation. The modular relation $\widetilde S^{\,2}=\mathcal C$ of Section~\ref{sub:modular} defines a permutation of the five labels. It fixes $\mathbf1$, $\widehat X$ and $\widehat\Phi_{1,3}$, whose modules are self-conjugate, and exchanges $\Phi_+$ with $\Phi_-$, which are conjugate to each other because $\check S^{\,2}=-1$. We look for an operator on the state space that realizes this permutation. Any such operator acts on
the two chiralities together, because an action on the left alone would
map the sector $\Phi_+\otimes\tilde\Phi_+$ to $\Phi_-\otimes\tilde\Phi_+$,
which the diagonal invariant does not contain. Such an operator can map $\Phi_+\otimes\tilde\Phi_+$ to $\Phi_-\otimes\tilde\Phi_-$ and each of the other three sectors to itself. We denote the operator by $C$,
after the conjugation $\mathcal C$ of labels that it realizes.

The example of $\omega$ shows where to look for $C$. We seek an
automorphism of the chiral algebra whose action on the extended modules
realizes the conjugation $\mathcal C$. Let us consider the sign change
$A:(U,W)\to(-U,-W)$, which fixes the stress tensor and the
supercurrent. For $A$ to preserve the
operator algebra, every structure constant with an odd total number of $U$
and $W$ currents must vanish. Most of them vanish for simple reasons. A structure constant with a single $U$ or $W$ current vanishes, because $(U,W)$ form a supermultiplet distinct from the identity and therefore cannot appear in the product of two
super-Virasoro descendants of the identity. Among the structure constants
with three such currents, fermion parity forbids a nonzero $C_{UUU}$ and
a nonzero $C_{UWW}$, since each has an odd number of fermionic currents.
The two structure constants that remain are $C_{UUW}$ and $C_{WWW}$, but the exchange
symmetry of identical operators makes both vanish.

To see this, note that exchanging the two identical currents $V$ in $\langle VVB\rangle$
multiplies the correlator by $(-1)^{h_B-2h_V}$, and a fermionic $V$
contributes one further sign $-1$. A nonzero structure constant requires
the product of these signs to be $+1$. For $V=W$, which is bosonic with $h_V=3$, the requirement is that
$h_B$ be even. For $V=U$, which is fermionic with $h_V=\frac52$, the
statistics sign and the half-integer conformal weight compensate each other, and
the requirement is again that $h_B$ be even. Both $C_{UUW}$ and $C_{WWW}$ have $B=W$, and $h_W=3$ is odd, so
they vanish,
\begin{equation}
C_{UUW}=0\,,\qquad C_{WWW}=0\,.
\label{eq:noself}
\end{equation}
Since every structure constant with an odd total number of $U$ and $W$ currents vanishes, $A$ is an order-two automorphism of $SW(3/2,5/2)$
\cite{Honecker:1992kz}. We also checked $A$ directly against the operator
products of $SW(3/2,5/2)$ at $c=\frac{10}7$, given in component form
in \cite{Ahn:1990nr}, and found every one of them invariant.\footnote{The vanishing \eqref{eq:noself} follows from the spins and the statistics of the two currents, so any algebra with this generator content admits the automorphism. With the spins $\frac32,2,\frac52,3$ and no further generators, associativity restricts the central charge to two values
\cite{Inami:1988xy}, but only $c=\frac{10}7$ is unitary \cite{Inami:1988xy,Hornfeck:1990zw}.}

The coset realization makes $A$ concrete. In the free-fermion form of the currents given in
\cite{Ahn:1990nr}, the eight numerator fermions form the adjoint of $su(3)$. The stress
tensor is built with $\delta^{ab}$, the supercurrent with the structure
constants $f^{abc}$, and $U$ and $W$ with the symmetric invariant $d^{abc}$. The denominator of
\eqref{eq:coset} is the $so(3)$ subalgebra of the $su(3)$, and under the $so(3)$ the adjoint
splits into the three fermions of $\widehat{su}(2)_2$ and the five of $\widehat{so}(5)_1$.
Each current of the denominator is the sum of an $\widehat{su}(2)_{10}$ current
bilinear in the five and an $\widehat{su}(2)_2$ current bilinear in the three,
and the levels add to the twelve of $\widehat{su}(2)_{12}$. Because the outer automorphism of $su(3)$ fixes $\delta^{ab}$ and $f^{abc}$ but reverses $d^{abc}$, on the currents it
acts as $(T,G,U,W)\mapsto(T,G,-U,-W)$. This is $A$. On the fermions it is a sign, $+1$ on the three and $-1$ on the five, which appears squared in each current of the denominator. The denominator is therefore fixed pointwise, and $A$
descends to the coset.

Two distinct notions have now appeared: the conjugation $\mathcal C$ of the
five module labels and the automorphism $A$ of the currents. To define the
symmetry, we need a third, the lift of $A$ to the states, defined below. Let $V$ denote any of the four currents
and $AV$ its image under $A$. Because $A$ preserves the operator algebra, the vector space of an extended module
becomes a second extended module when each mode $V_n$ is taken
to act as $(AV)_n$. The second module is again one of the five, and $A$
induces the permutation that sends the label of the first to the label of
the second.

Because $A$ fixes the stress tensor and reverses $W$, the second module has
the same $L_0$ eigenvalues and the opposite spin-3 charge. An extended module whose lowest conformal
weight appears once in the spectrum can therefore be mapped only to itself, and $w=-w$ gives
$w=0$. This accounts for $\mathbf1$, $\widehat X$ and
$\widehat\Phi_{1,3}$, whose conformal weights $0$, $\frac1{14}$ and $\frac5{14}$
are nondegenerate. The modules $\Phi_+$ and $\Phi_-$ share $h=\frac17$
and take the opposite spin-3 charges $\pm w\neq0$ recorded in Section~\ref{sub:coset}, which requires $A$ to exchange them.
The permutation induced by $A$ is therefore the conjugation $\mathcal C$.

To act on the states themselves, $A$ has to be \emph{lifted}, that is,
realized by an operator $\hat A$ with $\hat A\,V_n\,\hat A^{-1}=(AV)_n$, which fixes the modes of $T$ and $G$ and reverses those of $U$ and $W$. A lift maps each extended module to its
image under $\mathcal C$. On an extended module that $A$ maps to itself, the lift
does not act by a single overall sign. Instead, it is constant on each super-Virasoro tower, whose states are connected by modes of $T$ and $G$ that commute with $\hat A$.
A mode of $U$ or $W$ maps a state to one of opposite $\hat A$-eigenvalue, so the sign depends on the tower, as computed in Section~\ref{sub:spinning}. The defining relation leaves an overall constant undetermined, which $\hat A^2=1$ restricts to a sign. We stress that the lift as a whole is defined
only up to one overall sign per module.

A lift is not automatically a
symmetry of the theory. The defining relation involves only the modes of the currents, so it does not ensure that
$\hat A$ preserves the correlators of the theory, as a symmetry must. For the chiral fermion parity, the graded fusion ring
provided the conservation law through the multiplicities $N^\pm_{abc}$, but we
know of no such grading for $A$. We therefore construct the operator $C$ in
the free realization of the coset, where the correlators can be computed directly.

We recall from Section~\ref{sub:coset} that with the $E_6$ invariant the numerator of the coset is $\widehat{so}(5)_1\times\widehat{su}(2)_2$, the theory of eight free Majorana fermions. In a free-fermion theory, an orthogonal transformation applied to
the left and the right fermions simultaneously is an exact symmetry,
because every correlator reduces by Wick contractions to the
propagators $\langle\psi^a\psi^b\rangle\propto\delta^{ab}$, and an
orthogonal transformation preserves them. The outer automorphism is such a transformation.

Because the transformation fixes every current of the denominator, on the left and on the right, it maps the commutant of the denominator to itself, and in the Goddard--Kent--Olive realization \cite{Goddard:1984vk,Goddard:1986ee} the commutant is the coset. The transformation is a symmetry of the coset theory, since it maps coset operators to coset operators and preserves all their correlators. On the coset currents it acts as $A$, so it is a lift of $A$ on each chirality. On the sectors it exchanges $\Phi_+\otimes\tilde\Phi_+$ with $\Phi_-\otimes\tilde\Phi_-$ and maps each of the other three to itself. This is the operator $C$ that realizes the conjugation $\mathcal C$.

On a sector that $C$ maps to itself, $C$ acts by a lift of $A$ on the left module and by another lift of $A$ on the right. Each lift is defined only up to an overall sign, but in the free realization the lowest state of the sector is the same combination of fermion modes on the left and on the right, and $C$ acts on both by the same sign. It therefore multiplies the lowest state by the square of a sign, that is, by $+1$. On a state built from one left tower and one right tower, $C$ is then the product of the two tower signs, computed in Section~\ref{sub:spinning}.

We have now obtained the three symmetries, the two chiral fermion parities
$\omega$ and $\tilde\omega$ and the conjugation $C$. They commute, because $\omega$ and $\tilde\omega$ take equal values
on $\Phi_+$ and $\Phi_-$, which share both $\omega_a$ and the levels of their towers.
Together they generate the group
$\mathbb Z_2\times\mathbb Z_2\times\mathbb Z_2$. Two of its elements,
\begin{equation}
R=\omega\,C\,,\qquad \tilde R=\tilde\omega\,C\,,
\label{eq:Rop}
\end{equation}
are the R-parities of the theory. Since $C$ is $+1$ on the lowest state of a sector that it maps to itself, as shown above, the R-parities on the supermultiplet of that lowest state reduce to the chiral fermion parities \eqref{eq:grading}, $R=\omega_a(-1)^{2\ell}$ and $\tilde R=\omega_a(-1)^{2\tilde\ell}$. The lowest state of the $\widehat X$ sector, where $\omega_{\widehat X}=-1$, is thus a boson that is odd under both. The fermion reached by $G_{-1/2}$ and the top component, at $\ell=\frac12$, are even under $R$, while the fermion reached by $\tilde G_{-1/2}$, at $\ell=0$, is odd. These agree with the $\ZR$ charges of $\phi_X$, $\psi_X$, $F_X$ and $\tilde\psi_X$ in \eqref{eq:Rparity}. The two R-parities agree on every boson,
differ in sign on every fermion, and are exchanged by the
reflection that interchanges $h$ with $\tilde h$.

It remains to identify these symmetries with the discrete symmetries of
the LG theory of Section~\ref{sec:model}. The chiral fermion parities alone cannot
match the $\ZR$ transformation
\eqref{eq:Rparity}, because they do not distinguish $\Phi_+$ from
$\Phi_-$. In the dictionary of Section~\ref{sec:fixedpoint}, the two Hermitian combinations $\frac{1}{\sqrt 2}(\Phi_++\Phi_-)$ and $\frac{1}{i\sqrt 2}(\Phi_+-\Phi_-)$ of the pair correspond to $P$ and $Y$ and have opposite $\ZR$ charges, while $\omega$ and $\tilde\omega$ give them equal charges. The operator $C$ provides the missing sign. It is $+1$ on the first combination and $-1$ on the second, which are also the $\Zex$ charges of $P$ and $Y$. We therefore identify $C$ with $\Zex$ and $R=\omega C$ with $\ZR$. The choice of $R$ over $\tilde R$ fixes which Grassmann coordinate the reflection in \eqref{eq:Rparity} acts on. With this identification, the lowest state of $\widehat X$ has the charges $(-,+)$ that the
dictionary of Section~\ref{sec:fixedpoint} gives to $\phi_X$. Note that the identification of $\Zex$ with conjugation holds at $\ed$, where the two modules that $\Zex$ exchanges are conjugate to each other. It does not hold at $(D_6,E_6)$, where the resolved pair consists of $[X]\otimes[\mathbf1]$
and $[\mathbf1]\otimes[X]$ of $SM(3)^{\otimes2}$, each self-conjugate, so conjugation is
the identity on the pair. There $\Zex$ still exchanges the pair, because it exchanges
the two factors.

\phantomsection\label{sub:selection}%
Let us now derive the selection rules from the graded fusion ring. By \eqref{eq:npm}, the
grading makes seven three-point functions of lowest states vanish. A
subscript $-$ in Table~\ref{tab:fusion} marks a channel with $\omega_a\omega_b\omega_c=-1$,
whose even structure is absent and whose correlator of lowest states
therefore vanishes. To pass from a channel to a correlator, recall that
$[c]$ appears in $a\times b$ exactly when the three-point function of
$a$, $b$ and the conjugate of $c$ can be nonzero, and that $\mathcal C$
exchanges $\Phi_+$ with $\Phi_-$. One of the seven is $\bigl\langle\widehat X_{\rm low}\widehat X_{\rm low}\widehat X_{\rm low}\bigr\rangle$ of
\eqref{eq:XXX}.
The others are $\langle\widehat\Phi_{1,3}\widehat\Phi_{1,3}\widehat\Phi_{1,3}\rangle$,
$\langle\widehat\Phi_{1,3}\widehat\Phi_{1,3}\widehat X\rangle$,
$\langle\widehat\Phi_{1,3}\widehat X\widehat X\rangle$,
$\langle\widehat\Phi_{1,3}\Phi_+\Phi_+\rangle$,
$\langle\widehat\Phi_{1,3}\Phi_-\Phi_-\rangle$ and
$\langle\widehat X\Phi_+\Phi_-\rangle$, all taken between the lowest
states of the three modules. The operator $C$ exchanges
$\langle\widehat\Phi_{1,3}\Phi_+\Phi_+\rangle$ with
$\langle\widehat\Phi_{1,3}\Phi_-\Phi_-\rangle$ and leaves
$\langle\widehat X\Phi_+\Phi_-\rangle$ invariant, which reduces the seven to six independent ones. These are exactly the three-point functions that the ordinary fusion rules allow but the $\ZR$ and $\Zex$ charges of the LG fields forbid. The vanishing in
\eqref{eq:XXX} identifies $\widehat X$ with the $\ZR$-odd order parameter $\phi_X$.

The signs $\omega_a$ are the only ones that produce these selection rules. Suppose a second assignment $q'$ produced the same selection rules, with $q'_aq'_bq'_c=\omega_a\omega_b\omega_c$ on every channel of Table~\ref{tab:fusion}. Then $p_a=q'_a\omega_a$ would satisfy $p_ap_bp_c=1$ on every channel, so $p$ would be a grading of the ordinary fusion ring. We showed above that the ordinary fusion ring admits no nontrivial grading, which gives $p_a=1$ and $q'_a=\omega_a$. The chiral fermion parities are thus uniquely determined.

\subsection{The spinning primaries and their charges}
\label{sub:spinning}

A diagonal modular invariant of the super-Virasoro algebra has only
spinless primaries. The exceptional invariant has spinning primaries because it is
diagonal for the \emph{extended} algebra, whose
modules, listed in Table~\ref{tab:modules}, are reducible under the
super-Virasoro algebra. The vacuum module, $\widehat X$ and
$\widehat\Phi_{1,3}$ decompose into four super-Virasoro towers per
chirality, and each of $\Phi_\pm$ into two. The lowest conformal weights $h_i$ of the towers are
those of Table~\ref{tab:modules}, $\{0,\frac52,\frac{11}2,15\}$ for
the vacuum module,
$\{\frac1{14},\frac{15}{14},\frac{11}7,\frac{46}7\}$ for
$\widehat X$, $\{\frac5{14},\frac67,\frac{20}7,\frac{145}{14}\}$ for
$\widehat\Phi_{1,3}$, and $\{\frac17,\frac{51}{14}\}$ for each of
$\Phi_\pm$. We number the towers by increasing $h_i$ within each extended module, starting from $i=0$.
Since $\hat\chi^{\rm NS}_a=\sum_i\chi^{\rm NS}_{h_i}$,
the term $|\hat\chi^{\rm NS}_a|^2$ of \eqref{eq:ZNS} contains every cross
term $\chi^{\rm NS}_{h_i}\overline{\chi^{\rm NS}_{h_j}}$. Each cross term pairs a left tower $i$ with a right tower
$j$ of the same extended module, and the lowest state of the sector it counts
is a super-Virasoro primary of the theory.
We label the primary $(h_i,\tilde h_j)$, whose spin is $h_i-h_j$.

The chiral fermion parity and the lift of $A$, both from Section~\ref{sub:rparity},
assign charges to the towers of an extended module. The primary of
the $i$-th tower has the chiral fermion parity $\omega_a(-1)^{2(h_i-h_a)}$ by \eqref{eq:grading}, with the $h_a$
of \eqref{eq:omegarule}, and inside the tower
the parity-odd generator $G$ reverses the fermion parity at every half-integer level. On the three
extended modules that $A$ fixes, $\mathbf1$, $\widehat\Phi_{1,3}$ and $\widehat X$, the lift $\hat A$
acts on each tower by a single sign and is defined up to one overall sign per
extended module, as we showed in Section~\ref{sub:rparity}. We denote by $C_i$ the sign on the
$i$-th tower. Only the ratios of the $C_i$ are therefore intrinsic. We set $C_0=+1$ in every extended module and determine the remaining $C_i$ below.

The chiral fermion parities of the operator at $(h_i,\tilde h_j)$ are
$\omega=\omega_a(-1)^{2(h_i-h_a)}$ and
$\tilde\omega=\omega_a(-1)^{2(h_j-h_a)}$. Their product is the fermion parity
$(-1)^{2(h_i-h_j)}$ of the operator, because $\omega_a$ and $h_a$ cancel. In the three
sectors that $C$ fixes, $C$ is $+1$ on the lowest state, as we showed in Section~\ref{sub:rparity}, so the overall signs of the two lifts cancel and the eigenvalue of $C$ is $C_iC_j$. The R-parities of \eqref{eq:Rop} are then
$R=\omega\,C_iC_j$ and $\tilde R=\tilde\omega\,C_iC_j$. On the resolved pair $\Phi_\pm$, $C$ is not diagonal, and its eigenstates there are constructed below.

Two of the signs, $C_1$ and $C_2$ of the vacuum module, follow from the modes that build the towers, since only
the modes of $U$ and $W$ are $A$-odd. The primary of the second tower of the
vacuum module is the current $U=U_{-5/2}|0\rangle$, which the lift reverses, so
$C_1=-1$ there. The tower at $h=\frac{11}2$ is built on $U_{-5/2}W_{-3}$ acting
on the vacuum \cite{Schoutens:1990xg}. Since both modes are $A$-odd, the tower
is $A$-even and $C_2=+1$.

The coset realization fixes the remaining ratios. The NS Fock space
$\mathcal H$ of the eight numerator fermions decomposes under the
denominator $\widehat{su}(2)_{12}$ as
\begin{equation}
\mathcal H=\bigoplus_s\, B_s\otimes\mathcal M_s\,,
\label{eq:isotypic}
\end{equation}
where $\mathcal M_s$ is the integrable representation with label $s$ and $B_s$ its multiplicity space, a module of the extended algebra.
Because the three fermions of $\widehat{su}(2)_2$ have isospin one under the
denominator and the five of $\widehat{so}(5)_1$ isospin two, every state of $\mathcal H$
has integer isospin and only odd $s$ appear. The orbit $\{1,13\}$ of \eqref{eq:ZD8} belongs to the vacuum module, $\{3,11\}$ to $\widehat\Phi_{1,3}$, $\{5,9\}$ to $\widehat X$ and $\{7\}$ to $\Phi_\pm$. For $s\neq7$, $B_s$ is a copy of the extended module $B_a$ of Section~\ref{sub:rparity} whose orbit contains $s$, so $B_s$ and $B_{14-s}$ are two copies of the same extended module. At $s=7$, $B_7$ is $\Phi_+\oplus\Phi_-$.

As in Section~\ref{sub:rparity}, $A$ is $-1$ on the five fermions,
$+1$ on the three, and fixes every current of the denominator. Its lift $\hat A$ to
$\mathcal H$ acts on the fermion modes by those signs. It commutes with the denominator, so it preserves each summand $B_s\otimes\mathcal M_s$ of \eqref{eq:isotypic}. Since $\mathcal M_s$ is irreducible, $\hat A$ acts on the summand as an operator on $B_s$. For $s\neq7$, this operator is a lift of $A$ to the extended module $B_s$ and acts on the $i$-th tower by $\varepsilon_sC_i$, with the $C_i$ defined above and an overall sign $\varepsilon_s=\pm1$. The two copies $B_s$ and $B_{14-s}$ have the same $C_i$, but their overall signs $\varepsilon_s$ and $\varepsilon_{14-s}$ are independent. With $C_0=+1$ fixed above, the $C_i$ are relative signs and do not depend on $\varepsilon_s$. On $B_7$, the operator maps $\Phi_+$ to $\Phi_-$ and $\Phi_-$ to $\Phi_+$, since, by Section~\ref{sub:rparity}, $A$ exchanges the two resolved modules. Its trace over $B_7$ therefore vanishes.

Taking the trace of $\hat A\,q^{L_0-c/24}$ over \eqref{eq:isotypic} then gives
\begin{equation}\label{eq:twistedbr}
    q^{-\frac16}\prod_{r\in\mathbb N-\frac12}(1-q^r)^5(1+q^r)^3=\sum_{s\neq7} \varepsilon_s\left(\sum_i C_i\,\chi^{\rm NS}_{h_i}(q)\right)\,\chi^{(12)}_s(q)\,.
\end{equation}
On the left, each mode of the five $A$-odd fermions contributes $(1-q^r)$ and each mode of the three $A$-even ones $(1+q^r)$. The prefactor
is $q^{-c/24}$ with $c=4$ for the eight fermions, the sum $\frac{10}7+\frac{18}7$ of the central charges of the coset and of $\widehat{su}(2)_{12}$. On the right, $\varepsilon_s$ times the bracket is the trace of $\hat A$ over $B_s$, and the bracket is the sum
of the tower characters of the extended module weighted by the signs $C_i$, not the ordinary sum
$\hat\chi^{\rm NS}_a$ of \eqref{eq:ZNS}. The sum runs over the six labels $s\neq7$ rather than over the three orbits, because $\varepsilon_s$ and $\varepsilon_{14-s}$ are independent. The unknowns are the three sets of $C_i$ and the six signs $\varepsilon_s$.

Without $\hat A$, the same trace gives the untwisted branching identity of the Fock
space, $q^{-\frac16}\prod_r(1+q^r)^8$ on the left and the ordinary sums
$\hat\chi^{\rm NS}_a$ on the right, with $\hat\chi^{\rm NS}_{\Phi_+}+\hat\chi^{\rm NS}_{\Phi_-}$ at $s=7$. We verify this identity through level
fifteen.\footnote{Two $q$-series agree through level $n$ when their
coefficients match at every power $q^k$ with $k\le n$ above the leading
one, in steps of $\frac12$.} It contains no $C_i$ and no $\varepsilon_s$, so it confirms the other three ingredients of \eqref{eq:twistedbr}: the tower content of the extended modules,
the characters $\chi^{(12)}_s$, and the assignment of orbits to extended modules
that Section~\ref{sub:coset} obtained from the Kac weights.

Solving \eqref{eq:twistedbr} to the same level determines all the $C_i$ and all the $\varepsilon_s$ uniquely. The $C_i$ are
\begin{equation}
\begin{array}{l@{\qquad}l@{\qquad}l}
\text{vacuum} & h-h_a=0,\ \frac52,\ \frac{11}2,\ 15 & C=+,-,+,-\,,\\
\widehat\Phi_{1,3} & h-h_a=0,\ \frac12,\ \frac52,\ 10 & C=+,-,+,-\,,\\
\widehat X & h-h_a=0,\ 1,\ \frac32,\ \frac{13}2 & C=+,-,-,+\,.
\end{array}
\label{eq:Ctable}
\end{equation}
The signs $\varepsilon_s$ are $+$ for $s=1,3,9$ and $-$ for $s=5,11,13$.
Level fifteen is the lowest level at which all the $C_i$ appear, since
the highest tower of any extended module is the one at $h-h_a=15$ in the
vacuum module. The modes that build the towers already fixed $C_1$
and $C_2$ in the vacuum module, and the solution reproduces them.

Each tower now has two signs, the relative fermion parity $(-1)^{2(h-h_a)}$ and the sign $C_i$. The relative fermion parity reads $+,-,-,+$ along the vacuum and $\widehat\Phi_{1,3}$ lines of \eqref{eq:Ctable} and $+,+,-,-$ along the $\widehat X$ line. In each extended module, the four towers realize the four combinations of these two signs once each, so the two charges identify a tower within its extended module. 

The identity \eqref{eq:twistedbr} does not constrain $\Phi_+$ and $\Phi_-$, which are absent from it. Their charges follow instead from the exchange. At each conformal weight pair $(h_i,\tilde h_j)$, $C$ exchanges the operator on $\Phi_+$ with the operator on $\Phi_-$. Up to the sign $\eta$ introduced below, their symmetric combination is $C$-even and their antisymmetric combination is $C$-odd.

Our analysis leaves one charge of the resolved pair $\Phi_\pm$
undetermined. The two extended modules have the towers $\phi^\pm_{5,7}$ at $h=\frac17$ and
$\phi^\pm_{1,7}$ at $h=\frac{51}{14}$, where $\phi_{r,s}$ denotes the super-Virasoro primary of Kac label $(r,s)$ and the superscript labels the member of the pair. Since $A$ exchanges the two extended modules and the lift is
constant on each tower, $\hat A$ has one sign per tower. We use the freedom in the overall sign of the lift to set the sign on the lower tower to $+1$,
\begin{equation}
\hat A\,\phi^\pm_{5,7}\,\hat A^{-1}=\phi^\mp_{5,7}\,,\qquad
\hat A\,\phi^\pm_{1,7}\,\hat A^{-1}=\eta\,\phi^\mp_{1,7}\,,\qquad \eta=\pm1\,.
\label{eq:pairlift}
\end{equation}
Since the pair drops out of \eqref{eq:twistedbr}, the sign $\eta$ is left
undetermined. It appears only
in operators built with $\phi^\pm_{1,7}$, which are not in Table~\ref{tab:web}.

Every other charge in the theory is now fixed. Let us work out the charges of one operator as an example. Pairing the second tower $\phi_{5,3}$ of $\widehat\Phi_{1,3}$ on the left with its first
tower $\phi_{1,3}$ on the right gives the operator $\phi_{5,3}\tilde\phi_{1,3}$ at $(h,\tilde h)=(\frac67,\frac5{14})$, whose
spin is $\frac12$. Its chiral fermion parities are
$\omega=\omega_a(-1)^{2\cdot\frac12}=+1$ at level $\frac12$ and
$\tilde\omega=\omega_a=-1$ at level zero. Their product $-1$ makes it a
fermion, as the spin requires. Under $C$, its eigenvalue is
$C_iC_j=(-)(+)=-1$ by \eqref{eq:Ctable}. Its R-parities are
$R=\omega\,C_iC_j=-1$ and $\tilde R=\tilde\omega\,C_iC_j=+1$. The two are opposite, as
on every fermion. We list it in Table~\ref{tab:web} as the two-dimensional
counterpart of the LG composite $\chi_-$ of Section~\ref{sub:spin3}, whose entry in the column $(R,C)$ shows the charges $(-,-)$
just computed.

\section{Fixed points and RG flows in $d=4-\eps$}
\label{sec:fixedpoint}

\subsection{Anomalous dimensions and the operator dictionary}
\label{sub:fp}

The $\mathcal N=1$ LG theory with superpotential 
\eqref{eq:W}, treated as a GNY model, has interacting
fixed points in $d=4-\eps$. In components, we have the two real scalars $\phi_X$ and
$\phi_Y$ and the two Majorana fermions $\Psi_X$ and $\Psi_Y$ of
\eqref{eq:spinors}, labeled by the superfield flavors $X$ and $Y$,
in the basis in which $\Zex$ is diagonal.\footnote{\label{fn:spinors}In $d=3$, the fermion of a real $\mathcal N=1$ superfield is a two-component Majorana spinor, and in $d=2$, the superfield is the $(1,1)$ multiplet. Because there is no four-dimensional multiplet of one real scalar and one Majorana fermion, we fix the fermion content in $d=3$ and continue the loop integrals in $d$. To the order we
compute, the diagrams involve the spinor space only through the
anticommutator $\{\gamma_\mu,\gamma_\nu\}=2\delta_{\mu\nu}$ and through
one factor of $\operatorname{tr}\mathbf1$ for each closed fermion loop,
so we continue by evaluating the integrals in $d=4-\eps$ at $\operatorname{tr}\mathbf1=2$. The super-Ising model is treated in this
way in \cite{Fei:2016sgs}, as the case $N=1$ of the GNY models with $N$
two-component Majorana fermions studied there. Note that the RG functions \eqref{eq:beta1} that we take from \cite{Liendo:2021wpo} use two-component spinors throughout (see
also \cite{Jack:2024sjr}), so they are evaluated at the same
$\operatorname{tr}\mathbf1$.} Supersymmetry determines both interactions from ${\cal W}$: the Yukawa interaction $\frac12\,h_{ijk}\phi_i\bar\Psi_j\Psi_k$ with the constant symmetric tensor $h_{ijk}=\partial_i\partial_j\partial_k {\cal W}$, and the quartic potential $\frac12\sum_i(\partial_i {\cal W})^2$. Thus the Euclidean Lagrangian is\footnote{We may define the $O(n)$-symmetric generalization of this GNY model with four-component Majorana fermions
$\Psi_i^a$, where $i=X,Y$ and $a=1,\ldots,n$. We then formally continue
$n\rightarrow\frac12$ and find emergent supersymmetry, analogously to
\cite{Fei:2016sgs}.}
\begin{equation}
\begin{aligned}
&\mathcal L=\frac12(\partial\phi_X)^2+\frac12(\partial\phi_Y)^2+\frac12\bar\Psi_X\slashed\partial\Psi_X+\frac12\bar\Psi_Y\slashed\partial\Psi_Y\\
&+\frac12g_2\phi_X\bar\Psi_X\Psi_X+\frac12g_1\phi_X\bar\Psi_Y\Psi_Y+g_1\phi_Y\bar\Psi_X\Psi_Y+\frac18\left(g_2\phi_X^2+g_1\phi_Y^2\right)^2+\frac12g_1^2\phi_X^2\phi_Y^2\,.
\end{aligned}
\label{eq:LGNY}
\end{equation}
The $\Zex$ symmetry makes $h_{XXY}$ and $h_{YYY}$ vanish, so the Yukawa tensor
keeps two independent couplings, $h_{XXX}=g_2$ and $h_{XYY}=h_{YXY}=h_{YYX}=g_1$.

One comment about $\ZR$ in $d=4-\eps$ is in order. The candidate for its action on the fermions
is the chiral rotation by $\frac\pi2$, under which the kinetic term is even and $\bar\Psi\Psi$ odd. The rotation exists in four dimensions but not, in the strict
sense, in $4-\eps$ dimensions. To the order we compute, however, the prescription of footnote~\ref{fn:spinors} formally preserves the chiral rotation. In three dimensions, the chiral rotation is
absent altogether, since there is no chirality matrix. No internal rotation of
the Majorana fermions replaces the chiral one, because a real matrix $O$ that reversed $\bar\Psi\Psi$ would require $OO^{\rm T}=-1$. Instead, $\ZR$ is a spacetime
reflection.

At one loop, the RG functions of the superpotential
couplings are those of \cite{Liendo:2021wpo},
\begin{equation}\label{eq:beta1}
    \begin{aligned}
    \beta_{ijk}&=-\frac\eps2 h_{ijk}+\frac12(h_{ijm}h_{kpq}h_{mpq}+\text{2 perm.})+2h_{imp}h_{jpq}h_{kqm}\,,\\
    \gamma_{ij}&=\frac12\,h_{imn}h_{jmn}\,,
\end{aligned}
\end{equation}
in units where $16\pi^2=1$. Explicitly, for the superpotential \eqref{eq:W}, they are
\begin{equation}\label{eq:beta1LG}
    \begin{aligned}
    \beta_{g_1}&=-\frac\eps2 g_1+\frac{9}{2}g_1^3+2g_1^2g_2+\frac12 g_1g_2^2\,,&\qquad
    \gamma_X&=\frac12\bigl(g_1^2+g_2^2\bigr)\,,\\
    \beta_{g_2}&=-\frac\eps2 g_2+2g_1^3+\frac{3}{2}g_1^2g_2+\frac{7}{2}g_2^3\,,&\qquad
    \gamma_Y&=g_1^2\,.
\end{aligned}
\end{equation}
When both couplings are nonvanishing, the zeros of the beta functions in
\eqref{eq:beta1LG} fix the ratio $g_1/g_2$ to $1$, $-1$ or $3/2+\mathcal O(\eps)$. The fixed point we study is the one at $g_1/g_2=3/2+\mathcal O(\eps)$, the ratio at which the two fields remain coupled and no supersymmetry beyond $\mathcal N=1$ appears.\footnote{This fixed point appears in the survey \cite{Liendo:2021wpo} of $\mathcal N=1$ supersymmetric fixed points in $d=4-\eps$, as the $\mathbb Z_2$-symmetric entry with two coupled superfields.} A fourth fixed point lies at $g_1=0$, where the superpotential is $\frac{g_2}{6}X^3$ and $Y$ is free. We list the four fixed points in Table~\ref{tab:fps} and
identify the theories at $g_1=g_2$ and $g_1=-g_2$ in Section~\ref{sub:flow}.

We compute the anomalous dimensions below in components,
where the Yukawa and quartic couplings are independent. With $\Zex$ imposed, there are three Yukawa couplings and three quartic couplings. The supersymmetric theories form the surface parametrized by $g_1$ and $g_2$ inside the six-coupling space, and supersymmetry
is emergent when the fixed point of the six couplings lies on the surface.

Note that beyond leading order the coordinates of the fixed point depend on the
scheme, while its anomalous dimensions do not. We compute the anomalous dimensions to two loops from the general RG functions of a Yukawa theory \cite{Machacek:1983tz,Machacek:1983fi,Machacek:1984zw,Fei:2016sgs}, which we record in Appendix~\ref{app:twoloop}.\footnote{For the GNY model of one scalar and $N$ fermions, which contains the single-superfield theory with ${\cal W}\propto X^3$, the RG functions are known to five loops \cite{Gracey:2025aoj}.} The result for the two elementary fields $X$ and $Y$ is
\begin{equation}\label{eq:gammas}
    \gamma_X=\frac{13}{218}\eps+\frac{14836}{109^3}\eps^2+\mathcal O(\eps^3)\,,\qquad
\gamma_Y=\frac{9}{109}\eps+\frac{72753}{2\cdot109^3}\eps^2+\mathcal O(\eps^3)\,.
\end{equation}
In the component formulation, we compute the scalar and the fermion anomalous dimension of each superfield separately, but they agree,
$\gamma_{\phi_X}=\gamma_{\Psi_X}=\gamma_X$ and
$\gamma_{\phi_Y}=\gamma_{\Psi_Y}=\gamma_Y$. At second order, the two computations differ term by term, and the scalar one receives contributions from the quartic couplings that the fermion one does not. The equality nevertheless holds because of a nontrivial cancellation. The fixed point of the six couplings also lies on the supersymmetric surface, so supersymmetry is emergent.

The Pad\'e continuation of $\Delta=1-\frac{\eps}{2}+\gamma$ to $\eps=2$ gives $\Delta_X\approx0.163$ and $\Delta_Y\approx0.264$, compared with the exact
$\Delta_X=\frac17=0.143$ and $\Delta_Y=\frac27=0.286$ of the $\ed$
superconformal minimal model.\footnote{Unless stated otherwise, we use the one-sided $[1,1]$ Pad\'e of the dimension itself rather than of the
anomalous dimension.} To estimate the accuracy of such a continuation, we apply the same procedure to the single-superfield theory with ${\cal W}\propto X^3$, where the Pad\'e gives $\Delta_X\approx0.217$, as in \cite{Fei:2016sgs}, compared with the exact $\Delta_X=\frac15$ of $SM(3)$. The error is
nearly nine percent.

Let us now propose the dictionary between the LG fields and
the $\ed$ primaries in Table~\ref{tab:map}. We compare the $\Delta_{4-\eps}$ column, continued to $\eps=2$, with the $\Delta_{2d}$ column. The deviation ranges from 7.5 percent for $Y$ to nearly 30 percent for the composites $P$ and
$XP$.

Two rows of the table require comment. The first is the $P$ row, where one of
the two Hermitian combinations of the pair $\Phi_\pm$ is identified with a quadratic
composite rather than an elementary field. Among the quadratics, the
$\Zex$-even $X^2$ and $Y^2$ mix, and the supersymmetric equation of motion turns one of their two eigenoperators, $\partial_X {\cal W}\propto2X^2+3Y^2$, into a descendant of $X$ with dimension
$\Delta_X+1$. The other is a genuine primary, whose leading form is
\begin{equation}\label{eq:P}
    P=3X^2-2Y^2+\mathcal{O}(\eps)\,.
\end{equation}
From here on, a composite of superfields and its bottom component are denoted
by the same symbol. We compute the dimension of $P$ and give the $\mathcal O(\eps)$ correction \eqref{eq:vP} in Section~\ref{sub:deg}. The $\Zex$ charge determines which combination each field corresponds to. Conjugation exchanges $\Phi_+$ with $\Phi_-$, so their symmetric combination is $C$-even and their antisymmetric combination is $C$-odd. The $\Zex$-even composite $P$ therefore corresponds to the symmetric combination and the $\Zex$-odd field $\phi_Y$ to the antisymmetric one.

The second is the $XP$ row, whose two-dimensional counterpart is $\widehat\Phi_{1,3}$. To drive the RG flow of Section~\ref{sub:flow}, the $F$-component of
$\widehat\Phi_{1,3}$ must be a superpotential deformation. Among the deformations of the fixed point within the plane of the two superpotential couplings, exactly one has $y>0$.\footnote{The
stability matrix is $M_{ij}=\partial\beta_i/\partial g_j$. We report its
eigenvalues as the RG eigenvalues $y=-\,\mathrm{eig}(M)$,
so that the operator dimension is $\Delta=d-y$ and a relevant direction
has $y>0$.} Its eigenvalue, to the order of \eqref{eq:gammas}, is
\begin{equation}\label{eq:yrel}
    y_{\rm rel}=\frac{15}{109}\eps+\frac{1909918}{31\cdot109^3}\eps^2+\mathcal O(\eps^3)\,,
\end{equation}
and at one loop, the deformation is the cubic $\delta {\cal W}\propto3X^3-2XY^2$,
a direction that the two-loop terms shift at order $\eps$. The cubic
factorizes as $X\cdot P$, a product of two fields the dictionary
has already fixed. Among such products, only
$XP$ is $\ZR$-odd and $\Zex$-even, as $\widehat\Phi_{1,3}$ requires. The $F$-term of the cubic has dimension
$\Delta_F=d-y_{\rm rel}$, and the bottom component of its multiplet has
dimension $\Delta_F-1$, the value quoted for $XP$ in Table~\ref{tab:map}.

\begin{table}[!htbp]
\caption{\label{tab:map}Proposed dictionary between the LG fields and the $\ed$ primaries. In the column $(R,C)$, the first sign is $R$ of \eqref{eq:Rop} and the second is the $\Zex$ charge, the eigenvalue of $C$. The last column is the dimension in $d=4-\eps$ to two loops, from \eqref{eq:gammas} for $\phi_X$ and $\phi_Y$, from \eqref{eq:gammaP} for $P$, and from \eqref{eq:yrel} for $XP$. The rows $P$ and $XP$ show the leading representatives of the primary \eqref{eq:P} and of its product with $X$.}
\centering
\small
\begin{tabular}{lcccc}
\toprule
LG & $d=2$ field & $\Delta_{2d}$ & $(R,C)$ & $\Delta_{4-\eps}$ \\
\midrule
$\phi_X$ & $\widehat X$ & $\frac17$ & $(-,+)$ &
  $1-\frac{48}{109}\eps+\frac{14836}{109^3}\eps^2$ \\
$\phi_Y$ & $\frac1{i\sqrt2}(\Phi_+{-}\Phi_-)$ & $\frac27$ & $(-,-)$ &
  $1-\frac{91}{218}\eps+\frac{72753}{2\cdot109^3}\eps^2$ \\
$P=3\phi_X^2{-}2\phi_Y^2$ & $\frac1{\sqrt2}(\Phi_+{+}\Phi_-)$ & $\frac27$ & $(+,+)$ &
  $2-\frac{100}{109}\eps+\frac{12342}{109^3}\eps^2$ \\
$XP=3\phi_X^3{-}2\phi_X\phi_Y^2$ & $\widehat\Phi_{1,3}$ & $\frac57$ & $(-,+)$ &
  $3-\frac{124}{109}\eps-\frac{1909918}{31\cdot109^3}\eps^2$ \\
\bottomrule
\end{tabular}
\end{table}
Our identification of the LG field corresponding to $\widehat\Phi_{1,3}$ differs
from that of Li \cite{Li:1988pj}, who identified it with a quadratic composite of his
two superfields. In $d=4-\eps$, our identification is fixed, because the only relevant cubic superpotential deformation of the fixed point is $X\cdot P$.

\subsection{The spin-3 degeneracy and its two-loop resolution}
\label{sub:deg}

In Section~\ref{sub:coset}, we saw that the two members of the pair
$\Phi_\pm$ have the same conformal weight and differ only in the
eigenvalue $\pm w$ of $W_0$. Near four dimensions, however, no spin-3 current is conserved, as we show in Section~\ref{sub:spin3}, so the spin-3 charge is not defined there and the degeneracy is unprotected. The
LG description realizes one of the two Hermitian
combinations of $\Phi_\pm$ as the elementary field $Y$ and the other as the quadratic
composite $P$ of \eqref{eq:P}. Although no symmetry of the
Lagrangian relates an elementary field to a composite, at one loop the
two have the same anomalous dimension,
$\gamma_P=\gamma_Y=\frac9{109}\eps$. At this order, $\Delta_P-\Delta_Y$ is therefore only the classical difference $\frac{d-2}{2}$, which vanishes at $d=2$, where $\Phi_\pm$ are degenerate. The equality
$\gamma_P=\gamma_Y$ is unprotected, however, and we ask whether it persists at two loops or is an accident of the leading order.

We obtain the two-loop mixing of the quadratic composites
from the running of the scalar mass matrix, adding a spectator scalar coupled
to $X^2$ and $Y^2$ so that the quartic beta function of the component
theory \cite{Machacek:1984zw} produces the mixing at the fixed-point couplings
of Section~\ref{sub:fp}. The same anomalous-dimension matrix follows from the general result for quadratic operators derived in \cite{Pernici:1999kk} and given in \cite{Fei:2016sgs}. The two derivations therefore check each other,
and we record in Appendix~\ref{app:twoloop} the symmetrization and the basis convention that the comparison requires. At the fixed point, in the
basis $(X^2,Y^2)$ of the $\Zex$-even quadratics, the anomalous-dimension matrix is
\begin{equation}\label{eq:kernelXY}
    \Gamma_{\{X^2,Y^2\}}=\frac{\eps}{109}\begin{pmatrix}
    25&24\\
    24&45
    \end{pmatrix}+\frac{\eps^2}{109^3}\begin{pmatrix}
        -40418&-27692\\
        -14612&67596
    \end{pmatrix}+\mathcal{O}(\eps^3)\,.
\end{equation}
Note that the $\mathcal O(\eps^2)$ matrix is not symmetric, because the basis $(X^2,Y^2)$
is orthonormal only at leading order.

The $\Zex$-odd quadratic $XY$ does not mix with $X^2$ and $Y^2$. The three eigenoperators are, therefore, $\partial_X {\cal W}$, $\partial_Y {\cal W}$ and $P$. The dimensions of the first two follow from supersymmetry, and that of $P$ from the remaining eigenvalue of \eqref{eq:kernelXY}. Because $\gamma_Y$ is already known from \eqref{eq:gammas}, the coincidence of $\Delta_P$ and $\Delta_Y$ at $d=2$ is determined by $\gamma_P$. The eigenoperator
$\partial_X {\cal W}\propto2X^2+3Y^2$ has dimension $\Delta_X+1$ through order $\eps^2$,
as a supersymmetric equation of motion requires. The $\Zex$-odd eigenoperator $\partial_Y {\cal W}\propto XY$ has dimension $\Delta_Y+1$ at the same order. Here $\Delta_X$ and $\Delta_Y$ are the two-loop values built from \eqref{eq:gammas}.\footnote{For a single superfield with ${\cal W}\propto X^3$, the same computation gives $\Delta_{X^2}=2-\frac37\eps+\frac1{49}\eps^2$, in agreement with \cite{Fei:2016sgs}. Subtracting
the classical dimension $d-2$ leaves
$\gamma_{X^2}=\frac47\eps+\frac1{49}\eps^2$, whose one-loop
coefficient is the composite anomalous dimension reported for this
supersymmetric fixed point in
\cite{Jack:2024sjr}.}

Beyond leading order, the two $\Zex$-even eigenoperators receive $\mathcal O(\eps)$ corrections. In the basis
$(X^2,Y^2)$, the right eigenvectors of \eqref{eq:kernelXY} are
\begin{align}
v_{\partial_X {\cal W}}&=(2,3)-\frac{148}{13\cdot109}\,\eps\,(3,-2)+\mathcal{O}(\eps^2)\,,
\label{eq:vdXW}\\
v_P&=(3,-2)+\frac{118}{13\cdot109}\,\eps\,(2,3)+\mathcal{O}(\eps^2)\,.
\label{eq:vP}
\end{align}
The first is the equation-of-motion direction $\partial_X {\cal W}=\frac{1}{2}(g_{2,\star}X^2+g_{1,\star}Y^2)$. Its components in \eqref{eq:vdXW} therefore give the two-loop shift of the fixed-point ratio,
\begin{equation}\label{eq:ratioNLO}
    \frac{g_{1,\star}}{g_{2,\star}}=\frac32+\frac{37}{109}\eps+\mathcal{O}(\eps^2)\,.
\end{equation}
The second eigenoperator is $P$, and \eqref{eq:vP} gives the $\mathcal O(\eps)$ correction to its leading form $3X^2-2Y^2$. The eigenvalue of $P$ is
\begin{equation}
    \gamma_P=\frac{9}{109}\eps+\frac{12342}{109^3}\eps^2+\mathcal{O}(\eps^3)\,.
\label{eq:gammaP}
\end{equation}
The difference from the two-loop $\gamma_Y$ of \eqref{eq:gammas} lifts the degeneracy,
\begin{equation}
    \Delta_P-\Delta_Y=\frac{d-2}{2}-\frac12\Bigl(\frac{21}{109}\Bigr)^{\!2}\eps^2+\mathcal{O}(\eps^3)\,.
\label{eq:collision}
\end{equation}

The equality $\gamma_P=\gamma_Y$ is therefore an accident of the leading order. At $\eps=2$, the classical term $\frac{d-2}{2}$ of \eqref{eq:collision} vanishes, and the
exact degeneracy then constrains the sum of the remaining series rather than each of
its terms. The two-loop term is small, but it moves $\Delta_P$ toward the exact
$\Delta_P=\frac27$, from $\Delta_P=0.165$ in the one-loop truncation to
$\Delta_P=0.202$ in the two-loop Pad\'e.

\subsection{The RG flows to $(D_6,E_6)$ and $(A_3,D_4)$}
\label{sub:flow}

A faithful LG description must reproduce the
RG flows between the $\mathcal N=1$ superconformal minimal models.
In two dimensions, the $F$-component of $\phi_{1,3}$ drives the RG flow $SM(m)\to SM(m-2)$ \cite{Poghossian:1987ngr,Kastor:1988ef,Poghosyan:2014jia}. At $m=12$, the perturbation survives the extension that defines the $\ed$ superconformal minimal model, and the leading irrelevant operator of the infrared theory likewise survives the extension that defines $(D_6,E_6)$. The RG flow therefore connects two exceptional
theories,
\begin{equation}
\ed\big|_{m=12}\ \longrightarrow\ (D_6,E_6)\big|_{m=10}\,,
\label{eq:Eflow}
\end{equation}
and the central charge decreases from $10/7$ to $7/5$. The infrared theory factorizes as $SM(3)\otimes SM(3)$. Because $SM(3)$ is the LG theory of a single cubic superfield, the $(D_6,E_6)$ superconformal minimal model is described by \eqref{eq:W} at the decoupled fixed point $g_1=g_2$.

\begin{figure}[!ht]
\centering
\includegraphics[width=0.62\columnwidth]{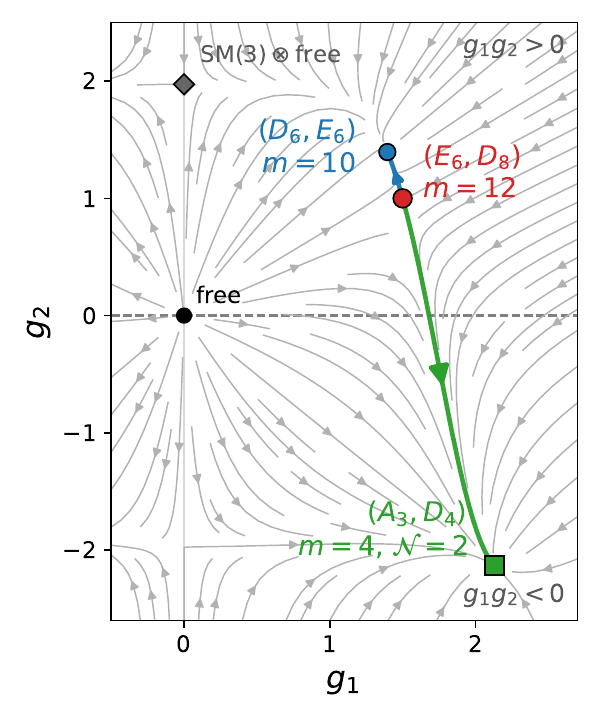}
\caption{\label{fig:rgflow}One-loop RG flow in the plane of the two superpotential couplings. Arrows point toward the infrared, and the thick curves are the two separatrices of the one relevant direction at $\ed$. The couplings are in units where the coupled fixed point is at $(g_1,g_2)=(3/2,1)$. The half-plane $g_1<0$ is omitted, since $(X,Y)\rightarrow(-X,-Y)$ maps it to the half shown.}
\end{figure}

The beta functions in $d=4-\eps$ reproduce the RG flow \eqref{eq:Eflow}. Through two loops, and in
the scheme of Appendix~\ref{app:twoloop}, the RG flow of the component theory preserves the supersymmetric surface of
Section~\ref{sub:fp}, and on the surface the beta functions are
\begin{equation}\label{eq:betag}
    \begin{aligned}
    \beta_{g_1}&=-\frac\eps2 g_1+\frac{9}{2}g_1^3+2g_1^2g_2+\frac12 g_1g_2^2-\frac{g_1}2\bigl(34g_1^4+28g_1^3g_2+13g_1^2g_2^2+8g_1g_2^3+g_2^4\bigr)\,,\\
    \beta_{g_2}&=-\frac\eps2 g_2+2g_1^3+\frac{3}{2}g_1^2g_2+\frac{7}{2}g_2^3-\frac32\bigl(8g_1^5+6g_1^4g_2+4g_1^3g_2^2+3g_1^2g_2^3+7g_2^5\bigr)\,.
    \end{aligned}
\end{equation}
Their cubic terms are those of \eqref{eq:beta1LG}, and their coupled fixed
point reproduces the ratio \eqref{eq:ratioNLO}.
We show the one-loop RG flow in Figure~\ref{fig:rgflow}. The free theory at the origin is the ultraviolet fixed point. The coupled fixed point $\ed$ is a saddle with a single relevant direction in the plane of the two superpotential couplings. At the decoupled fixed point $(D_6,E_6)$, both anomalous dimensions equal the single-superfield super-Ising value $\gamma_\Phi=\eps/14$ \cite{Fei:2016sgs}. This fixed point is infrared stable within the space of cubic superpotential couplings, as its identification with the endpoint of \eqref{eq:Eflow} requires.

We integrate the one-loop RG flow numerically, starting from $\ed$ and moving along the relevant direction $\delta {\cal W}\propto3X^3-2XY^2$. Only one sign of the perturbation realizes the RG flow \eqref{eq:Eflow}, the one for which the RG flow keeps both couplings positive and ends at the decoupled fixed point. For the opposite sign of the perturbation, the RG flow crosses the line $g_2=0$, enters the region $g_1g_2<0$ and ends at $g_1=-g_2$, the fixed point identified below as the $m=4$ model $(A_3,D_4)$.

\begin{table}[!htbp]
\caption{\label{tab:fps}Fixed points of the GNY model \eqref{eq:LGNY} in $d=4-\eps$.
The stability eigenvalues are the RG eigenvalues $y=d-\Delta$ of Section~\ref{sub:fp} in the plane of the two superpotential couplings.}
\centering
\begin{tabular}{lccl}
\toprule
$g_1/g_2$ & $(\gamma_X,\gamma_Y)/\eps$ & $y/\eps$ & identification\\
\midrule
$3/2+\mathcal O(\eps)$ & $\frac{13}{218},\frac9{109}$ & $\frac{15}{109},-1$
  & $SM(12)_{(E_6,D_8)}$ ($c=\frac{10}{7}$)\\
$1$ & $\frac1{14},\frac1{14}$ & $-\frac17,-1$
  & $SM(3)^{\otimes2}=SM(10)_{(D_6,E_6)}$ ($c=\frac{7}{5}$)\\
$0$ & $\frac1{14},0$ & $\frac37,-1$ & $SM(3)\otimes\text{free}$ ($c=\frac{11}{5}$)\\
$-1$ & $\frac16,\frac16$ & $-\frac53,-1$ & $SM(4)_{(A_3,D_4)}=A_2^{\mathcal{N}=2}$ ($c=1$ $S_3$ super-Potts)\\
\bottomrule
\end{tabular}
\end{table}

The component computation gives the full stability spectrum at the $\ed$
fixed point. At one loop, the stability matrix on the six-coupling space of
Section~\ref{sub:fp} is block triangular, because the Yukawa beta functions do not involve the quartic couplings. The Yukawa block and the quartic block then give
\begin{align}
\frac{y}{\eps}&=\frac{15}{109}+0.0476\,\eps,\ \
-\frac{30}{109}-0.0801\,\eps,\ \ -1+0.4240\,\eps,
\label{eq:spectrum2}\\
\frac{y}{\eps}&=0.3021+0.1094\,\eps,\ \
-0.5830+0.2409\,\eps,\ \ -1.8383+1.9229\,\eps\,.
\label{eq:spectrum2b}
\end{align}
The one-loop coefficients of the Yukawa block are rational, while those of the quartic block are the three roots of the irreducible cubic $1295029u^3+2744511u^2+440687u-419355=0$ in $u=y/\eps$.

The two directions of Table~\ref{tab:fps} appear in \eqref{eq:spectrum2} with $y=\frac{15}{109}\eps$ and $y=-\eps$. The remaining four directions in
\eqref{eq:spectrum2}--\eqref{eq:spectrum2b} are not supersymmetric, and near four dimensions exactly one of them is relevant. We return to the behavior of these
eigenvalues at $\eps=1$ in Section~\ref{sub:threed}.

Since each eigenvalue gives an operator dimension through $\Delta=d-y$, both
theories in \eqref{eq:Eflow} can be tested against exact data. Let us first consider the ultraviolet theory $\ed$. The relevant direction \eqref{eq:yrel} determines the dimension of the $F$-component that drives the RG flow. Truncated at one loop and continued to $\eps=2$, this dimension is
$\Delta_F(\phi_{1,3})=\frac{188}{109}=1.72$, compared with the exact
$\Delta_F(\phi_{1,3})=\frac{12}7=1.71$ at $m=12$. At two loops,
with the same Pad\'e resummation as in Section~\ref{sub:fp}, the
continuation gives $\Delta_F(\phi_{1,3})=1.52$. The one-loop agreement for this operator is therefore partly accidental.

Let us now turn to the infrared theory $(D_6,E_6)$. There the leading irrelevant operator couples the two copies of $SM(3)$ and is again a superpotential direction. At one loop, the deformation is the cubic $\delta {\cal W}\propto X^3-XY^2=\sqrt2\,\Phi_1\Phi_2(\Phi_1+\Phi_2)$ in the flavors of \eqref{eq:flavor}. By the equation of motion, $\Phi_i^2\propto\partial {\cal W}/\partial\Phi_i$ is the $F$-component of $\Phi_i$, so the bottom component of the cubic has conformal weights $(h,\tilde h)=(\frac1{10}+\frac35,\frac1{10}+\frac35)$, those of $\phi_{3,1}\tilde\phi_{3,1}$. The $F$-term of the cubic therefore corresponds to the $F$-component of the super-Virasoro primary $\phi_{3,1}$ of $SM(10)$. The one-loop continuation gives
$\Delta_F(\phi_{3,1})=\frac{16}7=2.29$, compared with the exact $\Delta_F(\phi_{3,1})=\frac{12}5=2.4$ at $m=10$.

The superpotential also fixes the Witten index \cite{Witten:1982df}, $\mathcal I_W=\operatorname{Tr}_{\mathcal H_R}(-1)^F e^{-\beta H}$.
After a generic linear perturbation that lifts the degenerate critical
point at the origin, the Witten index is the signed count of the critical
points of ${\cal W}$,
\begin{equation}\label{eq:wittencrit}
    \mathcal I_W=\sum_{\partial_i {\cal W}(p)=0}
    \operatorname{sgn}\det\bigl(\partial_i\partial_j {\cal W}(p)\bigr)\,.
\end{equation}
Perturbing the superpotential \eqref{eq:W} by $\delta {\cal W}=aX+bY$, we find that for $g_1g_2>0$ the number of real critical points is zero or four, depending on the values of $a$ and $b$. If there are four, their signs in \eqref{eq:wittencrit} cancel in pairs, so the Witten index vanishes in both cases. When $g_1g_2<0$, there are two critical points, which contribute to the Witten index with equal signs.
The Witten index therefore vanishes in the chamber $g_1g_2>0$ and equals $\pm2$ in the
chamber $g_1g_2<0$. The Witten index changes across the lines $g_1=0$ and $g_2=0$, where two critical points escape to infinity. The RG flow from $\ed$ to $(A_3,D_4)$ crosses the line $g_2=0$. This RG flow therefore does not preserve the Witten index.\footnote{RG flows that change the Witten index in this way, with a critical point moving to infinity and its ground state decoupling, were argued to connect some $\mathcal N=1$ superconformal minimal models \cite{Kastor:1988ef}. The existence of such additional vacua at infinity may play an important role in matching the non-invertible lines along the RG flow \cite{Ambrosino:2026kfu}.} Both exceptional theories lie in the chamber where the Witten index vanishes. A vanishing Witten index allows supersymmetry breaking, and such breaking was demonstrated in models of zero Witten index
\cite{Kastor:1988ef}.

The fixed point at $g_1=-g_2$ has more symmetry than we impose, with supersymmetry enhanced to $\mathcal N=2$ and $\Zex$ enhanced to $S_3$. There, the superpotential is ${\cal W}=\frac{g_2}{6}(X^3-3XY^2)$, which in terms of the complex superfield $Z=X+iY$ is the real part of the holomorphic cubic superpotential ${\cal W}_{A_2^{\mathcal N=2}}=\frac{g_2}{6}Z^3$.
The phase rotation of $Z$ is then a $U(1)_R$ symmetry, which the theory at a generic ratio of couplings does not possess. In components, the Lagrangian is that of the Wess--Zumino model \cite{Wess:1973kz}. The Witten index of the chamber $g_1g_2<0$ agrees with the count of critical points of ${\cal W}_{A_2^{\mathcal N=2}}$.
In general, for the $\mathcal N=2$ superconformal minimal model $A_{k+1}$, which has central charge $\frac{3k}{k+2}$, the superpotential is ${\cal W}_{A_{k+1}^{\mathcal N=2}}\propto Z^{k+2}$ \cite{Martinec:1988zu,Vafa:1988uu,Lerche:1989uy}. Because an $\mathcal N=2$ superpotential is not renormalized, its dimension is $d-1$, and ${\cal W}_{A_{k+1}^{\mathcal N=2}}$ then gives $\Delta_Z=\frac{d-1}{k+2}$, which at $k=1$ is
\begin{equation}\label{eq:deltaZ}
    \Delta_Z=\frac{d-1}{3}\,.
\end{equation}
Expanding \eqref{eq:deltaZ} in $d=4-\eps$ gives $\gamma_Z=\eps/6$. At this ratio of couplings, Table~\ref{tab:fps} gives $\gamma_X=\gamma_Y=\eps/6$ at one loop. The one-loop computation thus reproduces the exact result \eqref{eq:deltaZ}.
In $d=2$, the relation \eqref{eq:deltaZ} gives $\Delta_Z=\frac13$, the value in the $\mathcal N=2$ superconformal minimal model at $k=1$. Unprotected quantities in this theory can be studied with the $4-\eps$ expansion \cite{Giombi:2014xxa,Fei:2016sgs} applied to the Wess--Zumino model, whose RG functions are now known to five loops \cite{Gracey:2021yvb}.

The theory at $g_1=-g_2$ is the $S_3$ super-Potts model \cite{Liendo:2021wpo}. The semidirect product $S_3=\mathbb{Z}_3\rtimes\mathbb{Z}_2$ is generated by
\begin{equation}
    \mathbb{Z}_3:Z\rightarrow e^{2\pi i/3}Z,\qquad \mathbb{Z}_2:Z\rightarrow\bar Z\,.
\end{equation}
The construction of Section~\ref{sub:coset} confirms the identification with the $m=4$ model $(A_3,D_4)$. There
the second label belongs to $\widehat{su}(2)_4$, whose simple current has
$h_J=1$. Since $h_J$ is an integer, $D_4$ is again an integer-spin extension, and the fixed point of its simple current
is at $s=3$ \cite{Intriligator:1989zw}. Only odd $s$ survives, and since NS labels have $r-s$ even, only $r\in\{1,3\}$ appears in the NS sector. At $m=4$, the Kac weights \eqref{eq:kac} give one module containing the conformal weights $0$ and $1$, and two modules from the resolved fixed point, each containing the conformal weight $\frac16$. The state at conformal weight $1$ in the
vacuum module is the $U(1)$ current of the $\mathcal N=2$ algebra, and its superpartner at $h=\frac32$ is the second supercurrent, which extends $\mathcal N=1$ to $\mathcal N=2$. The resolved pair is $Z$ and $\bar Z$ at $\Delta=2h=\frac13$, distinguished by the sign of the $U(1)_R$ charge exactly as $\Phi_\pm$ are distinguished by the sign of $w$.

\subsection{The twist-two operators}
\label{sub:spin3}

Let us now turn to the anomalous dimensions of twist-two operators. The twist of an operator is its dimension minus its spin, $\Delta-s$. A composite of two free fields has
twist $d-2$, which is two at $d=4$. At
every spin from $\frac12$ upward, the LG theory \eqref{eq:LGNY}
has twist-two operators that mix with one another. They are built from its two scalars $\phi_X$ and $\phi_Y$ and its two fermions $\Psi_X$ and $\Psi_Y$. For instance, the $\Zex$-odd spin-$\frac12$ composites are $\chi_\pm=\phi_X\Psi_Y\pm\phi_Y\Psi_X$. The symmetric one, however, is not an independent operator. The fermion equation of motion gives $\slashed\partial\Psi_Y=-g_1\,\chi_+$, so $\chi_+$ is a
descendant of the elementary fermion $\Psi_Y$, of dimension
$\Delta_{\Psi_Y}+1=\Delta_Y+\frac32$. The antisymmetric combination $\chi_-$, by contrast, is a genuine
primary. At spin one, the $\Zex$-odd bilinear currents are
\begin{equation}
J^\phi_{1\mu}=\phi_X\overset{\leftrightarrow}{\partial}_\mu\phi_Y\,,\qquad
J^\Psi_{1\mu}=\bar\Psi_X\gamma_\mu\Psi_Y\,,
\label{eq:J1def}
\end{equation}
which follow the definitions \eqref{eq:Jdef} and \eqref{eq:Jpsidef} of Appendix~\ref{app:currents} at $s=1$, with the two fields of different flavor. Their spin-2 analogs $J^\phi_2$ and $J^\Psi_2$ have one further
symmetrized derivative, and the one-flavor versions $J^{\phi,a}_2$ and
$J^{\Psi,a}_2$, with $a\in\{X,Y\}$, follow the same definitions at $s=2$ with both fields of identical flavor $a$.

At half-integer spin, a twist-two operator is
built from one scalar and one fermion instead,
\begin{equation}
S^{ab}_{s\,\mu_1\ldots\mu_{s-1/2}}=\sum_{k=0}^{s-1/2}d_k\,
\partial_{\mu_1}\!\!\cdots\partial_{\mu_k}\phi_a\;
\partial_{\mu_{k+1}}\!\!\cdots\partial_{\mu_{s-1/2}}\Psi_b\;-\;\text{traces}\,,
\label{eq:Sdef}
\end{equation}
symmetrized in the $s-\frac12$ vector indices, with the spinor
index of $\Psi_b$ left uncontracted, so the two fields share $s-\frac12$ derivatives and the spinor index provides the remaining half unit of spin. The traces to be removed include the gamma traces $\gamma^{\mu_i}S^{ab}_{s\,\mu_1\ldots\mu_{s-1/2}}$, not only the $\delta_{\mu_i\mu_j}$ traces, since a tensor-spinor is irreducible only when both vanish. The $\Zex$-odd operators are those with the two mixed flavor
assignments $(a,b)=(X,Y)$ and $(Y,X)$. At spin $\frac32$ and at spin $\frac52$, the two
assignments give two independent combinations each, which we call $\Sigma_{3/2}$, $\Sigma'_{3/2}$, $\Sigma_{5/2}$ and $\Sigma'_{5/2}$ and distinguish by their anomalous
dimension. Throughout, a prime marks the top component of a supermultiplet.

At each integer spin, the scalar and the fermion current mix under
renormalization.
The norms of their divergences then give the mixing matrix $\hat\gamma_s$ by the standard recombination relation
\cite{Giombi:2016hkj,Giombi:2017rhm}. We report the details of the computation in
Appendix~\ref{app:currents}. In the $\Zex$-odd sector, we obtain $\hat\gamma_1$ and $\hat\gamma_2^{\rm odd}$,
\begin{equation}\label{eq:gamma12}
\begin{aligned}
&\hat\gamma_1=\frac{\eps}{218}
\begin{pmatrix}31&6\sqrt2\\[2pt]6\sqrt2&37\end{pmatrix}\,, &\qquad &\gamma_1^{\pm}=\frac{43}{218}\eps,\ \frac{25}{218}\eps\,,\\
&\hat\gamma_2^{\rm odd}=\frac{\eps}{218}
\begin{pmatrix}31&-10\sqrt6\\[2pt]-10\sqrt6&21\end{pmatrix}\,, &\qquad &\gamma_2^{{\rm odd},\pm}=\frac{51}{218}\eps,\ \frac1{218}\eps\,.
\end{aligned}
\end{equation}
Similarly, at spin three, we obtain $\hat\gamma_3$,
\begin{equation}\label{eq:gamma3}
\hat\gamma_3=\frac{\eps}{218}
\begin{pmatrix}31&2\sqrt3\\[2pt]2\sqrt3&32\end{pmatrix},\qquad
\gamma_3^{\pm}=\frac{35}{218}\eps,\ \frac{14}{109}\eps\,.
\end{equation}
We denote by $I_3$ the eigenoperator with $\gamma_3^-$ and by $I'_3$ the one with $\gamma_3^+$.
Every mixing matrix is written in the basis $(\hat J^\phi_s,\hat J^\Psi_s)$, where the hat denotes unit normalization. All six eigenvalues are positive. A conserved spin-$s$ current would lie at the unitarity bound $\Delta=d-2+s$, which none of the six reaches.

Near four dimensions, only twist-two operators can come close to the dimension $d+1$ of a conserved spin-3 current, because an operator built from $n$ fields has twist $n(d-2)/2$ up to corrections of order $\eps$. We have just seen that the $\Zex$-odd ones are not conserved. The $\Zex$-even ones would be same-flavor bilinears, which vanish at spin three. The fixed point therefore has no conserved spin-3 current at all.

Before matching the twist-two operators to the extended modules in Section~\ref{sub:web},
we group them into supermultiplets. A supermultiplet is a superconformal primary at spin $s$ and
its supercharge descendant at spin $s+\frac12$, with the same anomalous dimension. These are the bottom and top components of Section~\ref{sub:warmup}. Whether an operator is a bottom or a top component determines its two-dimensional counterpart, since a bottom component maps to a quasi-primary and a top component to a $G_{-1/2}$ descendant of a quasi-primary.

To understand the supermultiplet structure, we use the free supercharge at
leading order. At $\eps=0$, the
fixed point is free, and the supercharge acts by $Q\phi=\Psi$ and
$Q\Psi=\slashed\partial\phi$.\footnote{With the spinor index restored, the transformations with a constant Majorana parameter $\xi$ are $\delta\phi=\bar\xi\Psi$ and $\delta\Psi=\gamma^\mu(\partial_\mu\phi)\,\xi$. The gamma-matrix structure reduces to the collinear derivative on the
null line of \eqref{eq:Qray} below.} On a twist-two quasi-primary of spin $s$, the supercharge gives a twist-two operator of spin $s+\frac12$ together with total derivatives. In unit-normalized bases $\mathcal O^{(s)}_m$ at spin $s$ and $\mathcal O^{(s+1/2)}_n$ at spin $s+\frac12$, the supercharge acts as a matrix $\mathcal G$,
\begin{equation}
Q\,\mathcal O^{(s)}_m=\sum_n\mathcal G_{nm}\,\mathcal O^{(s+1/2)}_n
+\partial(\cdots)\,.
\label{eq:Gdef}
\end{equation}
At one loop, every eigenoperator of $\hat\gamma_s$ is a combination of free
operators, so $\mathcal G$ is the leading form of the supercharge on the
eigenoperators of the interacting theory. Supersymmetry sends an
eigenoperator to an eigenoperator of the same anomalous dimension,
because the superconformal algebra fixes the dimension of $Q\,\mathcal O$ to be exactly one half above that of $\mathcal O$. In terms of $\hat\gamma_s$ and $\hat\gamma_{s+1/2}$, this reads
\begin{equation}
\hat\gamma_{s+1/2}\,\mathcal G=\mathcal G\,\hat\gamma_{s}\,.
\label{eq:intertwine}
\end{equation}

The matrix $\mathcal G$ can be computed explicitly on the null line. Every twist-two operator is a two-field operator, and the collinear fields that build it have the collinear conformal spin $j=\frac12(\Delta+s)$, where $s$ is the spin along the line. The spin $j$ is $\frac12$ for a scalar and $1$ for the component $\lambda$ of a fermion that
survives the light-cone projection. The supercharge acts on them by
\begin{equation}
Q\,\phi_X=\lambda_X\,,\qquad Q\,\lambda_X=\partial \phi_X\,,
\label{eq:Qray}
\end{equation}
so that $Q^2=\partial$ along the null line. An operator of spin $s$ is fixed by the constants that distribute its
derivatives between the two fields, the $a_k$, $c_k$ and $d_k$ of \eqref{eq:Jdef},
\eqref{eq:Jpsidef} and \eqref{eq:Sdef}. The action \eqref{eq:Qray} is linear in these constants, so the removal of the total derivative in \eqref{eq:Gdef} reduces to linear algebra. The two currents at a given spin are both twist-two, with two-point functions of the same form \eqref{eq:JJ} that differ only in the constant $C_J$. Unit normalization divides each current by $\sqrt{C_J}$, and
passing to the null line multiplies both constants $C_J$ by a common kinematic factor. Because the factor cancels in the ratio of the two unit-normalized components,
a ratio computed on the null line is identical to the four-dimensional one, and we can
compare it with the eigenvectors of $\hat\gamma_s$.

Let us take the lowest pair of spins as an example. On the null line,
the $\Zex$-odd spin-$\frac12$ operators are $\chi_\pm=\phi_X\lambda_Y\pm\phi_Y\lambda_X$,
and the spin-$1$ quasi-primaries are $J^\phi=\phi_X\partial\phi_Y-\partial\phi_X\,\phi_Y$
and $J^\Psi=\lambda_X\lambda_Y$, which are \eqref{eq:J1def} restricted to
the null line. Applying \eqref{eq:Qray} gives $Q\chi_+=\partial(\phi_X\phi_Y)$, a total derivative, so $\chi_+$ is the direction that $\mathcal G$ annihilates, and
\begin{equation}
Q\chi_-=J^\phi+2\,\lambda_X\lambda_Y\,.
\label{eq:Qchi}
\end{equation}
In the normalization \eqref{eq:Qray}, which fixes the collinear scalar and
fermion relative to each other, the squared norms of $J^\phi$ and $J^\Psi$
are in the ratio $2:1$, so that the image \eqref{eq:Qchi} has components proportional to $(\frac1{\sqrt2},1)$ in unit-normalized bases. The image, built in the free theory, is the eigenoperator
$\hat J^\Psi_1+\frac1{\sqrt2}\hat J^\phi_1$ of the one-loop $\hat\gamma_1$ in \eqref{eq:gamma12},
at the eigenvalue $\gamma_1^+=\frac{43}{218}\eps$. By \eqref{eq:intertwine}, it is the top component of the supermultiplet whose bottom component $\chi_-$ shares its anomalous dimension $\frac{43}{218}\eps$. The other eigenoperator of $\hat\gamma_1$, at $\gamma_1^-=\frac{25}{218}\eps$, is the
combination $\hat J^\Psi_1-\sqrt2\,\hat J^\phi_1$ of Table~\ref{tab:web}. It lies
outside the image and is therefore a bottom component, whose top component appears at spin
$\frac32$ with the same eigenvalue.

Together, the action \eqref{eq:Gdef} and the relation \eqref{eq:intertwine} distinguish a top component from a bottom component.
In the $\Zex$-odd sector, the space of twist-two quasi-primaries is
two-dimensional at every spin, and the structure of the lowest pair repeats at every
higher spin. The action of $Q$ on a top component is the action of $Q^2$, a derivative, on its bottom component. At each spin above the lowest, $\mathcal G$ then annihilates the top component of the supermultiplet below, and the eigenoperator orthogonal to it is the bottom component of the next supermultiplet. On every pair of adjacent spins in the $\Zex$-odd sector, $\mathcal G$ therefore
has rank one.

We now evaluate $\mathcal G$ at the remaining pairs. At integer spin, the mixing matrix is known, and the relation \eqref{eq:intertwine} requires the image to be one of its eigenvectors. At half-integer spin, we did not compute a mixing matrix,\footnote{Here, the divergence contains three-fermion operators, which make the computation more complicated.} so the supercharge alone assigns the eigenoperators and their anomalous dimensions.

At half-integer spin, the two $\Zex$-odd operators are $S^{XY}_s$ and $S^{YX}_s$ of \eqref{eq:Sdef}. The supercharge acts on $X$ and $Y$ in the same way, and, as in Appendix~\ref{app:currents}, the spin-one currents change sign under $X\leftrightarrow Y$ while the spin-two currents do not. The image of $\mathcal G$ is therefore $S^{XY}_{3/2}-S^{YX}_{3/2}$ at spin $\frac32$ and $S^{XY}_{5/2}+S^{YX}_{5/2}$ at spin $\frac52$.

The difference $S^{XY}_{3/2}-S^{YX}_{3/2}$ is $\Sigma'_{3/2}$. Since $\mathcal G$ annihilates the spin-one top component, $\Sigma'_{3/2}$ is the top component of $\hat J^\Psi_1-\sqrt2\,\hat J^\phi_1$ at $\gamma_1^-=\frac{25}{218}\eps$, and the sum $S^{XY}_{3/2}+S^{YX}_{3/2}$ is the bottom component $\Sigma_{3/2}$. From spin $\frac32$ to spin $2$, the image has fermion and scalar components in the ratio $\frac{\sqrt6}2$. This is the ratio of the eigenvector $\hat J^\Psi_2+\frac{\sqrt6}3\hat J^\phi_2$ of $\hat\gamma_2^{\rm odd}$ at $\gamma_2^{{\rm odd},-}=\frac1{218}\eps$. Since the eigenvector is the top component of $\Sigma_{3/2}$, $\Sigma_{3/2}$ has the anomalous dimension $\frac1{218}\eps$ as well, and the second eigenvector $\hat J^\Psi_2-\frac{\sqrt6}2\hat J^\phi_2$ at $\gamma_2^{{\rm odd},+}=\frac{51}{218}\eps$ is the bottom component at spin two.

The sum $S^{XY}_{5/2}+S^{YX}_{5/2}$ is $\Sigma'_{5/2}$, the top component of $\hat J^\Psi_2-\frac{\sqrt6}2\hat J^\phi_2$, and the difference $S^{XY}_{5/2}-S^{YX}_{5/2}$ is the bottom component $\Sigma_{5/2}$. From spin $\frac52$ to spin $3$, the image has fermion and scalar components in the ratio $\frac2{\sqrt3}$. This is the ratio of $I'_3$, the eigenvector of $\hat\gamma_3$ at $\gamma_3^+=\frac{35}{218}\eps$. Since $I'_3$ is the top component of $\Sigma_{5/2}$, $\Sigma_{5/2}$ has the anomalous dimension $\frac{35}{218}\eps$ as well, and $I_3$, at $\gamma_3^-=\frac{14}{109}\eps$, is the bottom component at spin three. The top component of $I_3$ should appear at spin $\frac72$. At every pair above the lowest, $\mathcal G$ has rank one and annihilates the top component, as predicted above.

The $\Zex$-even sector is organized in the same way. It has no twist-two operators at spins one and three, where the same-flavor bilinears vanish. At spin $\frac32$, it has exactly two, the conserved supercurrent $S$ and the relative supercurrent $S_X-S_Y$, and both are bottom components, because a top component at spin $\frac32$ would be the image of a spin-one operator. At spin two, it has the four one-flavor currents $J^{\phi,a}_2$ and $J^{\Psi,a}_2$, and their mixing matrix has the eigenvalues
\begin{equation}\label{eq:g2even}
    \gamma_2^{\rm even}=\left\{\,0,\ \frac{18}{109},\ \frac{119\pm\sqrt{1921}}{654}\right\}\eps\,.
\end{equation}
On the pairs of spins from $\frac32$ to $2$ and from $2$ to $\frac52$, $\mathcal G$ has rank two. At leading order, the supercharge sends the supercurrent of each flavor to the stress tensor of the same flavor. The top component of $S$ is then the stress tensor $T$, at the zero eigenvalue, and the top component of $S_X-S_Y$ is the relative stress tensor $K=T_X-T_Y$, at the eigenvalue $\frac{18}{109}\eps$ of \eqref{eq:g2even}. Note that $K$ is a quasi-primary of the conformal algebra but not a superconformal primary. The two remaining spin-two eigenoperators are bottom components. Their top components at spin $\frac52$ have the eigenvalues $\frac{119\pm\sqrt{1921}}{654}\eps$, which are predictions of supersymmetry.

\subsection{Operator spectrum and matching}
\label{sub:web}

Upon continuation to $d=2$, each operator of the LG theory becomes an operator of the $\ed$ superconformal minimal model. We propose the two-dimensional counterpart of every low-lying operator in Table~\ref{tab:web}, together with the discrete charges, the one-loop coefficients of the anomalous dimensions, and the dimensions continued to $d=3$ and $d=2$. The proposal is tested by comparing the continued dimensions in the column $\Delta_{d=2}$ with the exact dimensions in the last column. Some operators map to the spinning primaries of Section~\ref{sub:spinning}. Most of the others map to quasi-primaries at higher level in the towers of the extended modules. The exception is $\chi_+$, a descendant on both sides.

We use the following notation in the table. A pair $\phi_{r,s}\tilde\phi_{r',s'}$ is the product of the holomorphic primary of Kac label $(r,s)$ and the antiholomorphic primary of label $(r',s')$, with the conformal weights of \eqref{eq:kac}. The reflection $(r,s)\simeq(12-r,14-s)$ gives every primary a second Kac label, but we use the one with the smaller $r$. While every operator has a conjugate with the two sides exchanged, we list only the one with non-negative spin. An entry with modes, such as $G_{-\frac32}\phi_{5,3}\tilde\phi_{1,3}$, denotes a quasi-primary by its leading term. The row $\chi_+$ is the exception, and its entry contains the only right mode in the table, $\tilde G_{-\frac12}$. The positions and multiplicities of the quasi-primaries below, and their separation into bottom and top components, follow from the super-Virasoro characters of the towers in Table~\ref{tab:modules}.

\begin{table}[!ht]
\caption{\label{tab:web}LG description of the $\ed$ universality class. The low-lying operators, their dimensions continued to $d=3$ and $d=2$, and the exact dimensions in the $\ed$ superconformal minimal model. The blocks are the scalars, the $\Zex$-odd operators and the $\Zex$-even ones. The column $(R,C)$ is as in Table~\ref{tab:map}. A fermion row contains $\psi_a$, the upper component of the two-component Majorana spinor $\Psi_a$ of \eqref{eq:spinors}. A fermion reached by the left $G_{-\frac12}$ has the $R$ opposite to that of the boson of its supermultiplet, while the right $\tilde G_{-\frac12}$ in the row $\chi_+$ keeps the $R$ of the boson. The two $\Delta$ columns continue each row to $d=3$ and $d=2$, by the one-sided $[1,1]$ Pad\'e where the row has two-loop information and by the one-loop truncation otherwise.}
\centering
\resizebox{\ifdim\width>\textwidth\textwidth\else\width\fi}{!}{%
\begin{tabular}{llllllll}
\toprule
operator & $s$ & $(R,C)$ & $\gamma^{(1)}/\eps$ & $\Delta_{d=3}$ &
$\Delta_{d=2}$ & $d=2$ field & $\Delta_{2d}$ exact \\
\midrule
$\phi_X$ & $0$ & $(-,+)$ & $\frac{13}{218}$ & $0.571$ & $0.163$ & $\phi_{5,5}\tilde\phi_{5,5}$ &
  $\frac17=0.143$ \\
$\phi_Y$ & $0$ & $(-,-)$ & $\frac{9}{109}$ & $0.609$ & $0.264$ & $i(\phi^-_{5,7}\tilde\phi^-_{5,7}{-}\phi^+_{5,7}\tilde\phi^+_{5,7})$ &
  $\frac27=0.286$ \\
$P=3\phi_X^2{-}2\phi_Y^2$ & $0$ & $(+,+)$ & $\frac{9}{109}$ & $1.092$ & $0.202$ & $\phi^+_{5,7}\tilde\phi^+_{5,7}{+}\phi^-_{5,7}\tilde\phi^-_{5,7}$ &
  $\frac27=0.286$ \\
$XP=3\phi_X^3{-}2\phi_X\phi_Y^2$ & $0$ & $(-,+)$ & $\frac{79}{218}$ & $1.813$ & $0.517$ & $\phi_{1,3}\tilde\phi_{1,3}$ &
  $\frac57=0.714$ \\
\midrule
$\psi_Y$ & $\frac12$ & $(+,-)$ & $\frac{9}{109}$ & $1.109$ & $0.764$ &
  $G_{-\frac12}i(\phi^-_{5,7}\tilde\phi^-_{5,7}{-}\phi^+_{5,7}\tilde\phi^+_{5,7})$ & $\frac{11}{14}=0.786$ \\
$\chi_+=\phi_X\psi_Y{+}\phi_Y\psi_X$ & $\frac12$ & $(-,-)$ & $\frac{127}{218}$ & $2.109$ & $1.764$ &
  $\partial\tilde G_{-\frac12}i(\phi^-_{5,7}\tilde\phi^-_{5,7}{-}\phi^+_{5,7}\tilde\phi^+_{5,7})$ & $\frac{25}{14}=1.786$ \\
$\chi_-=\phi_X\psi_Y{-}\phi_Y\psi_X$ & $\frac12$ & $(-,-)$ & $\frac{43}{218}$ & $1.697$ & $0.894$ &
  $\phi_{5,3}\tilde\phi_{1,3}$ & $\frac{17}{14}=1.214$ \\
$\hat J^\Psi_1{+}\frac1{\sqrt2}\hat J^\phi_1$ & $1$ & $(+,-)$ & $\frac{43}{218}$ &
  $2.197$ & $1.394$ & $G_{-\frac12}\phi_{5,3}\tilde\phi_{1,3}$ & $\frac{12}7=1.714$ \\
$\hat J^\Psi_1{-}\sqrt2\,\hat J^\phi_1$ & $1$ & $(+,-)$ & $\frac{25}{218}$ & $2.115$ & $1.229$ &
  $\phi_{5,9}\tilde\phi_{5,5}$ & $\frac87=1.143$ \\
$\Sigma'_{3/2}$ & $\frac32$ & $(-,-)$ & $\frac{25}{218}$ & $2.615$ & $1.729$ &
  $G_{-\frac12}\phi_{5,9}\tilde\phi_{5,5}$ & $\frac{23}{14}=1.643$ \\
$\Sigma_{3/2}$ & $\frac32$ & $(-,-)$ & $\frac{1}{218}$ & $2.505$ & $1.509$ &
  $\phi_{1,5}\tilde\phi_{5,5}$ & $\frac{23}{14}=1.643$ \\
$\hat J^\Psi_2{+}\frac{\sqrt6}3\hat J^\phi_2$ & $2$ & $(+,-)$ & $\frac{1}{218}$ &
  $3.005$ & $2.009$ & $G_{-\frac12}\phi_{1,5}\tilde\phi_{5,5}$ & $\frac{15}7=2.143$ \\
$\hat J^\Psi_2{-}\frac{\sqrt6}2\hat J^\phi_2$ & $2$ & $(+,-)$ & $\frac{51}{218}$ &
  $3.234$ & $2.468$ & $G_{-\frac32}\phi_{5,3}\tilde\phi_{1,3}$ &
  $\frac{19}7=2.714$ \\
\midrule
$\psi_X$ & $\frac12$ & $(+,+)$ & $\frac{13}{218}$ & $1.071$ & $0.663$ &
  $G_{-\frac12}\phi_{5,5}\tilde\phi_{5,5}$ & $\frac9{14}=0.643$ \\
$\psi_P=2(3\phi_X\psi_X{-}2\phi_Y\psi_Y)$ & $\frac12$ & $(-,+)$ & $\frac{9}{109}$ & $1.592$ & $0.702$ &
  $G_{-\frac12}(\phi^+_{5,7}\tilde\phi^+_{5,7}{+}\phi^-_{5,7}\tilde\phi^-_{5,7})$ & $\frac{11}{14}=0.786$ \\
$S_X{-}S_Y$ & $\frac32$ & $(-,+)$ & $\frac{18}{109}$ & $2.665$ & $1.830$ &
  $G_{-\frac32}(\phi^+_{5,7}\tilde\phi^+_{5,7}{+}\phi^-_{5,7}\tilde\phi^-_{5,7})$ & $\frac{25}{14}=1.786$ \\
$K$ & $2$ & $(+,+)$ & $\frac{18}{109}$ & $3.165$ & $2.330$ & $G_{-\frac12}G_{-\frac32}(\phi^+_{5,7}\tilde\phi^+_{5,7}{+}\phi^-_{5,7}\tilde\phi^-_{5,7})$ &
  $\frac{16}7=2.286$ \\
\bottomrule
\end{tabular}%
}
\end{table}

Not every anomalous dimension in Table~\ref{tab:web} is computed directly. The dimension of $XP$ comes from the RG eigenvalue \eqref{eq:yrel} as in Table~\ref{tab:map}, the dimension of $\chi_+$ is the $\Delta_{\Psi_Y}+1$ of Section~\ref{sub:spin3}, and the values for $\chi_-$, $\Sigma'_{3/2}$, $\Sigma_{3/2}$ and $S_X{-}S_Y$ are those of their supermultiplet partners. The scalar rows, and the fermion rows $\psi_X$, $\psi_Y$, $\psi_P$ and $\chi_+$, have two-loop anomalous dimensions from \eqref{eq:gammas}, \eqref{eq:gammaP} and \eqref{eq:yrel}, resummed by the Pad\'e of Section~\ref{sub:fp}. Every other row, $\chi_-$ included, is known only at one loop and is quoted as the one-loop truncation.

We match a row by its spin, its charges $(R,C)$, and its position in a supermultiplet. Since $R$ and $\tilde R$ agree on a boson and differ on a fermion, we compare the charges of a supermultiplet on its bosonic member. On the LG side, a composite of $n$ scalars has R-parity $(-1)^n$. On a fermion bilinear, $\ZR$ acts in four dimensions as the chiral rotation of Section~\ref{sub:fp}, whose $\gamma_5$ anticommutes with every $\gamma^\mu$ and commutes with every derivative, so the R-parity counts the gamma matrices. The bilinear $\bar\Psi\Psi$ is $\ZR$-odd, while the currents of \eqref{eq:Jpsidef} are $\ZR$-even at every spin. Accordingly, a twist-two operator at integer spin, a bilinear of two scalars or of two fermions, is $\ZR$-even in either $\Zex$ sector. On the two-dimensional side, the charges of a state follow from \eqref{eq:Rop} with \eqref{eq:grading} and \eqref{eq:Ctable}. We recall from Section~\ref{sub:spin3} that a bottom component maps to a quasi-primary and a top component to the $G_{-\frac12}$ descendant of the two-dimensional counterpart of its bottom component. Finally, the lightest LG operator with given spin and charges maps to the lightest operator of the $\ed$ superconformal minimal model with the same spin and charges.

The four scalar rows are the dictionary of Section~\ref{sub:fp}, with the degeneracy of the pair $\Phi_\pm$ lifted in Section~\ref{sub:deg}. The fermion rows of their supermultiplets follow. The two-dimensional counterpart of $\psi_X$ is $G_{-\frac12}\phi_{5,5}\tilde\phi_{5,5}$. The two-dimensional counterparts of $\psi_Y$ and $\psi_P$ are the $G_{-\frac12}$ descendants of those of $\phi_Y$ and $P$, half a unit above $\phi^\pm_{5,7}$, at $\Delta=\frac{11}{14}$. Note that the fermionic current $U$ does not double these states, because a null vector of \cite{Schoutens:1990xg} sets one combination of $G_{-\frac12}$ and $U_{-\frac12}$ to zero on $\phi^\pm_{5,7}$. Each of the two modules therefore contains a single state at $h=\frac9{14}$. The equation of motion of Section~\ref{sub:spin3} makes the spin-$\frac12$ operator $\chi_+$ a descendant of $\Psi_Y$, with upper component $\partial\tilde\psi_Y$, whose two-dimensional counterpart is the $\partial\tilde G_{-\frac12}$ descendant of that of $\phi_Y$, at $\Delta=\frac{25}{14}$.

Let us now turn to the $\Zex$-odd twist-two operators, which Section~\ref{sub:spin3} organized into supermultiplets. On the two-dimensional side, the spinning primaries of Section~\ref{sub:spinning} pair a left tower and a right tower of the same extended module, and exactly one of them is $\Zex$-odd at each of the spins $\frac12$, $1$ and $\frac32$: the fermion $\phi_{5,3}\tilde\phi_{1,3}$ at $\Delta=\frac{17}{14}$, the vector $\phi_{5,9}\tilde\phi_{5,5}$ at $\Delta=\frac87$, and the $\phi_{1,5}\tilde\phi_{5,5}$ of $\widehat X$ at $\Delta=\frac{23}{14}$. The bottom component $\chi_-$ maps to $\phi_{5,3}\tilde\phi_{1,3}$, the lightest spinning fermion of the $\ed$ superconformal minimal model, and the bottom component $\hat J^\Psi_1-\sqrt2\,\hat J^\phi_1$ maps to the vector $\phi_{5,9}\tilde\phi_{5,5}$. Their top components, the rows $\hat J^\Psi_1+\frac1{\sqrt2}\hat J^\phi_1$ and $\Sigma'_{3/2}$ of the table, map to the $G_{-\frac12}$ descendants $G_{-\frac12}\phi_{5,3}\tilde\phi_{1,3}$ and $G_{-\frac12}\phi_{5,9}\tilde\phi_{5,5}$.

At spin two, the $\Zex$-odd sector has two rows, the top component of $\Sigma_{3/2}$ and the bottom component $\hat J^\Psi_2-\frac{\sqrt6}2\hat J^\phi_2$. Both have charges $(R,C)=(+,-)$, and as twist-two operators they are the lightest spin-two operators with these charges. In the $\ed$ superconformal minimal model, the two lightest spin-two quasi-primaries with charges $(+,-)$ are the top component $G_{-\frac12}\phi_{1,5}\tilde\phi_{5,5}$ at $\Delta=\frac{15}7$, the descendant of the spinning primary at $\Delta=\frac{23}{14}$, and the bottom component $G_{-\frac32}\phi_{5,3}\tilde\phi_{1,3}$ at $\Delta=\frac{19}7$. The dictionary therefore maps the top component of $\Sigma_{3/2}$ to $G_{-\frac12}\phi_{1,5}\tilde\phi_{5,5}$, $\Sigma_{3/2}$ itself to $\phi_{1,5}\tilde\phi_{5,5}$, and $\hat J^\Psi_2-\frac{\sqrt6}2\hat J^\phi_2$ to $G_{-\frac32}\phi_{5,3}\tilde\phi_{1,3}$. Note that at $\Delta=\frac{23}{14}$ two quasi-primaries have the same charges, the primary $\phi_{1,5}\tilde\phi_{5,5}$ and the descendant $G_{-\frac12}\phi_{5,9}\tilde\phi_{5,5}$. The descendant is the two-dimensional counterpart of the top component $\Sigma'_{3/2}$, so the primary is that of the bottom component $\Sigma_{3/2}$.

\begin{sloppypar}
At spin $\frac52$ and at spin three, the $\Zex$-odd operators are $\Sigma_{5/2}$ and $\Sigma'_{5/2}$, with charges $(R,C)=(-,-)$, and $I'_3$ and $I_3$, with charges $(+,-)$. The lightest spin-$\frac52$ quasi-primary with charges $(-,-)$ is the bottom component $G_{-\frac32}\phi_{5,9}\tilde\phi_{5,5}$ at $\Delta=\frac{37}{14}$, and the three lightest spin-3 quasi-primaries with charges $(+,-)$ lie at $\Delta=\frac{22}7$: the $G_{-\frac12}$ descendant of $G_{-\frac32}\phi_{5,9}\tilde\phi_{5,5}$, the only top component there, and the two bottom components $G_{-\frac32}\phi_{1,5}\tilde\phi_{5,5}$ and $L_{-2}\phi_{5,9}\tilde\phi_{5,5}$. The dictionary therefore maps $\Sigma_{5/2}$ to $G_{-\frac32}\phi_{5,9}\tilde\phi_{5,5}$, $I'_3$ to $G_{-\frac12}G_{-\frac32}\phi_{5,9}\tilde\phi_{5,5}$, and $I_3$ to one of the two bottom components at $\Delta=\frac{22}7$, which the charges do not distinguish. The top component $\Sigma'_{5/2}$ of $\hat J^\Psi_2-\frac{\sqrt6}2\hat J^\phi_2$ maps to $G_{-\frac12}G_{-\frac32}\phi_{5,3}\tilde\phi_{1,3}$ at $\Delta=\frac{45}{14}$.
\end{sloppypar}

In the $\Zex$-even sector, the conserved operators $T$ and $S$ are the stress tensor and the supercurrent of the $\ed$ superconformal minimal model, and their dimensions $d$ and $d-\frac12$ are exact. We match the supermultiplet of $S_X-S_Y$ and $K$ of Section~\ref{sub:spin3} on its bosonic member $K$. We recall that within a tower the eigenvalue of $C$ does not change, while the chiral fermion parity reverses at every half-integer level. A state at an integer number of levels above the two-dimensional counterpart of a scalar therefore has the charges of that scalar. Of the four scalars, only $P$ has the charges $(+,+)$ of $K$, so the lightest spin-two states with charges $(+,+)$ outside the vacuum module lie two levels above the two-dimensional counterpart of $P$, at $\Delta=\frac{16}7$. This level contains two quasi-primaries, a superconformal quasi-primary and the $G_{-\frac12}$ descendant of the single quasi-primary at level $\frac32$. Since $K$ is a top component, its two-dimensional counterpart is the descendant, and the quasi-primary at level $\frac32$ is that of the bottom component $S_X-S_Y$, at $\Delta=\frac{25}{14}$.

Note that no twist-two operator maps to the extended currents. The current $W$ lies at $(h,\tilde h)=(3,0)$, half a level above the current $U$ at $h=\frac52$ in the vacuum module, and $U$ and $W$ form one supermultiplet, whose bosonic member is $W$. It is $\Zex$-odd, since the tower of $U$ has $C_i=-1$ in \eqref{eq:Ctable} and the tower of the identity has $C_j=+1$. At levels $3$ and $0$ above the lowest state of the vacuum module, both chiral fermion parities \eqref{eq:grading} of $W$ are $+1$, and the R-parities \eqref{eq:Rop} make it $\ZR$-odd. No twist-two supermultiplet has the charges $(-,-)$ of $W$, because a twist-two operator at integer spin is $\ZR$-even.

For $\ed$, every row of Table~\ref{tab:web} thus has a two-dimensional counterpart with matching charges and supermultiplet structure. We can repeat the matching for $(D_6,E_6)$, the endpoint of the RG flow \eqref{eq:Eflow}. Since the
two copies of $SM(3)$ decouple there, every dimension is a sum of
two single-superfield dimensions and every conformal weight is a sum of two
$SM(3)$ weights. We compare the two sums in Table~\ref{tab:m10} for the
low-lying operators, using the two-loop series $\gamma_\Phi=\frac\eps{14}+\frac{\eps^2}{49}$ of each flavor $\Phi_i$ and the components \eqref{eq:wcomponents} of $SM(3)$. Here, we label
the two-dimensional side by the Kac indices of $SM(10)$. In this table, $R$ is the chiral fermion parity $\omega$ of \eqref{eq:grading} without the factor $C$ of \eqref{eq:Rop}, because $\omega$ alone reverses the sign of $\phi_X$ and $\phi_Y$ at $m=10$, where both $\phi^\pm_{5,5}$ have $\omega=-1$.

Two properties of the decoupled fixed point have no analog at $m=12$. Because the
two copies are identical, the elementary fields $\phi_X$ and $\phi_Y$ have equal
dimensions at every order, whereas the Hermitian combinations of $\Phi_\pm$ appear in the LG description of $\ed$ as one elementary
field and one composite, whose dimensions separate at two loops
(Section~\ref{sub:deg}). Because each copy keeps its own stress tensor and
supercurrent, the differences $T_1-T_2$ and $S_1-S_2$ are conserved, and the
vacuum module of $(D_6,E_6)$ contains them as the supermultiplet of the primary at $h_{1,5}=\frac32$. In their place, the vacuum module of $\ed$ contains
the current $U$ at $h_{5,1}=\frac52$. The $\ed$ operators with the spins and charges of $S_1-S_2$ and $T_1-T_2$ are $\Sigma_{3/2}$ and its top component, whose anomalous dimension $\frac1{218}\eps$ is the smallest value in Table~\ref{tab:web}.

\begin{table}[!htbp]
\caption{\label{tab:m10}LG description of the $(D_6,E_6)$ theory at $m=10$, the decoupled fixed point $g_1=g_2$ of \eqref{eq:W}. Here $T_i$ and $S_i$ are the stress tensor and the supercurrent of the flavor $\Phi_i$ of \eqref{eq:flavor}. The two-dimensional column is written in the Kac indices of $SM(10)$, with \eqref{eq:kac} at $m=10$ and the reflection $(r,s)\simeq(10-r,12-s)$. The extended modules are the $D_6$ orbits $\{r,10-r\}$ with $s\in\{1,5,7,11\}$ (Section~\ref{sub:gradedring}), and $\phi^\pm_{5,5}$ are the two modules into which the fixed point $r=5$ of the simple current resolves, exchanged by $\Zex$. The operator $K=T_X-T_Y$ is that of Section~\ref{sub:spin3}. The column $C$ is the $\Zex$ charge and $R$ is $\omega$, as explained in the text. The blocks are the scalars, the $\Zex$-odd operators and the $\Zex$-even ones. The scaling dimensions in $d=3$ and $d=2$ are from one-sided $[1,1]$ Pad\'e extrapolations, except for the conserved rows $T_1{-}T_2$ and $S_1{-}S_2$, whose dimensions are exact.}
\centering
\resizebox{\ifdim\width>\textwidth\textwidth\else\width\fi}{!}{%
\begin{tabular}{llllllll}
\toprule
operator & $s$ & $(R,C)$ & $\gamma^{(1)}/\eps$ & $\Delta_{d=3}$ &
$\Delta_{d=2}$ & $d=2$ field & $\Delta_{2d}$ exact \\
\midrule
$\phi_X$ & $0$ & $(-,+)$ & $\frac1{14}$ & $0.591$ & $0.217$ & $\phi^+_{5,5}\tilde\phi^+_{5,5}{+}\phi^-_{5,5}\tilde\phi^-_{5,5}$ & $\frac15=0.2$ \\
$\phi_Y$ & $0$ & $(-,-)$ & $\frac1{14}$ & $0.591$ & $0.217$ & $\phi^+_{5,5}\tilde\phi^+_{5,5}{-}\phi^-_{5,5}\tilde\phi^-_{5,5}$ & $\frac15=0.2$ \\
$\phi_X^2{-}\phi_Y^2$ & $0$ & $(+,+)$ & $\frac17$ & $1.182$ & $0.435$ & $\phi_{3,5}\tilde\phi_{3,5}$ & $\frac25=0.4$ \\
$\partial_X {\cal W}\propto\phi_X^2{+}\phi_Y^2$ & $0$ & $(+,+)$ & $\frac47$ & $1.591$ & $1.217$ & $G_{-\frac12}\tilde G_{-\frac12}(\phi^+_{5,5}\tilde\phi^+_{5,5}{+}\phi^-_{5,5}\tilde\phi^-_{5,5})$ & $\frac65=1.2$ \\
$\partial_Y {\cal W}\propto\phi_X\phi_Y$ & $0$ & $(+,-)$ & $\frac47$ & $1.591$ & $1.217$ & $G_{-\frac12}\tilde G_{-\frac12}(\phi^+_{5,5}\tilde\phi^+_{5,5}{-}\phi^-_{5,5}\tilde\phi^-_{5,5})$ & $\frac65=1.2$ \\
\midrule
$\psi_Y$ & $\frac12$ & $(+,-)$ & $\frac1{14}$ & $1.091$ & $0.717$ & $G_{-\frac12}(\phi^+_{5,5}\tilde\phi^+_{5,5}{-}\phi^-_{5,5}\tilde\phi^-_{5,5})$ & $\frac7{10}=0.7$ \\
$S_1{-}S_2$ & $\frac32$ & $(-,-)$ & $0$ & $\frac52$ & $\frac32$ & $\phi_{1,5}\tilde\phi_{1,1}$ & $\frac32$ \\
$T_1{-}T_2$ & $2$ & $(+,-)$ & $0$ & $3$ & $2$ & $G_{-\frac12}\phi_{1,5}\tilde\phi_{1,1}$ & $2$ \\
\midrule
$\psi_X$ & $\frac12$ & $(+,+)$ & $\frac1{14}$ & $1.091$ & $0.717$ & $G_{-\frac12}(\phi^+_{5,5}\tilde\phi^+_{5,5}{+}\phi^-_{5,5}\tilde\phi^-_{5,5})$ & $\frac7{10}=0.7$ \\
$K$ & $2$ & $(+,+)$ & $\frac17$ & $3.182$ & $2.435$ & $G_{-\frac12}G_{-\frac32}\phi_{3,5}\tilde\phi_{3,5}$ & $\frac{12}5=2.4$ \\
\bottomrule
\end{tabular}%
}
\end{table}

\subsection{Continuation to $d=3$}
\label{sub:threed}

Continued to $\eps=1$, the fixed point of Section~\ref{sub:fp} is a
candidate interacting three-dimensional $\mathcal N=1$ superconformal theory
of two superfields.\footnote{Its single-superfield counterpart, with ${\cal W}\propto X^3$, is the super-Ising model, the minimal three-dimensional $\mathcal N=1$ theory \cite{Fei:2016sgs,Bashkirov:2013vya}.} Because the superpotential \eqref{eq:W} is $\Zex$ symmetric at every ratio of its two couplings, the $d=3$ theory has the $\Zex$ symmetry of Section~\ref{sec:model}. In Section~\ref{sub:web}, we tested the continuation at $\eps=2$ against the $\ed$ superconformal minimal model. At $\eps=1$, there is no such test, so the numbers below are predictions. We summarize them in Table~\ref{tab:d3}. Where a row also appears in Table~\ref{tab:web}, the two values differ. This is because the resummation here imposes the exact dimension at $\eps=2$ as a second boundary condition, while that of Table~\ref{tab:web} uses only the expansion around four dimensions.

Inside the six-coupling space of
Section~\ref{sub:fp}, two of the six resummed eigenvalues
\eqref{eq:spectrum2}--\eqref{eq:spectrum2b} are still positive at $\eps=1$.
One is the supersymmetric direction \eqref{eq:yrel}, at $y=0.21$. The
other, at $y=0.47$, is the first of \eqref{eq:spectrum2b} and is not
supersymmetric.\footnote{The numbers in
\eqref{eq:spectrum2}--\eqref{eq:spectrum2b} are $y/\eps$, while this subsection gives $y$ itself.
Because most of these eigenvalues belong to no operator with a tabulated dimension, we resum the series for $y$ rather than the series for a $\Delta$. The series for $y$ and for $\Delta$ are related by $y=4-\eps-\Delta$. Since the Pad\'e is not a linear operation on the series, the two resummations differ. Where both exist, they agree on the sign. For the supersymmetric direction, resumming $y$ gives $0.21$, while the $\Delta_{XP}=1.857$ of Table~\ref{tab:d3} gives $y=d-\Delta_{XP}-1=0.14$, since the deformation is the $F$-component of the $XP$ multiplet, one unit above $XP$.}
One of the four negative eigenvalues, the last of \eqref{eq:spectrum2b},
has a sign that is not determined at two loops, because its two terms are of comparable
size and opposite sign. Its two-loop truncation crosses zero at $\eps\simeq0.96$ and gives $y=+0.08$ at $\eps=1$, while the one-sided Pad\'e never changes sign and gives $y=-0.90$ instead. A third relevant direction is therefore not excluded at two loops, and a higher-loop computation would be necessary to determine whether it exists.

To estimate the dimensions of $X$ and $Y$ in $d=3$, we use their $\eps$-expansions together with the boundary conditions $\Delta_X=1/7$ and $\Delta_Y=2/7$ at $\eps=2$. The resulting two-sided $[2,1]$ Pad\'e gives the estimates
\begin{equation}
\Delta_X\approx0.567\,,\qquad \Delta_Y\approx 0.612\,,
\label{eq:d3}
\end{equation}
which are about eight percent apart. The one-sided $[1,1]$ Pad\'e of Section~\ref{sub:fp} gives $\Delta_X\approx0.571$ and
$\Delta_Y\approx0.609$, the values listed in Table~\ref{tab:web}. The boundary condition therefore shifts the estimates by less than one percent.

We can calibrate the two-sided $[2,1]$ Pad\'e on the super-Ising model with ${\cal W}\propto X^3$, where it gives the prediction $\Delta_X=10/17\approx0.588$ \cite{Fei:2016sgs}, close to the bootstrap value $\Delta_X=0.5844435(83)$ obtained later in \cite{Atanasov:2022bpi}. In this calibration, the error of the extrapolation is less than one percent. The separation in \eqref{eq:d3} is an order of magnitude larger than this error.

In $d=3$, the composite field $P$ of \eqref{eq:P} has $\Delta_P\approx 1.097$ in
Table~\ref{tab:d3}, far above $\Delta_Y$. In the $d=2$ $\ed$ superconformal minimal model, by contrast, $\Delta_P= \Delta_Y=\frac27$, because the automorphism $A$ of Section~\ref{sub:rparity} exchanges the pair $\Phi_\pm$, whose two Hermitian combinations are $P$ and $Y$. The automorphism requires a conserved spin-3 current, which the
Maldacena--Zhiboedov theorem \cite{Maldacena:2011jn,Alba:2015upa} forbids in an
interacting unitary theory in three dimensions. At $\eps=1$, therefore, neither the automorphism nor the degeneracy of $P$ and $Y$ is present.
In agreement with the theorem, the twist-two spin-3
operators $I'_3$ and $I_3$ both lie above the unitarity bound $d+1=4$ that a
conserved current would saturate. Their approximate dimensions at $\eps=1$, from the one-loop
truncation, are $4.16$ and $4.13$. Classically, the spin-3 operators of higher twist, including the twist-three multiplet that we match to $W$ in Section~\ref{sub:origin}, lie at least $1-\frac\eps2$ above the twist-two ones.

\begin{table}[!htbp]
\caption{\label{tab:d3}Estimated dimensions of selected low-lying operators in the candidate $d=3$ $\mathcal N=1$ theory of two superfields. All four operators are scalars. The column $(R,C)$ is as in Table~\ref{tab:map}. The estimates are two-loop series resummed by the two-sided $[2,1]$ Pad\'e of \eqref{eq:d3}.}
\centering
\begin{tabular}{lll}
\toprule
operator & $(R,C)$ & $\Delta_{d=3}$ \\
\midrule
$\phi_X$ & $(-,+)$ & $0.567$ \\
$\phi_Y$ & $(-,-)$ & $0.612$ \\
$P$ & $(+,+)$ & $1.097$ \\
$XP$ & $(-,+)$ & $1.857$ \\
\bottomrule
\end{tabular}
\end{table}

A conformal bootstrap of the two-superfield theory could test these predictions. In the single-superfield theory with ${\cal W}\propto X^3$, $\Delta_X$ is bounded below by $0.565$ \cite{Bashkirov:2013vya}, a bound obtained by taking the Ising bound on the first scalar in $X\times X$ as a function of $\Delta_X$ and intersecting it with the supersymmetric line $\Delta_{X^2}=\Delta_X+1$. The bound assumes that the first scalar in $X\times X$ is the equation-of-motion operator.

For two superfields, the assumption fails, because the first scalar in $X\times X$ is $P$ and not $\partial_X {\cal W}$. The two are separated by
\begin{equation}\label{eq:sep}
    \Delta_{\partial_X {\cal W}}-\Delta_P=\frac{52}{109}\eps+\frac{2494}{109^3}\eps^2+\mathcal O(\eps^3)\,,
\end{equation}
which is positive at both orders. At $\eps=2$, the exact dimensions are
$\Delta_P=\frac27$ and $\Delta_{\partial_X {\cal W}}=\frac87$, and no bound on
$\Delta_X$ follows. Because $P$ and $\partial_X {\cal W}$ have the same charges $(+,+)$, the discrete symmetries do not distinguish them. No symmetry assumption can promote $\partial_X {\cal W}$ to the first scalar of $X\times X$.

A bootstrap of the
two-superfield theory would therefore have to be set up anew. The closest
existing study is the GNY archipelago
\cite{Erramilli:2022kgp}, which isolates the $O(N)$ GNY fixed points by bootstrapping a mixed system of bosonic and fermionic four-point functions. It does not assume the supersymmetry expected to emerge at $N=1$. The isolated region lies on the supersymmetric line, and the lowest spin-$\frac32$ operator there is close to the unitarity bound $\Delta=\frac52$, which a conserved supercurrent saturates. The four-point function of $O(N)$ fermions was bootstrapped earlier in \cite{Iliesiu:2017nrv}, and the same mixed system was later used in \cite{Mitchell:2024hix} to bound the irrelevant operators of those fixed points.

Our theory is a two-superfield
GNY fixed point with the $\Zex$ symmetry. The
analogous mixed system would have to impose $\Zex$ as well. It
could then test the emergent supersymmetry in the same way and resolve the
eight-percent separation of $\Delta_X$ from $\Delta_Y$ in \eqref{eq:d3}.
Supersymmetry may instead be assumed from the start, as in the
single-superfield super-Ising studies
\cite{Bashkirov:2013vya,Atanasov:2018kqw,Rong:2018okz,Atanasov:2022bpi}.

\section{Discussion}
\label{sec:discussion}

\subsection{Origin and fate of the $W$ current}
\label{sub:origin}

The $\ed$ superconformal minimal model has the conserved spin-3 current $W$, but we
showed in Section~\ref{sub:spin3} that no spin-3 current is conserved near four
dimensions, so $W$ must emerge at $d=2$. To find its origin, we look for the lightest
spin-3 operator near four dimensions with the charges $(R,C)=(-,-)$ of $W$. A scalar is
$\ZR$-odd and a leading-twist fermion bilinear is $\ZR$-even, so a leading-twist
operator of $n$ fields has R-parity $(-1)^n$. The twist-two operators of
\eqref{eq:gamma3} are thus $\ZR$-even and are excluded, as we saw in
Section~\ref{sub:web}. Since the R-parity of a fermion bilinear counts its gamma matrices, a two-fermion operator with the R-parity of $W$ has no gamma matrix, and the fermion
equation of motion reduces it, up to total derivatives, to operators of three fields.
The lightest spin-3 primaries with the charges of $W$ therefore contain three fields, with an
odd number of $Y$ fields as the $\Zex$ charge requires. With $\partial$ denoting the contraction $\zeta\!\cdot\!\partial$ with the
null vector $\zeta$ of Appendix~\ref{app:currents}, there are eight of them,
\begin{equation}
\begin{aligned}
&\partial^{a}\phi_X\,\partial^{b}\phi_X\,\partial^{c}\phi_Y\ (2)\,,\;
\partial^{a}\phi_Y\,\partial^{b}\phi_Y\,\partial^{c}\phi_Y\ (1)\,, &a+b+c&=3\,,\\
&\partial^{a}\phi_X\,\partial^{b}\bar\Psi_X\slashed\zeta\,\partial^{c}\Psi_Y\ (3)\,,\;
\partial^{a}\phi_Y\,\partial^{b}\bar\Psi_X\slashed\zeta\,\partial^{c}\Psi_X\ (1)\,,\;
\partial^{a}\phi_Y\,\partial^{b}\bar\Psi_Y\slashed\zeta\,\partial^{c}\Psi_Y\ (1)\,, &a+b+c&=2\,.
\end{aligned}
\label{eq:tw3basis}
\end{equation}
The numbers in parentheses count the operators of each type that are not total
derivatives of spin-two operators.\footnote{For the five types in \eqref{eq:tw3basis}, the powers $a$, $b$ and $c$ can be chosen in
$6$, $3$, $6$, $2$ and $2$ ways, respectively, of which $4$, $2$, $3$, $1$ and $1$ are total derivatives
of the spin-two operators with the same fields, hence descendants. The last two counts
are already halved by the Majorana flip of Appendix~\ref{app:currents}, which requires
the powers $b$ and $c$ on the two fermions to differ.} These eight operators have
classical dimension $6-\frac32\eps$, which is above the dimension $5-\eps$ of a conserved current with spin $3$.  
The operators with two additional scalars lie $2-\eps$ higher still. On the two-dimensional side, the lightest spin-3 operator with the charges of $W$ is $W$ itself. Matching the lightest operator on each side identifies $W$ with the lightest of the eight. The identification requires the dimension of this operator to reach the conserved value $3$ exactly at $\eps=2$, an endpoint constraint of the same kind as the coincidence of $\Delta_Y$ and $\Delta_P$ in Section~\ref{sub:deg}.\footnote{The eight-by-eight mixing matrix for \eqref{eq:tw3basis} requires a Feynman diagram computation, because the recombination method does not apply to three-field operators. We do not attempt the full computation here.
}

Conservation that appears only at $d=2$, however, poses a recombination puzzle. At
$d>2$, the operator identified with $W$ lies in a long multiplet. By \eqref{eq:recomb} of
Appendix~\ref{app:currents}, the squared norm of its divergence is proportional to the
excess of its dimension over the conserved value, and the identification sends this
excess to zero at $d=2$. The divergence itself therefore vanishes at $d=2$. Its
unit-normalized form, however, exists at every $d>2$, and we assume that it has a limit
as $d\to2$. The limit is then not a descendant of $W$ but an independent operator of
the $d=2$ theory, the partner that the recombination requires. The divergence has spin
two and dimension $\Delta_W+1$, which tends to four. In two-dimensional terms, the
limit lies at $(h,\tilde h)=(3,1)$.

If the $d\to2$ limit of the $\eps$-expansion is the $\ed$ superconformal minimal model, the
partner must be one of its operators at $(3,1)$. No primary or descendant of a
non-vacuum module lies at $(3,1)$, as Table~\ref{tab:modules} shows. In the vacuum
module, the only states at $(3,1)$ would be the $\bar\partial$ derivatives of the
holomorphic currents, such as $\bar\partial W$, and these are precisely what
conservation sets to zero. The $\ed$ superconformal minimal model thus has no operator
at $(3,1)$. The same puzzle appears in the bosonic theory with the same spin-3
structure, the three-state Potts model $M(5,6)$.

Three possibilities could resolve the puzzle. The first is that
conservation is restored above $d=2$. A related recombination was studied
in \cite{DeCesare:2025ukl,DeCesare:2026dwm} for the $O(N)$ nonlinear sigma model in $d=2+\eps$. The sigma model has a protected
operator of dimension $N-1$ that the Wilson--Fisher fixed point lacks.
Lifting it by multiplet recombination in $d>2$ requires a partner of dimension $N$. Since at leading order the available partners lie instead at $N+4$, the recombination cannot proceed at small $\eps$. Whether the dimension of one of them
decreases to $N$ at some finite $\eps$ is studied at one loop in
\cite{DeCesare:2026dwm} for $N=3$ and $N=4$, with a negative conclusion,
because the dimensions of the candidates grow with $\eps$.

The same question arises here. In the first scenario, conservation would be restored at some $d_\star$
and persist from there down to $d=2$. An interacting theory would then have an exactly conserved spin-3 current over a whole range of dimensions.
The Maldacena--Zhiboedov theorem states that a unitary conformal theory
with a unique stress tensor and a conserved current of spin above two has
free correlators \cite{Maldacena:2011jn,Alba:2015upa}. Because the Maldacena--Zhiboedov theorem is proved in $d=3$ and in integer $d>3$, it forbids $d_\star\ge3$. At non-integer $d$, its unitarity assumption fails \cite{Hogervorst:2015akt}, and at $d=2$ its conclusion fails, since the unitary $W$-algebra minimal models are interacting but have conserved higher-spin currents \cite{Bouwknegt:1992wg}. The $\ed$ superconformal minimal model is one
of them. This scenario survives only inside the
window $2<d<3$, where the Maldacena--Zhiboedov theorem is unproved.

The second possibility is that the $\ed$ superconformal minimal model is only a subsector of the $d\to2$
limit of the $\eps$-expansion. The analogous puzzle for the continuation of the
Wilson--Fisher fixed point to the two-dimensional Ising model is treated in \cite{Zan:2026oyb}.
There the spin-4 current becomes conserved at $d=2$, as the Virasoro
symmetry of the Ising model requires. Its divergence would then have to
become an independent spin-3 operator of dimension five, which in
two-dimensional notation lies at $(h,\tilde h)=(4,1)$. The Ising model
contains no such operator, exactly as $\ed$ contains none at $(3,1)$. The
two obstructions have the same origin. The divergence of a holomorphic
spin-$s$ current always lies at $\tilde h=1$, a value that neither model realizes. The resolution proposed in \cite{Zan:2026oyb} takes the Ising model to be a subsector of
a larger $d\to2$ limit. There the operators beyond the Ising model still contribute to the correlators of Ising operators, but their contributions cancel against those of operators in $SO(d)$ representations whose multiplicities turn negative at integer $d$, so the Ising correlators are unchanged. The same mechanism could apply here.

Under the subsector scenario, two statements about the $\ed$ superconformal minimal model must be distinguished. One is
that the LG theory at $d=2$ is $\ed$. The other is that the
$d\to2$ limit of the $\eps$-expansion is $\ed$. The scenario excludes the
second by construction, since the limit then contains operators beyond $\ed$. It does not affect the first, because some of the operators beyond $\ed$ lie in representations with negative multiplicities at $d=2$, which a unitary two-dimensional theory does not allow. At non-integer $d$, the continued theory is not unitary in any case \cite{Hogervorst:2015akt}. The
comparisons of Section~\ref{sub:fp} and Section~\ref{sub:web} concern the dimensions of operators of $\ed$ itself. The mechanism of \cite{Zan:2026oyb} does not change those dimensions, since the contributions of the operators beyond $\ed$ cancel.

The third possibility is that a further sector provides the missing
partner. In \cite{Zhou:2022pah}, the $SU(N)_1$
Wess--Zumino--Witten model is continued to $d=2+\eps$ through its dual gauge
description, where the divergence of the broken current is proportional
to the gauge field strength. The field strength decouples at $d=2$ but not at $d=2+\eps$, where the gauge sector provides the partner and the recombination proceeds. The $d\to2$
theory is then the Wess--Zumino--Witten model itself, with no operators beyond it to
remove. Here, however, we know of no free or gauge sector that provides a partner at $\tilde h=1$.

It would be interesting to see whether any of the three scenarios resolves the puzzle. A first step would be to compute the anomalous dimension of the three-field operator identified with $W$ above. This computation requires the eight-by-eight mixing matrix of the operators \eqref{eq:tw3basis}. The continuation of the resulting anomalous dimension would indicate whether this operator can become conserved above $d=2$, as the first scenario requires.

\subsection{Further directions}
\label{sub:further}

This work can be extended in several directions.
First, the mixing matrices at half-integer spin remain to be computed. In Section~\ref{sub:spin3}, we obtained the $\Zex$-odd eigenoperators at spins $\frac32$ and $\frac52$ from supersymmetry alone, through the action of the supercharge between adjacent spins. We obtained the two $\Zex$-even spin-$\frac52$ eigenvalues in the same way from \eqref{eq:g2even}. A direct computation of the three mixing matrices would check these results.

Second, it would be interesting to construct a microscopic lattice realization of the $\ed$ superconformal minimal model. In \cite{Lahtinen_2014}, two coupled tricritical Blume--Capel chains have a critical point described by $(D_6,E_6)$ at $m=10$. A lattice realization of $\ed$ would allow finite-size tests of the operator spectrum and of the extended currents, and a lattice study of the RG flows of Section~\ref{sub:flow}. Such a realization would provide independent nonperturbative evidence for the LG description proposed here.

Third, the LG description can also be tested at finite temperature, on $S^1\times\mathbb R^{d-1}$. The thermal free energy, masses, and one-point functions of the fixed point are computable in the $\eps$-expansion and can be continued by Pad\'e for comparison with the exact two-dimensional data. The comparison was recently carried out for $M(2,5)$ and the $D$-series model $M(3,8)$ \cite{Diatlyk:2026ptt}, and it would be interesting to extend it to the two-superfield GNY fixed points.

Fourth, the fuzzy sphere method \cite{Zhu:2022gjc} applies directly in $d=3$ and gives the operator spectrum through the state--operator correspondence. It has been applied to the super-Ising model \cite{Tang:2025wtj} and other fermionic CFTs \cite{Zhou:2025kng}. While the fixed point $g_1=g_2$ of Section~\ref{sub:flow} consists of two decoupled copies of the super-Ising model, a realization of the coupled fixed point on the fuzzy sphere would test the predictions of Table~\ref{tab:d3}. The fuzzy sphere method can be used to calculate the sphere free energy $F$ \cite{Hu:2024pen}, which is also computable in the $\eps$-expansion \cite{Giombi:2014xxa,Fei:2015oha,Fei:2016sgs}, so the two values of $F$ for the coupled fixed point can be compared. The $F$-theorem \cite{Jafferis:2011zi,Casini:2012ei} requires $F$ to decrease along the RG flows of Section~\ref{sub:flow} continued to $\eps=1$, and the values of $F$ at their endpoints provide a test of these RG flows.

Fifth, the super-$W_3$ algebra is associative at another value of the central charge, $c=-\frac52$, where its representations are not unitary \cite{Inami:1988xy}, and one may look for an LG description there. No ordinary minimal model of the superconformal or super-$W_3$ algebra lies at $c=-\frac52$ \cite{Nam:1990bt}. In the $\mathcal N=1$ superconformal minimal series, the logarithmic member with $m=1$ has $c=-\frac52$,\footnote{In the notation of \cite{Pearce:2013bea}, it is called $LSM(2,3)\equiv LM(1,3;2)$.} and it describes superconformal dense polymers \cite{Pearce:2013bea}, as the Virasoro analog $LM(1,2)$ with $c=-2$ describes critical dense polymers \cite{Pearce:2006sz}. An LG description is difficult to construct even for the non-unitary $\mathcal N=1$ theories that are not logarithmic \cite{Nakayama:2021zcr}. The RG flows of non-unitary theories are also less well understood. Recently, however, infinitely many RG flows between non-unitary Virasoro minimal models were proposed in \cite{Tanaka:2024igj} using non-invertible symmetries, and the corresponding RG flows in the non-unitary $\mathcal N=1$ superconformal minimal models were studied in \cite{Gaberdiel:2026sfg}. An RG flow of this kind that ends at $c=-\frac52$ could provide an LG description.

\section*{Acknowledgments}

We thank M. Gaberdiel, K. Intriligator, A. M. Polyakov, S. Pufu, and G. Tarnopolsky for useful discussions.
YN would like to thank PCTS for its hospitality during his
visit to Princeton in February 2025, when this project was initiated.
YN is supported in part by JSPS KAKENHI Grant Numbers 21K03581 and
26K00699. 
IRK and AK are supported in part by the National Science Foundation grant PHY-2609862, Simons Foundation grant 917464 and Simons Foundation International grant SFI-MPS-QCD-00014867-01 (Simons Collaboration on Confinement and QCD Strings). IRK's work was performed in part at the Aspen Center for Physics, which is supported by the National Science Foundation grant PHY-2210452.
ZS is supported by the U.S. Department of Energy grant DE-SC0009988 and the
Sivian Fund.

Large language models, Anthropic Fable 5/5.1 and OpenAI GPT 5.6 Sol/6 Astra, were used for editing, for writing numerical code, and for independent checks of equations. The ancillary files are available upon request.

\appendix

\section{Free twist-two currents and their recombination}\label{app:currents}

In this appendix, we compute the mixing matrices of Section~\ref{sub:spin3}. A current that is conserved in the free theory can acquire an anomalous dimension only by recombining with the operator that appears in its divergence. To leading order, the mixing matrix $\hat\gamma_{ij}$ is fixed by the norms of the divergences relative to those of the currents \cite{Giombi:2016hkj,Giombi:2017rhm}. Explicitly, if we denote the divergences at spin $s$ by $\mathcal O_i=\partial\!\cdot\!J_s^{(i)}$, it is given, for currents that are orthogonal in the free theory, by
\begin{equation}\label{eq:recomb}
    \hat\gamma_{ij}=\frac{C_{\mathcal O_i\mathcal O_j}}{2\,\kappa_s\sqrt{C_{J_i}C_{J_j}}}\,,
\end{equation}
where $C_{\mathcal O_i\mathcal O_j}$ and $C_{J_i}$ are the coefficients of
the two-point functions of the divergences and of the currents, and
$\kappa_s$ is the kinematic constant we discuss in Section~\ref{sub:norms}.

\subsection{The currents and their divergences}\label{sub:currents}

A twist-two spin-$s$ current is a sum of bilinears in two fields, with the derivatives acting on either field \cite{Craigie:1983fb}. For two scalars,
\begin{equation}\label{eq:Jdef}
    J^\phi_s=\sum_{k=0}^{s}a_k\partial_{\mu_1}\!\!\cdots\partial_{\mu_k}\phi_1\partial_{\mu_{k+1}}\!\!\cdots\partial_{\mu_s}\phi_2-\text{traces}\,,
\end{equation}
symmetrized in $\mu_1\ldots\mu_s$, and for two fermions
\begin{equation}\label{eq:Jpsidef}
    J^\Psi_s=\sum_{k=0}^{s-1}c_k\partial_{\mu_2}\!\!\cdots\partial_{\mu_{k+1}}\bar\Psi_1\;\gamma_{\mu_1}\partial_{\mu_{k+2}}\!\!\cdots\partial_{\mu_s}\Psi_2-\text{traces}\,,
\end{equation}
where $\gamma_{\mu_1}$ provides one of the $s$ indices and the derivatives the other $s-1$. The problem is to find the $a_k$ and the $c_k$ for which these are conserved.

Contracting every index with a single null vector $\zeta$ turns conservation into a condition on a polynomial \cite{Dobrev:1977qv,Costa:2011mg}. The subtracted traces contain $\delta_{\mu_i\mu_j}$, so they vanish when contracted with $\zeta^{\mu_i}\zeta^{\mu_j}$, and we omit them. In momentum space, a derivative on the first field contributes $A=\zeta\!\cdot\!k_1$ and one on the second contributes $B=\zeta\!\cdot\!k_2$, where $k_1$ and $k_2$ are the momenta of the two fields. The two currents then reduce to polynomials in $A$ and $B$,
\begin{equation}\label{eq:polys}
    J(A,B)=\sum_{k=0}^{s}a_kA^kB^{s-k}\,,\qquad\mathcal P(A,B)=\sum_{k=0}^{s-1}c_kA^kB^{s-1-k}\,,
\end{equation}
the fermionic one multiplying $\bar u\slashed\zeta u$, where $\bar u$ and $u$ are the plane-wave spinors.

To take the divergence, we have to undo the contraction on one index. This requires more than $\partial/\partial\zeta^\mu$,
which restores an index but not the trace subtraction. The Todorov operator does both \cite{Bargmann:1977gy,Dobrev:1977qv},
\begin{equation}\label{eq:todorovop}
    D_\mu=(\delta+\zeta\!\cdot\!\partial_\zeta)\,\partial_\zeta^\mu-\frac12\,\zeta_\mu\,\partial_\zeta^2\,,\qquad\delta=\frac d2-1\,.
\end{equation}
Both $J^\phi_s$ and $J^\Psi_s$ are bilinear in the fields, so in momentum space $\partial_\mu$ acts on them as $k_1+k_2$. To discuss their divergences, we therefore work with $(k_1+k_2)^\mu D_\mu J$ throughout and use the notation $\partial\!\cdot\!J$ for it. On a polynomial of degree $s$, the operator \eqref{eq:todorovop} restores one index with a factor $s(\delta+s-1)$, so $\partial\!\cdot\!J$ is $s(\delta+s-1)$ times the divergence $\partial^\mu J_{\mu\mu_2\ldots\mu_s}$. On shell, where $k_1^2=k_2^2=0$, $\partial\!\cdot\!J$ reduces to $(k_1\!\cdot\!k_2)$ times
\begin{equation}\label{eq:conscond}
    \mathcal C[F]=(\delta+s-1)\,(\partial_1+\partial_2)F-(A+B)\,\partial_1\partial_2F\,,\qquad\partial_1=\partial_A\,,\quad\partial_2=\partial_B\,,
\end{equation}
so the scalar current is conserved when $\mathcal C[J]=0$ and the fermionic one
when $\mathcal C[\mathcal P]=0$. Since the solution of $\mathcal C[J]=0$ obeys
$a_{s-k}=(-1)^sa_k$, at odd $s$ the scalar current is antisymmetric
under exchange of the two fields and only the flavor-antisymmetric
combination is nonzero. For example, at $s=3$ and $d=4$ the two conserved polynomials are
$J=\frac1{36}A^3-\frac14A^2B+\frac14AB^2-\frac1{36}B^3$ and
$\mathcal P=A^2-3AB+B^2$.

The divergence of each current follows from the equations of motion. Because the scalar equation of motion is second order, we first isolate the terms of $\partial\!\cdot\!J$ that contain the Laplacian. Let us then introduce $\sigma_n=k_n^2$, which is minus the Laplacian acting on the $n$-th field. Although it vanishes on shell in the free theory, the interaction turns it into a source. Keeping $\sigma_n$ in $(k_1+k_2)^\mu D_\mu J$, with $D_\mu$ of \eqref{eq:todorovop}, adds to $(k_1\!\cdot\!k_2)\,\mathcal C[J]$ the source terms
\begin{equation}\label{eq:divoff}
    \partial\!\cdot\!J\big|_{\rm source}=\sum_{n=1,2}\sigma_n\,\mathcal K_nJ\,,\qquad\mathcal K_n=(\delta+s-1)\,\partial_n-\frac12(A+B)\,\partial_n^2\,.
\end{equation}

To the order needed here, the scalar equation of motion of \eqref{eq:LGNY} is $\Box\phi_a=\frac12S_a$, with $S_a=h_{abc}\,\bar\Psi_b\Psi_c$. Substituting it into \eqref{eq:divoff} at $s=3$ and $d=4$ gives 
\begin{equation}\label{eq:divJ3} 
    \partial\!\cdot\!J_3^\phi= \frac12\left(-\frac16A^2+\frac43AB-B^2\right)\,S_X\,\phi_Y \;-\;(X\leftrightarrow Y)\,,
\end{equation}
in which each factor of $A$ is a derivative on $S_X$ and each factor of $B$ a
derivative on $\phi_Y$. Since the divergence of a spin-3 current has spin two, the polynomial has degree two. We denote by $\mathcal V$ an operator built from one
scalar field and one fermion bilinear, multiplied by a Yukawa coupling of \eqref{eq:LGNY}.
Both terms of \eqref{eq:divJ3} are of the form $\mathcal V$.

Keeping the off-shell terms in the divergence of the fermion current gives
\begin{equation}
    \partial\cdot\!J^\Psi=\bar u\,\slashed k_1[(\mathcal K_1\mathcal P)\,\slashed k_1\slashed\zeta+\alpha_0]u+\bar u[(\mathcal K_2\mathcal P)\,\slashed\zeta\slashed k_2+\beta_0]\slashed k_2\,u+(k_1\!\cdot\!k_2)\,\mathcal C[\mathcal P]\,\bar u\slashed\zeta u\,,
\label{eq:divpsi}
\end{equation}
with $\mathcal K_n$ of \eqref{eq:divoff}, $\mathcal C$ of \eqref{eq:conscond},
$\alpha_0=(\delta+s-1)\mathcal P-(A+B)\,\partial_1\mathcal P$ and
$\beta_0=(\delta+s-1)\mathcal P-(A+B)\,\partial_2\mathcal P$. The conserved $\mathcal P$ has $\mathcal C[\mathcal P]=0$, so the last term vanishes. Because the fermion equation of motion is first order, we now isolate the terms that contain the free Dirac operator rather than the Laplacian. In the first term of \eqref{eq:divpsi}, the Dirac operator is $\bar u\,\slashed k_1$, acting on $\bar\Psi_1$. In the second term, it is $\slashed k_2\,u$, acting on $\Psi_2$. The fermion equation of motion of \eqref{eq:LGNY} is $\slashed\partial\Psi_a=-h_{abc}\,\phi_b\Psi_c$. Its Majorana conjugate follows from $\bar\Psi=\Psi^TC$ with $C\gamma_\mu C^{-1}=-\gamma_\mu^T$ and has the opposite sign,
$\partial_\mu\bar\Psi_a\gamma^\mu=+h_{abc}\,\phi_b\bar\Psi_c$. Substituting the two equations of motion for $\bar u\,\slashed k_1$ and $\slashed k_2\,u$ turns both terms into operators of the form $\mathcal V$.

Both divergences are of the form $\mathcal V$, so the two currents mix. Every $\langle\mathcal O_i\mathcal O_j\rangle$ is then computed from the
Wick contractions with one scalar and two fermion propagators, on which the
derivatives of the two divergences act. We evaluate these contractions with computer
algebra. The code, together with a script for every rational
coefficient quoted in this paper, is available upon request.

\subsection{Norms and the mixing matrices}
\label{sub:norms}

To compute the mixing matrices, we need the two-point functions of the free theory, which we obtain by Wick contraction in four Euclidean dimensions, with Hermitian gamma matrices satisfying $\{\gamma_\mu,\gamma_\nu\}=2\delta_{\mu\nu}$ and with the Majorana condition $\bar\Psi=\Psi^TC$. Under this condition, a pair of bilinears $\bar\Psi_b\Psi_c$ contracts in two ways. Every closed fermion loop contributes the factor $\operatorname{tr}\mathbf1=2$ of the two-component spinors, as in footnote~\ref{fn:spinors}. We express the free propagators at $d=4$ as 
\begin{equation}\label{eq:props}
    \bigl\langle\phi(x)\phi(0)\bigr\rangle=\frac{c_\phi}{x^2}\,,\qquad\bigl\langle\Psi(x)\bar\Psi(0)\bigr\rangle=\frac{c_\Psi\,\slashed x}{(x^2)^2}\,.
\end{equation} 
The kinetic terms of \eqref{eq:LGNY} fix the two constants $c_\phi$ and $c_\Psi$. Since the scalar propagator is the Green function of $-\Box$, $c_\phi=\frac1{4\pi^2}$. Since the fermion propagator obeys $\slashed\partial\,\langle\Psi\bar\Psi\rangle=\delta^4(x)$, $c_\Psi=\frac1{2\pi^2}$.

The conservation condition is linear, so it fixes the $a_k$ of \eqref{eq:Jdef} and the $c_k$ of
\eqref{eq:Jpsidef} only up to an overall constant. The coefficient $C_J$ of \eqref{eq:JJ} below is
quadratic in them. Since rescaling a current multiplies $C_{\mathcal O_i\mathcal O_j}$
and $\sqrt{C_{J_i}C_{J_j}}$ by the same factor, the ratio \eqref{eq:recomb} does not
depend on the overall constant. To make the computation explicit, however, we quote in
\eqref{eq:CJ} below the values for the $s=3$ polynomials as normalized
in Section~\ref{sub:currents}.

Let us evaluate the norms in a particular kinematic configuration. Every two-point function below has two polarizations, $\zeta$ at $x$ and $\eta$ at the origin. Both are null and orthogonal to $x$. A conformal two-point function of spin-$s$ primaries has the tensor structure $(\zeta\!\cdot\!I(x)\!\cdot\!\eta)^{s}$, where $I_{\mu\nu}(x)=\delta_{\mu\nu}-2x_\mu x_\nu/x^2$. Since $\zeta$ and $\eta$ are orthogonal to $x$, the tensor structure reduces to $(\zeta\!\cdot\!\eta)^{s}$, and a two-point function of spin-$s$ twist-two currents
must take the form
\begin{equation}\label{eq:JJ}
    \bigl\langle J(x,\zeta)\,J(0,\eta)\bigr\rangle=\frac{C_J(\zeta\cdot\eta)^{s}}{(x^2)^{d-2+s}}\,.
\end{equation}
Wick contracting each current with itself and extracting the coefficient
of $(\zeta\!\cdot\!\eta)^{s}$ gives, at $s=3$ and $d=4$,
\begin{equation}\label{eq:CJ}
    C_{J_3^\phi}=\frac{40}9c_\phi^2\,,\qquad C_{J_3^\Psi}=480c_\Psi^2\,.
\end{equation}

The numerator of \eqref{eq:recomb} requires the two-point functions of the divergences. A divergence has spin $s-1$ and dimension $d-1+s$, so for the same
$\zeta$ and $\eta$ its two-point function is
$C_{\mathcal O_i\mathcal O_j}\,(\zeta\!\cdot\!\eta)^{s-1}/(x^2)^{d-1+s}$,
which defines the coefficients $C_{\mathcal O_i\mathcal O_j}$.
One current is built from scalars and the other from fermions, so
$\langle J_3^\phi J_3^\Psi\rangle=0$ and \eqref{eq:recomb} applies. The two-point
function of the divergences, in contrast, has an off-diagonal component, because both
divergences are of the form $\mathcal V$.

We have yet to determine one constant, the $\kappa_s$ of \eqref{eq:recomb}. For $\partial\!\cdot\!J$ normalized as in \eqref{eq:todorovop}, its value is
$s(s+\frac d2-2)(s+\frac d2-1)(s+d-3)$ \cite{Giombi:2017rhm}, which equals $144$ at $s=3$ and $d=4$.

The simplest case is the anomalous dimension of $\phi$, which follows from the
two-point function of $\Box\phi$. Differentiating \eqref{eq:props} with the
dimension shifted to $1+\gamma_\phi$ and keeping the leading term in $\gamma$
gives
\begin{equation}\label{eq:eomnorm}
    \bigl\langle\Box\phi\,\Box\phi\bigr\rangle=\frac{32\,\gamma_\phi\,c_\phi}{(x^2)^3}\,.
\end{equation}
For each flavor $a$, the two-point function of the fermion bilinear $S_a$ is $C_{S_a}/(x^2)^3$, which defines $C_{S_a}$. Substituting $\Box\phi_a=\frac12S_a$ into \eqref{eq:eomnorm} then gives
\begin{equation}\label{eq:gammafromS}
    32\gamma_ac_\phi=\frac14C_{S_a}\,. 
\end{equation}
Wick contracting the two fermions in $S_a=h_{abc}\bar\Psi_b\Psi_c$, with the two contractions and one trace, gives $C_{S_X}=4(g_1^2+g_2^2)\,c_\Psi^2$ and $C_{S_Y}=8g_1^2c_\Psi^2$, where the flavor factors follow from $h_{XXX}=g_2$ and $h_{XYY}=h_{YXY}=h_{YYX}=g_1$. With the canonical propagator constants, the relation \eqref{eq:gammafromS} becomes
\begin{equation}
    \gamma_X=\frac{c_\Psi^2}{32\,c_\phi}(g_1^2+g_2^2)=\frac{g_1^2+g_2^2}{32\pi^2}\,,\qquad \gamma_Y=\frac{c_\Psi^2}{16\,c_\phi}\,g_1^2=\frac{g_1^2}{16\pi^2}\,,
\end{equation}
which are the $\gamma_X$ and $\gamma_Y$ of \eqref{eq:beta1LG} and, at the fixed point, the one-loop terms of \eqref{eq:gammas}.

Let us now compute the spin-3 anomalous dimensions. Contracting the divergence \eqref{eq:divJ3} with itself, again with the two contractions and one trace, and dividing by $C_{J_3^\phi}$ and by $2\kappa_3=288$ as in \eqref{eq:recomb} gives the scalar component
\begin{equation}\label{eq:g3phiphi}
    \gamma_3^{\phi\phi}=\frac{c_\Psi^2}{32\,c_\phi}(3g_1^2+g_2^2)=\frac{31}{218}\eps\,,
\end{equation}
where in the last step we insert the fixed-point couplings of \eqref{eq:beta1LG}, $g_{2,\star}^2=\frac{64\pi^2}{109}\eps$ and $g_{1,\star}=\frac32g_{2,\star}$. Contracting the fermion divergence at $s=3$ with itself and with \eqref{eq:divJ3}, then dividing by the norms \eqref{eq:CJ} and by $2\kappa_3$, gives the other two components,
\begin{equation} \label{eq:g3fermion}
    \gamma_3^{\Psi\Psi}=\frac{c_\phi}{24}(10g_1^2-g_1g_2+3g_2^2)=\frac{16}{109}\eps\,,\quad\gamma_3^{\phi\Psi}=\frac{\sqrt3\,c_\Psi}{24}\,g_1(g_1-g_2)=\frac{\sqrt3}{109}\eps\,,  
\end{equation} 
where we again insert the fixed-point couplings in the last steps.

The construction is not specific to spin three. In the $\Zex$-odd sector at spins one and two, the operators are again one scalar current and one fermion current, and the formula \eqref{eq:recomb} gives the mixing matrices \eqref{eq:gamma12} in the unit-normalized basis $(\hat J^\phi_s,\hat J^\Psi_s)$ built from the currents \eqref{eq:J1def} and their spin-two analogs. The eigenvalues are $\frac{43}{218}\eps$ and $\frac{25}{218}\eps$ at spin one and $\frac{51}{218}\eps$ and $\frac1{218}\eps$ at spin two, with the eigenoperators that we list in Table~\ref{tab:web}. The $\Zex$-even sector at spin two has the four operators $J^{\phi,a}_2$ and $J^{\Psi,a}_2$ with $a\in\{X,Y\}$, and the same steps give the four eigenvalues \eqref{eq:g2even}. The zero among them belongs to the stress tensor.

\subsection{Consistency checks}
\label{app:verif}

The first check of the formalism of this appendix is the conservation of the stress tensor. Since conservation holds for any $g_1$ and $g_2$, the divergence of the stress tensor must vanish identically in the couplings and not only at the fixed point. To construct the stress tensor, we set $\phi_1=\phi_2=\phi_a$ in \eqref{eq:Jdef} and $\Psi_1=\Psi_2=\Psi_a$ in \eqref{eq:Jpsidef} at $s=2$, with the conserved polynomials $J=\frac14A^2-AB+\frac14B^2$ and $\mathcal P=B-A$. The resulting operators are the one-flavor currents $J^{\phi,a}_2$ and $J^{\Psi,a}_2$ of Section~\ref{sub:spin3}. We now consider the ansatz 
\begin{equation}\label{eq:Tcand} 
    T=\sum_{a\in\{X,Y\}}\bigl(J^{\phi,a}_2+r\,J^{\Psi,a}_2\bigr)\,, 
\end{equation}
and determine $r$ from conservation. The divergence of \eqref{eq:Tcand} follows from the source terms \eqref{eq:divoff} and their fermionic analog at $s=2$, and contracting it with itself gives
\begin{equation}\label{eq:Tgate}
    \langle\partial\!\cdot\!T\;\partial\!\cdot\!T\rangle\propto(3g_1^2+g_2^2)(8r+3)^2\,.
\end{equation} 
Only at $r=-\frac38$ does the right-hand side vanish for every value of the couplings. The Noether construction from the Lagrangian determines $r$ independently. Contracted with $\zeta$, the improved scalar part of the Noether stress tensor is $(\zeta\!\cdot\!\partial\phi_a)^2-\frac16(\zeta\!\cdot\!\partial)^2\phi_a^2$, and the fermion part is $\frac14\bigl[\bar\Psi_a\slashed\zeta\,(\zeta\!\cdot\!\partial)\Psi_a-(\zeta\!\cdot\!\partial\bar\Psi_a)\slashed\zeta\Psi_a\bigr]$. The Majorana flip relation $\bar\lambda\slashed\zeta\chi=-\bar\chi\slashed\zeta\lambda$, at $\lambda=(\zeta\!\cdot\!\partial)\Psi_a$ and $\chi=\Psi_a$, makes the two terms equal, so the fermion part is the canonical $\frac12\bar\Psi_a\slashed\zeta\,(\zeta\!\cdot\!\partial)\Psi_a$. In the polynomial notation, the two parts are $AB-\frac16(A+B)^2=-\frac23\,J$ and $\frac14(B-A)=\frac14\,\mathcal P$, so the stress tensor is $-\frac23$ times \eqref{eq:Tcand} with $r=-\frac38$. The two results agree.

The second check is the scalar diagonal component of the mixing matrix. The divergence of the scalar current contains only $S_X\phi_Y$ and $S_Y\phi_X$, and the norms of $S_X$ and $S_Y$ give $\gamma_X$ and $\gamma_Y$ in \eqref{eq:gammafromS}. The diagonal component must therefore be the sum $\gamma_X+\gamma_Y$. At $s=1$, $2$ and $3$, the formula \eqref{eq:recomb} indeed gives
\begin{equation}\label{eq:largespin}         
    \frac{c_\Psi^2}{32\,c_\phi}\,(3g_1^2+g_2^2)=\gamma_X+\gamma_Y\,, 
\end{equation}
identically in $g_1$ and $g_2$, which verifies the kinematic constant $\kappa_s$ together with the polynomials of the divergences.

The third check is the large-spin limit. There the interaction between the two fields of a current is suppressed, so the anomalous dimensions of the two currents approach the sums of those of their fields and the off-diagonal component vanishes. Since $\gamma_{\Psi_a}=\gamma_{\phi_a}$, both sums equal $\gamma_X+\gamma_Y$. We have computed the $\Zex$-odd mixing matrices through spin eight with the same code, and their eigenvalues are
\begin{equation}\label{eq:spinclosed}
\begin{aligned}
    &\Bigl\{31+\frac{12}{s},\ 31-\frac{12}{s+1}\Bigr\}\frac{\eps}{218}\,,&\qquad& s\ \text{odd}\,,\\
    &\Bigl\{31-\frac{60}{s},\ 31+\frac{60}{s+1}\Bigr\}\frac{\eps}{218}\,,&\qquad& s\ \text{even}\,.
\end{aligned}
\end{equation}
The two differ from $\gamma_X+\gamma_Y=\frac{31}{218}\eps$ at order $1/s$.

\section{Two-loop RG functions}
\label{app:twoloop}

The two-loop anomalous dimensions of Section~\ref{sub:fp}, the anomalous-dimension matrix of Section~\ref{sub:deg}, and the stability eigenvalues of Section~\ref{sub:flow} come from the two-loop RG functions of a general Yukawa theory. We record them in the component GNY form, following \cite{Machacek:1983tz,Machacek:1983fi,Machacek:1984zw} as given in \cite{Fei:2016sgs}. The general tensor form to three loops appears in \cite{Jack:2024sjr}, and the expressions below agree with its two-loop scalar and fermion anomalous dimensions.

We denote by $h_i$ the real symmetric Yukawa matrix of the scalar $\phi_i$, with $(h_i)_{jk}=h_{ijk}$, written as $\Gamma_i$ in \cite{Fei:2016sgs}, and by $g_{ijkl}$ the quartic tensor of $\frac1{4!}g_{ijkl}\phi_i\phi_j\phi_k\phi_l$. In \cite{Fei:2016sgs}, the fermion is normalized as $\bar\Psi\slashed\partial\Psi$ and the Yukawa interaction is $\phi_i\,\bar\Psi\,h_i\,\Psi$. Rescaling $\Psi\to\Psi/\sqrt2$ thus takes both to the normalization of \eqref{eq:LGNY}. The coefficients of the formulas below, however, are the same in the two normalizations. Since a closed fermion loop contains as many vertices as propagators, the factor $2$ from each propagator cancels the $\frac12$ from each vertex. Each closed loop also contains the trace $\operatorname{tr}\mathbf1=2$ of the two-component spinors and the factor $\frac12$ of a Majorana loop relative to a Dirac loop. The net factor per closed loop is thus one.\footnote{The factor of one is one quarter of the Dirac trace. It corresponds to $N_f=\frac14$ Dirac fermions, or $N=4N_f=1$ two-component Majorana fermions, the case in which the GNY model is supersymmetric \cite{Fei:2016sgs,Giombi:2017rhm}.} We stress that the traces used below are therefore flavor traces alone. With this convention, the scalar and fermion anomalous dimensions to two
loops are
\begin{equation}
    \begin{aligned}
        \gamma_{\phi,ij}&=\frac12\,\mathrm{tr}(h_ih_j)+\frac1{12}\,g_{iklm}g_{jklm}-\frac34\,\mathrm{tr}(h_ih_jh_kh_k)-\frac12\,\mathrm{tr}(h_ih_kh_jh_k)\,,\\
        \gamma_\Psi&=\frac12h_ih_i-\frac18h_ih_jh_jh_i-\frac38\mathrm{tr}(h_ih_j)\,h_ih_j\,,
    \end{aligned}
\end{equation}
in units where $16\pi^2=1$. At the fixed point of Section~\ref{sub:fp}, they give
\eqref{eq:gammas}.

In the same units, the Yukawa beta function is
\begin{equation}\label{eq:betaY}
    \beta_{h_i}=-\frac\eps2\,h_i+\gamma_{\phi,ij}h_j+\gamma_\Psi h_i+h_i\gamma_\Psi+V_i\,,
\end{equation}
where the wave-function terms are built from the anomalous dimensions above and the vertex part is
\begin{equation}
    \begin{aligned}
        V_i&=2\,h_jh_ih_j-2\,g_{ijkl}h_jh_kh_l-h_jh_ih_k\mathrm{tr}(h_jh_k)\\
        &-h_jh_ih_kh_kh_j-h_jh_kh_kh_ih_j -2\,h_jh_kh_ih_kh_j+2\,h_jh_kh_ih_jh_k\,.
    \end{aligned}
\end{equation}

The quartic beta function has the same form,
\begin{equation}\label{eq:betalam}
    \beta_{g_{ijkl}}=-\eps\,g_{ijkl}+(\gamma_{\phi,im}g_{mjkl}+\text{3 perm.})+V_{ijkl}\,.
\end{equation}
In its vertex part, $\sum_{\rm part}$ runs over the three ways to split $(ijkl)$ into two pairs $(ij)(kl)$, and $\sum_{\rm pair}$ over the six unordered pairs $(ij)$, with $(kl)$ the remaining pair. The sum $\sum_{\rm cyc}$ runs over the three cyclic orderings of $(jkl)$, and $\sum_{S_4}$ over all twenty-four
permutations. Then
\begin{equation}
    \begin{aligned}
        V_{ijkl}&=\sum_{\rm part}g_{ijmn}g_{klmn}-4\sum_{\rm cyc}\mathrm{tr}(h_ih_jh_kh_l)-\sum_{\rm pair}g_{ijmn}g_{kmpq}g_{lnpq}\\
        &-\sum_{\rm part}g_{ijmn}g_{klmp}\,\mathrm{tr}(h_nh_p)+2\sum_{\rm pair}g_{ijmn}\,\mathrm{tr}(h_mh_kh_nh_l)\\
        &+\sum_{S_4}(\mathrm{tr}(h_ih_mh_mh_jh_kh_l) +2\,\mathrm{tr}(h_ih_mh_jh_kh_lh_m) +\mathrm{tr}(h_ih_jh_mh_kh_lh_m))\,.
    \end{aligned}
\end{equation}

The quadratic composites $\mathcal O_{ij}=\phi_i\phi_j$ mix among themselves. Their anomalous-dimension matrix is the sum of the wave-function terms and a vertex part. To two loops, the vertex part is \cite{Pernici:1999kk,Fei:2016sgs}
\begin{equation}\label{eq:kernel}
    M_{ij,kl}=g_{ijkl}-g_{ikmn}g_{jlmn}-\mathrm{tr}(h_lh_m)g_{ijkm}+2\,\mathrm{tr}(h_ih_kh_jh_l)\,.
\end{equation}
To obtain the full anomalous-dimension matrix, let us define $W_{ij,kl}$ as the matrix acting on the coefficients $v_{kl}$ of a combination $\sum_{k,l}v_{kl}\,\mathcal O_{kl}$, so that the combination has dimension $d-2+\gamma$ when $\sum_{k,l}W_{ij,kl}\,v_{kl}=\gamma\,v_{ij}$. Here $k$ and $l$ are summed independently and a composite with $k\neq l$ is counted twice. Both index pairs of $W$ are symmetric, because $\mathcal O_{ij}=\mathcal O_{ji}$. Only $M_{ij,kl}$ symmetrized over both index pairs contributes. The symmetrized matrix
\begin{equation}\label{eq:Wkernel}
    W_{ij,kl}=\frac12\bigl(\gamma_{\phi,ik}\delta_{jl}+\gamma_{\phi,il}\delta_{jk}+\gamma_{\phi,jk}\delta_{il}+\gamma_{\phi,jl}\delta_{ik}\bigr)+\frac14\sum_{\substack{i\leftrightarrow j\\ k\leftrightarrow l}}M_{ij,kl}
\end{equation}
equals the anomalous-dimension matrix obtained from the spectator scalar in Section~\ref{sub:deg}. The equality holds identically in all six couplings. At the fixed point, $W_{ij,kl}$ gives \eqref{eq:gammaP}.

\bibliographystyle{JHEP}
\bibliography{refs}

\end{document}